%
%
%
\def\unredoffs{} \def\redoffs{\voffset=-.31truein\hoffset=-.48truein}
\def\speclscape{}
%
%
%
%
%
\newbox\leftpage \newdimen\fullhsize \newdimen\hstitle \newdimen\hsbody
\tolerance=1000\hfuzz=2pt
\catcode`\@=11 
\ifx\hyperdef\UNd@FiNeD\def\hyperdef#1#2#3#4{#4}\def\hyperref#1#2#3#4{#4}\fi
\def\bigans{b }
\def\answ{b }
%
\ifx\answ\bigans\message{(This will come out unreduced.}
\magnification=1200\unredoffs\baselineskip=16pt plus 2pt minus 1pt
\hsbody=\hsize \hstitle=\hsize 
\else\message{(This will be reduced.} \let\l@r=L
\magnification=1000\baselineskip=16pt plus 2pt minus 1pt \vsize=7truein
\redoffs \hstitle=8truein\hsbody=4.75truein\fullhsize=10truein\hsize=\hsbody
\output={\ifnum\pageno=0 
  \shipout\vbox{\speclscape{\hsize\fullhsize\makeheadline}
    \hbox to \fullhsize{\hfill\pagebody\hfill}}\advancepageno
  \else
  \almostshipout{\leftline{\vbox{\pagebody\makefootline}}}\advancepageno
  \fi}
\def\almostshipout#1{\if L\l@r \count1=1 \message{[\the\count0.\the\count1]}
      \global\setbox\leftpage=#1 \global\let\l@r=R
 \else \count1=2
  \shipout\vbox{\speclscape{\hsize\fullhsize\makeheadline}
      \hbox to\fullhsize{\box\leftpage\hfil#1}}  \global\let\l@r=L\fi}
\fi
%
\newcount\yearltd\yearltd=\year\advance\yearltd by -1900

\def\Title#1#2{\nopagenumbers\abstractfont\hsize=\hstitle\rightline{#1}%
\vskip 1in\centerline{\titlefont #2}\abstractfont\vskip .5in\pageno=0}
\def\Date#1{\vfill\leftline{#1}\tenpoint\supereject\global\hsize=\hsbody%
\footline={\hss\tenrm\hyperdef\hypernoname{page}\folio\folio\hss}}%
%

\def\draftmode{\message{ DRAFTMODE }\def\draftdate{{\rm preliminary draft:
\number\month/\number\day/\number\yearltd\ \ \hourmin}}%
\headline={\hfil\draftdate}\writelabels\baselineskip=20pt plus 2pt minus 2pt
 {\count255=\time\divide\count255 by 60 \xdef\hourmin{\number\count255}
  \multiply\count255 by-60\advance\count255 by\time
  \xdef\hourmin{\hourmin:\ifnum\count255<10 0\fi\the\count255}}}
\def\nolabels{\def\wrlabeL##1{}\def\eqlabeL##1{}\def\reflabeL##1{}}
\def\writelabels{\def\wrlabeL##1{\leavevmode\vadjust{\rlap{\smash%
{\line{{\escapechar=` \hfill\rlap{\sevenrm\hskip.03in\string##1}}}}}}}%
\def\eqlabeL##1{{\escapechar-1\rlap{\sevenrm\hskip.05in\string##1}}}%
\def\reflabeL##1{\noexpand\llap{\noexpand\sevenrm\string\string\string##1}}}
\nolabels
%
\global\newcount\secno \global\secno=0
\global\newcount\meqno \global\meqno=1
\def\s@csym{}
\def\newsec#1{\global\advance\secno by1%
{\toks0{#1}\message{(\the\secno. \the\toks0)}}%
\global\subsecno=0\eqnres@t\let\s@csym\secsym\xdef\secn@m{\the\secno}\noindent
{\bf\hyperdef\hypernoname{section}{\the\secno}{\the\secno.} #1}%
\writetoca{{\string\hyperref{}{section}{\the\secno}{\the\secno.}} {#1}}%
\par\nobreak\medskip\nobreak}
\def\eqnres@t{\xdef\secsym{\the\secno.}\global\meqno=1\bigbreak\bigskip}
\def\sequentialequations{\def\eqnres@t{\bigbreak}}\xdef\secsym{}
\global\newcount\subsecno \global\subsecno=0
\def\subsec#1{\global\advance\subsecno by1%
{\toks0{#1}\message{(\s@csym\the\subsecno. \the\toks0)}}%
\ifnum\lastpenalty>9000\else\bigbreak\fi
\noindent{\it\hyperdef\hypernoname{subsection}{\secn@m.\the\subsecno}%
{\secn@m.\the\subsecno.} #1}\writetoca{\string\quad
{\string\hyperref{}{subsection}{\secn@m.\the\subsecno}{\secn@m.\the\subsecno.}}
{#1}}\par\nobreak\medskip\nobreak}
\def\appendix#1#2{\global\meqno=1\global\subsecno=0\xdef\secsym{\hbox{#1.}}%
\bigbreak\bigskip\noindent{\bf Appendix \hyperdef\hypernoname{appendix}{#1}%
{#1.} #2}{\toks0{(#1. #2)}\message{\the\toks0}}%
\xdef\s@csym{#1.}\xdef\secn@m{#1}%
\writetoca{\string\hyperref{}{appendix}{#1}{Appendix {#1.}} {#2}}%
\par\nobreak\medskip\nobreak}
%
%
\def\checkm@de#1#2{\ifmmode{\def\f@rst##1{##1}\hyperdef\hypernoname{equation}%
{#1}{#2}}\else\hyperref{}{equation}{#1}{#2}\fi}
\def\eqnn#1{\DefWarn#1\xdef #1{(\noexpand\relax\noexpand\checkm@de%
{\s@csym\the\meqno}{\secsym\the\meqno})}%
\wrlabeL#1\writedef{#1\leftbracket#1}\global\advance\meqno by1}
\def\f@rst#1{\c@t#1a\em@ark}\def\c@t#1#2\em@ark{#1}
\def\eqna#1{\DefWarn#1\wrlabeL{#1$\{\}$}%
\xdef #1##1{(\noexpand\relax\noexpand\checkm@de%
{\s@csym\the\meqno\noexpand\f@rst{##1}}{\hbox{$\secsym\the\meqno##1$}})}
\writedef{#1\numbersign1\leftbracket#1{\numbersign1}}\global\advance\meqno by1}
\def\eqn#1#2{\DefWarn#1%
\xdef #1{(\noexpand\hyperref{}{equation}{\s@csym\the\meqno}%
{\secsym\the\meqno})}$$#2\eqno(\hyperdef\hypernoname{equation}%
{\s@csym\the\meqno}{\secsym\the\meqno})\eqlabeL#1$$%
\writedef{#1\leftbracket#1}\global\advance\meqno by1}
\def\xeqn{\expandafter\xe@n}\def\xe@n(#1){#1}
\def\xeqna#1{\expandafter\xe@n#1}
\def\eqns#1{(\e@ns #1{\hbox{}})}
\def\e@ns#1{\ifx\UNd@FiNeD#1\message{eqnlabel \string#1 is undefined.}%
\xdef#1{(?.?)}\fi{\let\hyperref=\relax\xdef\next{#1}}%
\ifx\next\em@rk\def\next{}\else%
\ifx\next#1\xeqn#1\else\def\n@xt{#1}\ifx\n@xt\next#1\else\xeqna#1\fi
\fi\let\next=\e@ns\fi\next}

\def\DefWarn#1{\ifx\UNd@FiNeD#1\else
\immediate\write16{*** WARNING: the label \string#1 is already defined ***}\fi}
%
\newskip\footskip\footskip14pt plus 1pt minus 1pt 
\def\footnotefont{\ninepoint}\def\f@t#1{\footnotefont #1\@foot}
\def\f@@t{\baselineskip\footskip\bgroup\footnotefont\aftergroup\@foot\let\next}
\setbox\strutbox=\hbox{\vrule height9.5pt depth4.5pt width0pt}
\global\newcount\ftno \global\ftno=0
\def\foot{\global\advance\ftno by1\def\foot@rg{\hyperref{}{footnote}%
{\the\ftno}{\the\ftno}\xdef\foot@rg{\noexpand\hyperdef\noexpand\hypernoname%
{footnote}{\the\ftno}{\the\ftno}}}\footnote{$^{\foot@rg}$}}
%
\newwrite\ftfile
\def\footend{\def\foot{\global\advance\ftno by1\chardef\wfile=\ftfile
\hyperref{}{footnote}{\the\ftno}{$^{\the\ftno}$}%
\ifnum\ftno=1\immediate\openout\ftfile=\jobname.fts\fi%
\immediate\write\ftfile{\noexpand\smallskip%
\noexpand\item{\noexpand\hyperdef\noexpand\hypernoname{footnote}
{\the\ftno}{f\the\ftno}:\ }\pctsign}\findarg}%
\def\footatend{\vfill\eject\immediate\closeout\ftfile{\parindent=20pt
\centerline{\bf Footnotes}\nobreak\bigskip\input \jobname.fts }}}
\def\footatend{}
%
%
\global\newcount\refno \global\refno=1
\newwrite\rfile
\def\ref{[\hyperref{}{reference}{\the\refno}{\the\refno}]\nref}
\def\nref#1{\DefWarn#1%
\xdef#1{[\noexpand\hyperref{}{reference}{\the\refno}{\the\refno}]}%
\writedef{#1\leftbracket#1}%
\ifnum\refno=1\immediate\openout\rfile=\jobname.refs\fi
\chardef\wfile=\rfile\immediate\write\rfile{\noexpand\item{[\noexpand\hyperdef%
\noexpand\hypernoname{reference}{\the\refno}{\the\refno}]\ }%
\reflabeL{#1\hskip.31in}\pctsign}\global\advance\refno by1\findarg}
\def\findarg#1#{\begingroup\obeylines\newlinechar=`\^^M\pass@rg}
{\obeylines\gdef\pass@rg#1{\writ@line\relax #1^^M\hbox{}^^M}%
\gdef\writ@line#1^^M{\expandafter\toks0\expandafter{\striprel@x #1}%
\edef\next{\the\toks0}\ifx\next\em@rk\let\next=\endgroup\else\ifx\next\empty%
\else\immediate\write\wfile{\the\toks0}\fi\let\next=\writ@line\fi\next\relax}}
\def\striprel@x#1{} \def\em@rk{\hbox{}}
\def\lref{\begingroup\obeylines\lr@f}
\def\lr@f#1#2{\DefWarn#1\gdef#1{\let#1=\UNd@FiNeD\ref#1{#2}}\endgroup\unskip}

\def\addref#1{\immediate\write\rfile{\noexpand\item{}#1}} 
\def\listrefs{\footatend\vfill\supereject\immediate\closeout\rfile\writestoppt
\baselineskip=\footskip\centerline{{\bf References}}\bigskip{\parindent=20pt%
\frenchspacing\escapechar=` \input \jobname.refs\vfill\eject}\nonfrenchspacing}
\def\startrefs#1{\immediate\openout\rfile=\jobname.refs\refno=#1}
\def\xref{\expandafter\xr@f}\def\xr@f[#1]{#1}
\def\refs#1{\count255=1[\r@fs #1{\hbox{}}]}
\def\r@fs#1{\ifx\UNd@FiNeD#1\message{reflabel \string#1 is undefined.}%
\nref#1{need to supply reference \string#1.}\fi%
\vphantom{\hphantom{#1}}{\let\hyperref=\relax\xdef\next{#1}}%
\ifx\next\em@rk\def\next{}%
\else\ifx\next#1\ifodd\count255\relax\xref#1\count255=0\fi%
\else#1\count255=1\fi\let\next=\r@fs\fi\next}
%

%
\newwrite\ffile\global\newcount\figno \global\figno=1
\def\fig{fig.~\hyperref{}{figure}{\the\figno}{\the\figno}\nfig}
\def\nfig#1{\DefWarn#1%
\xdef#1{fig.~\noexpand\hyperref{}{figure}{\the\figno}{\the\figno}}%
\writedef{#1\leftbracket fig.\noexpand~\xfig#1}%
\ifnum\figno=1\immediate\openout\ffile=\jobname.figs\fi\chardef\wfile=\ffile%
{\let\hyperref=\relax
\immediate\write\ffile{\noexpand\medskip\noexpand\item{Fig.\ %
\noexpand\hyperdef\noexpand\hypernoname{figure}{\the\figno}{\the\figno}. }
\reflabeL{#1\hskip.55in}\pctsign}}\global\advance\figno by1\findarg}
\def\listfigs{\vfill\eject\immediate\closeout\ffile{\parindent40pt
\baselineskip14pt\centerline{{\bf Figure Captions}}\nobreak\medskip
\escapechar=` \input \jobname.figs\vfill\eject}}
\def\xfig{\expandafter\xf@g}\def\xf@g fig.\penalty\@M\ {}
\def\figs#1{figs.~\f@gs #1{\hbox{}}}
\def\f@gs#1{{\let\hyperref=\relax\xdef\next{#1}}\ifx\next\em@rk\def\next{}\else
\ifx\next#1\xfig #1\else#1\fi\let\next=\f@gs\fi\next}
\def\figin{\epsfcheck\figin}\def\figins{\epsfcheck\figins}
\def\epsfcheck{\ifx\epsfbox\UNd@FiNeD
\message{(NO epsf.tex, FIGURES WILL BE IGNORED)}
\gdef\figin##1{\vskip2in}\gdef\figins##1{\hskip.5in}
\else\message{(FIGURES WILL BE INCLUDED)}%
\gdef\figin##1{##1}\gdef\figins##1{##1}\fi}
\def\DefWarn#1{}
\def\figinsert{\goodbreak\midinsert}
\def\ifig#1#2#3{\DefWarn#1\xdef#1{fig.~\noexpand\hyperref{}{figure}%
{\the\figno}{\the\figno}}\writedef{#1\leftbracket fig.\noexpand~\xfig#1}%
\figinsert\figin{\centerline{#3}}\medskip\centerline{\vbox{\baselineskip12pt
\advance\hsize by -1truein\noindent\wrlabeL{#1=#1}\footnotefont%
{\bf Fig.~\hyperdef\hypernoname{figure}{\the\figno}{\the\figno}:} #2}}
\bigskip\endinsert\global\advance\figno by1}
\newwrite\lfile
{\escapechar-1\xdef\pctsign{\string\%}\xdef\leftbracket{\string\{}
\xdef\rightbracket{\string\}}\xdef\numbersign{\string\#}}
\def\writedefs{\immediate\openout\lfile=\jobname.defs \def\writedef##1{%
{\let\hyperref=\relax\let\hyperdef=\relax\let\hypernoname=\relax
 \immediate\write\lfile{\string\def\string##1\rightbracket}}}}%
\def\writestop{\def\writestoppt{\immediate\write\lfile{\string\pageno
 \the\pageno\string\startrefs\leftbracket\the\refno\rightbracket
 \string\def\string\secsym\leftbracket\secsym\rightbracket
 \string\secno\the\secno\string\meqno\the\meqno}\immediate\closeout\lfile}}
\def\writestoppt{}\def\writedef#1{}
\def\seclab#1{\DefWarn#1%
\xdef #1{\noexpand\hyperref{}{section}{\the\secno}{\the\secno}}%
\writedef{#1\leftbracket#1}\wrlabeL{#1=#1}}
\def\subseclab#1{\DefWarn#1%
\xdef #1{\noexpand\hyperref{}{subsection}{\secn@m.\the\subsecno}%
{\secn@m.\the\subsecno}}\writedef{#1\leftbracket#1}\wrlabeL{#1=#1}}
\def\applab#1{\DefWarn#1%
\xdef #1{\noexpand\hyperref{}{appendix}{\secn@m}{\secn@m}}%
\writedef{#1\leftbracket#1}\wrlabeL{#1=#1}}
\newwrite\tfile \def\writetoca#1{}
\def\leaderfill{\leaders\hbox to 1em{\hss.\hss}\hfill}
\def\writetoc{\immediate\openout\tfile=\jobname.toc
   \def\writetoca##1{{\edef\next{\write\tfile{\noindent ##1
   \string\leaderfill {\string\hyperref{}{page}{\noexpand\number\pageno}%
                       {\noexpand\number\pageno}} \par}}\next}}}
\newread\ch@ckfile
\def\listtoc{\immediate\closeout\tfile\immediate\openin\ch@ckfile=\jobname.toc
\ifeof\ch@ckfile\message{no file \jobname.toc, no table of contents this pass}%
\else\closein\ch@ckfile\centerline{\bf Contents}\nobreak\medskip%
{\baselineskip=18pt
\parskip=2pt\catcode`\@=11\input\jobname.toc
\catcode`\@=12\bigbreak\bigskip}\fi}
\catcode`\@=12 
%
\edef\tfontsize{\ifx\answ\bigans scaled\magstep3\else scaled\magstep4\fi}
\font\titlerm=cmr10 \tfontsize \font\titlerms=cmr7 \tfontsize
\font\titlermss=cmr5 \tfontsize \font\titlei=cmmi10 \tfontsize
\font\titleis=cmmi7 \tfontsize \font\titleiss=cmmi5 \tfontsize
\font\titlesy=cmsy10 \tfontsize \font\titlesys=cmsy7 \tfontsize
\font\titlesyss=cmsy5 \tfontsize \font\titleit=cmti10 \tfontsize
\skewchar\titlei='177 \skewchar\titleis='177 \skewchar\titleiss='177
\skewchar\titlesy='60 \skewchar\titlesys='60 \skewchar\titlesyss='60
\def\titlefont{\def\rm{\fam0\titlerm}
\textfont0=\titlerm \scriptfont0=\titlerms \scriptscriptfont0=\titlermss
\textfont1=\titlei \scriptfont1=\titleis \scriptscriptfont1=\titleiss
\textfont2=\titlesy \scriptfont2=\titlesys \scriptscriptfont2=\titlesyss
\textfont\itfam=\titleit \def\it{\fam\itfam\titleit}\rm}
 \ifx\answ\bigans\else scaled\magstep1\fi
\ifx\answ\bigans\def\abstractfont{\tenpoint}\else
\font\absit=cmti10 scaled \magstep1
\font\abssl=cmsl10 scaled \magstep1
\font\absrm=cmr10 scaled\magstep1 \font\absrms=cmr7 scaled\magstep1
\font\absrmss=cmr5 scaled\magstep1 \font\absi=cmmi10 scaled\magstep1
\font\absis=cmmi7 scaled\magstep1 \font\absiss=cmmi5 scaled\magstep1
\font\abssy=cmsy10 scaled\magstep1 \font\abssys=cmsy7 scaled\magstep1
\font\abssyss=cmsy5 scaled\magstep1 \font\absbf=cmbx10 scaled\magstep1
\skewchar\absi='177 \skewchar\absis='177 \skewchar\absiss='177
\skewchar\abssy='60 \skewchar\abssys='60 \skewchar\abssyss='60
\def\abstractfont{\def\rm{\fam0\absrm}
\textfont0=\absrm \scriptfont0=\absrms \scriptscriptfont0=\absrmss
\textfont1=\absi \scriptfont1=\absis \scriptscriptfont1=\absiss
\textfont2=\abssy \scriptfont2=\abssys \scriptscriptfont2=\abssyss
\textfont\itfam=\absit \def\it{\fam\itfam\absit}\def\footnotefont{\tenpoint}%
\textfont\slfam=\abssl \def\sl{\fam\slfam\abssl}%
\textfont\bffam=\absbf \def\bf{\fam\bffam\absbf}\rm}\fi
\def\tenpoint{\def\rm{\fam0\tenrm}
\textfont0=\tenrm \scriptfont0=\sevenrm \scriptscriptfont0=\fiverm
\textfont1=\teni  \scriptfont1=\seveni  \scriptscriptfont1=\fivei
\textfont2=\tensy \scriptfont2=\sevensy \scriptscriptfont2=\fivesy
\textfont\itfam=\tenit \def\it{\fam\itfam\tenit}\def\footnotefont{\ninepoint}%
\textfont\bffam=\tenbf \def\bf{\fam\bffam\tenbf}\def\sl{\fam\slfam\tensl}\rm}
\font\ninerm=cmr9 \font\sixrm=cmr6 \font\ninei=cmmi9 \font\sixi=cmmi6
\font\ninesy=cmsy9 \font\sixsy=cmsy6 \font\ninebf=cmbx9
\font\nineit=cmti9 \font\ninesl=cmsl9 \skewchar\ninei='177
\skewchar\sixi='177 \skewchar\ninesy='60 \skewchar\sixsy='60
\def\ninepoint{\def\rm{\fam0\ninerm}
\textfont0=\ninerm \scriptfont0=\sixrm \scriptscriptfont0=\fiverm
\textfont1=\ninei \scriptfont1=\sixi \scriptscriptfont1=\fivei
\textfont2=\ninesy \scriptfont2=\sixsy \scriptscriptfont2=\fivesy
\textfont\itfam=\ninei \def\it{\fam\itfam\nineit}\def\sl{\fam\slfam\ninesl}%
\textfont\bffam=\ninebf \def\bf{\fam\bffam\ninebf}\rm}
%
%
\def\noblackbox{\overfullrule=0pt}
\hyphenation{anom-aly anom-alies coun-ter-term coun-ter-terms}
\def\inv{^{\raise.15ex\hbox{${\scriptscriptstyle -}$}\kern-.05em 1}}

\def\Dsl{\,\raise.15ex\hbox{/}\mkern-13.5mu D} 
\def\dsl{\raise.15ex\hbox{/}\kern-.57em\partial}

\def\lspace{\ifx\answ\bigans{}\else\qquad\fi}
\def\lbspace{\ifx\answ\bigans{}\else\hskip-.2in\fi} 
\def\boxeqn#1{\vcenter{\vbox{\hrule\hbox{\vrule\kern3pt\vbox{\kern3pt
	\hbox{${\displaystyle #1}$}\kern3pt}\kern3pt\vrule}\hrule}}}
\def\mbox#1#2{\vcenter{\hrule \hbox{\vrule height#2in
		\kern#1in \vrule} \hrule}}  
%

\def\darr#1{\raise1.5ex\hbox{$\leftrightarrow$}\mkern-16.5mu #1}

\def\roughly#1{\raise.3ex\hbox{$#1$\kern-.75em\lower1ex\hbox{$\sim$}}}

\input epsf
\noblackbox
\ifx\answ\bigans
\magnification=1200\baselineskip=14pt plus 2pt minus 1pt
\else\baselineskip=16pt 
\fi

\def\tb{type~$IIB$\ }
\def\ap{\alpha'}

\def\cf{{\it cf.\ }}
\def\ie{{\it i.e.\ }}
\def\eg{{\it e.g.\ }}
\def\eqq{{\it Eq.\ }}
\def\eqqs{{\it Eqs.\ }}
\def\th{\theta}

\def\al{\alpha}

\def\si{\sigma}
\def\Om{\Omega}

\def\Om{\Omega}

\def\CS{{\cal C}{\cal S}}
\newif\ifnref
\def\rrr#1#2{\relax\ifnref\nref#1{#2}\else\ref#1{#2}\fi}
\def\ldf#1#2{\begingroup\obeylines
\gdef#1{\rrr{#1}{#2}}\endgroup\unskip}
\def\nrf#1{\nreftrue{#1}\nreffalse}

\def\multrefv#1#2#3#4#5{\nrf{#1#2#3#4#5}\refs{#1{--}#5}}

\def\doubref#1#2{\refs{{#1},{#2} }}

\nreffalse

\def\lref{\ldf}

\newcount\figno
\figno=1
\def\fig#1#2#3{
\par\begingroup\parindent=0pt\leftskip=1cm\rightskip=1cm\parindent=0pt
\baselineskip=11pt \global\advance\figno by 1 \midinsert
\epsfxsize=#3 \centerline{\epsfbox{#2}} \vskip 12pt
\centerline{{\bf Figure \the\figno :}{\it ~~ #1}}\par
\endinsert\endgroup\par}
\def\figlabel#1{\xdef#1{\the\figno}}

\input epsf
\def\figin{\epsfcheck\figin}\def\figins{\epsfcheck\figins}
\def\epsfcheck{\ifx\epsfbox\UnDeFiNeD
\message{(NO epsf.tex, FIGURES WILL BE IGNORED)}
\gdef\figin##1{\vskip2in}\gdef\figins##1{\hskip.5in}
\else\message{(FIGURES WILL BE INCLUDED)}%
\gdef\figin##1{##1}\gdef\figins##1{##1}\fi}
\def\DefWarn#1{}
\def\figinsert{\goodbreak\midinsert}
\def\ifig#1#2#3{\DefWarn#1\xdef#1{ \the\figno}
\writedef{#1\leftbracket fig.\noexpand~\the\figno}%
\figinsert\figin{\centerline{#3}}\medskip\centerline{\vbox{\baselineskip12pt
\advance\hsize by -1truein\noindent\centerline{{\bf Fig.~\the\figno :~ } {\it #2}}}}
\bigskip\endinsert\global\advance\figno by1}

\def\appA{A}

\def\tilde{\widetilde}

\def\h {{1\over 2}}

\def\ov {\overline}
\def\o {\over}
\def\fc#1#2{{#1 \o #2}}

\def\IZ{ {\bf Z}}\def\IQ{{\bf Q}}
\def\IP{{\bf P}}\def\IC{{\bf C}}\def\IF{{\bf F}}
\def\IR{ {\bf R}}


\def\br{\hfill\break}

\def\det {{\rm det}}

\def\lf {\left}
\def\ri {\right}
\def\ra {\rightarrow}
\def\lra {\longrightarrow}
\def\re {{\rm Re}}
\def\im {{\rm Im}}
\def\p {\partial}

 \def\Dc {{\cal D}}

 \def\Gc{{\cal G}}
 \def\Oc {{\cal O}}

 \def\Jc {{\cal J}}
\def\Kc {{\cal K}} 

\lref\Mario{R. D'Auria, S.~Ferrara and M.~Trigiante,
``c-map,very special quaternionic geometry and dual K\"ahler spaces,''
  Phys.\ Lett.\ B {\bf 587}, 138 (2004)
  [arXiv:hep-th/0401161].
}

\lref\AM{I.~Antoniadis and T.~Maillard,
``Moduli stabilization from magnetic fluxes in type I string theory,''
  Nucl.\ Phys.\ B {\bf 716}, 3 (2005)
  [arXiv:hep-th/0412008].
}

\lref\GL{T.W.~Grimm and J.~Louis,
  ``The effective action of N = 1 Calabi-Yau orientifolds,''
  Nucl.\ Phys.\ B {\bf 699}, 387 (2004)
  [arXiv:hep-th/0403067];\br
T.W.~Grimm,
  ``The effective action of type II Calabi-Yau orientifolds,''
  Fortsch.\ Phys.\  {\bf 53}, 1179 (2005)
  [arXiv:hep-th/0507153]
}

\lref\JL{H.~Jockers and J.~Louis,
  ``The effective action of D7-branes in N = 1 Calabi-Yau orientifolds,''
  Nucl.\ Phys.\ B {\bf 705}, 167 (2005)
  [arXiv:hep-th/0409098];
``D-terms and F-terms from D7-brane fluxes,''
  Nucl.\ Phys.\ B {\bf 718}, 203 (2005)
  [arXiv:hep-th/0502059];\br
H.~Jockers,
 ``The effective action of D-branes in Calabi-Yau orientifold
  compactifications,''
  Fortsch.\ Phys.\  {\bf 53}, 1087 (2005)
  [arXiv:hep-th/0507042].
}

\lref\Linde{J.J.~Blanco-Pillado,  C.P. Burgess, J.M. Cline, C. Escoda, 
M. Gomez-Reino, R. Kallosh, A. Linde and F. Quevedo,
``Inflating in a better racetrack,''
  arXiv:hep-th/0603129.
}

\lref\Lars{L.~G\"orlich, S.~Kachru, P.K.~Tripathy and S.P.~Trivedi,
``Gaugino condensation and nonperturbative superpotentials in flux
compactifications,''
  JHEP {\bf 0412}, 074 (2004)
  [arXiv:hep-th/0407130].
}

\lref\LS{D.~L\"ust and S.~Stieberger,
``Gauge threshold corrections in intersecting brane world models,''
  arXiv:hep-th/0302221.
}

\lref\LMRS{D.~L\"ust, P.~Mayr, S.~Reffert and S.~Stieberger,
``F-theory flux, destabilization of orientifolds and soft terms on
D7-branes,''
  Nucl.\ Phys.\ B {\bf 732}, 243 (2006)
  [arXiv:hep-th/0501139].
}

\lref\Waldemar{Work in progress.}

\lref\Japan{H.~Abe, T.~Higaki and T.~Kobayashi,
``Remark on integrating out heavy moduli in flux compactification,''
  Phys.\ Rev.\ D {\bf 74}, 045012 (2006)
  [arXiv:hep-th/0606095];\br
H.X.~Yang,
``On moduli stabilization scheme in type IIB flux compactifications,''
  arXiv:hep-th/0608155;
``Moduli stabilization in type IIB flux compactifications,''
  Phys.\ Rev.\ D {\bf 73}, 066006 (2006)
  [arXiv:hep-th/0511030].
}

\lref\ChoiSX{
K.~Choi, A.~Falkowski, H.P.~Nilles, M.~Olechowski and S.~Pokorski,
``Stability of flux compactifications and the pattern of supersymmetry
  breaking,''
  JHEP {\bf 0411}, 076 (2004)
  [arXiv:hep-th/0411066].
}

\lref\LRSi{ D.~L\"ust, S.~Reffert and S.~Stieberger,
``Flux-induced soft supersymmetry breaking in chiral type IIb  orientifolds
with D3/D7-branes,''
  Nucl.\ Phys.\ B {\bf 706}, 3 (2005)
  [arXiv:hep-th/0406092]
}

\lref\IBANEZ{P.G.~Camara, L.E.~Ibanez and A.M.~Uranga,
``Flux-induced SUSY-breaking soft terms on D7-D3 brane systems,''
Nucl.\ Phys.\ B {\bf 708}, 268 (2005)
  [arXiv:hep-th/0408036].
}

\lref\LRSii{D.~L\"ust, S.~Reffert and S.~Stieberger,
``MSSM with soft SUSY breaking terms from D7-branes with fluxes,''
  Nucl.\ Phys.\ B {\bf 727}, 264 (2005)
  [arXiv:hep-th/0410074].
}

\lref\HPNii{K.~Choi, A.~Falkowski, H.P.~Nilles and M.~Olechowski,
``Soft supersymmetry breaking in KKLT flux compactification,''
  Nucl.\ Phys.\ B {\bf 718}, 113 (2005)
  [arXiv:hep-th/0503216].
}

\lref\HB{
  I.~Brunner and K.~Hori,
  ``Orientifolds and mirror symmetry,''
  JHEP {\bf 0411}, 005 (2004)
  [arXiv:hep-th/0303135].
}

\lref\DHVW{L.J.~Dixon, J.A.~Harvey, C.~Vafa and E.~Witten,
``Strings On Orbifolds,''
Nucl.\ Phys.\ B {\bf 261}, 678 (1985);
``Strings On Orbifolds. 2,''
Nucl.\ Phys.\ B {\bf 274}, 285 (1986).
}

\lref\DSW{
M.~Dine, N.~Seiberg and E.~Witten,
``Fayet-Iliopoulos Terms In String Theory,''
  Nucl.\ Phys.\ B {\bf 289}, 589 (1987).
}

\lref\ORI{B.~Acharya, M.~Aganagic, K.~Hori and C.~Vafa,
``Orientifolds, mirror symmetry and superpotentials,''
  arXiv:hep-th/0202208.
}

\lref\KKLT{S.~Kachru, R.~Kallosh, A.~Linde and S.~P.~Trivedi,
``De Sitter vacua in string theory,''
  Phys.\ Rev.\ D {\bf 68}, 046005 (2003)
  [arXiv:hep-th/0301240].
}

\lref\Lukas{B.~de Carlos, S.~Gurrieri, A.~Lukas and A.~Micu,
``Moduli stabilisation in heterotic string compactifications,''
  JHEP {\bf 0603}, 005 (2006)
  [arXiv:hep-th/0507173].
}

\lref\Duff{M.J.~Duff, B.E.W.~Nilsson and C.N.~Pope,
``Kaluza-Klein Supergravity,''
  Phys.\ Rept.\  {\bf 130}, 1 (1986).
}

\lref\Race{F.~Denef, M.R.~Douglas and B.~Florea,
``Building a better racetrack,''
  JHEP {\bf 0406}, 034 (2004)
  [arXiv:hep-th/0404257].
}

\lref\KachruA{O.~DeWolfe, A.~Giryavets, S.~Kachru and W.~Taylor,
 ``Enumerating flux vacua with enhanced symmetries,''
  JHEP {\bf 0502}, 037 (2005)
  [arXiv:hep-th/0411061].
}

\lref\first{D.~L\"ust, S.~Reffert, E.~Scheidegger, S.~Stieberger
``Resolved Toroidal Orbifolds and their Orientifolds,'' arXiv:hep-th/0609014.
}

\lref\Cand{P.~Candelas and X.~de la Ossa,
``Moduli Space Of Calabi-Yau Manifolds,''
  Nucl.\ Phys.\ B {\bf 355}, 455 (1991).
}

\lref\Rabadan{M.~Klein and R.~Rabadan,
``Z(N) x Z(M) orientifolds with and without discrete torsion,''
  JHEP {\bf 0010}, 049 (2000)
  [arXiv:hep-th/0008173];
``D = 4, N = 1 orientifolds with vector structure,''
  Nucl.\ Phys.\ B {\bf 596}, 197 (2001)
  [arXiv:hep-th/0007087];
``Orientifolds with discrete torsion,''
  JHEP {\bf 0007}, 040 (2000)
  [arXiv:hep-th/0002103].
}

\lref\FM{J.~Gomis, F.~Marchesano and D.~Mateos,
``An open string landscape,''
  JHEP {\bf 0511}, 021 (2005)
  [arXiv:hep-th/0506179].
}

\lref\BF{
P.~Breitenlohner and D.Z.~Freedman,
``Stability In Gauged Extended Supergravity,''
Annals Phys.\  {\bf 144}, 249 (1982).
}

\lref\Aspinwall{
  P.S.~Aspinwall and R.~Kallosh,
``Fixing all moduli for M-theory on K3 x K3,''
  JHEP {\bf 0510}, 001 (2005)
  [arXiv:hep-th/0506014].
}

\lref\LRSS{D.~L\"ust, S.~Reffert, W.~Schulgin and S.~Stieberger,
``Moduli stabilization in type IIB orientifolds. I,''
  arXiv:hep-th/0506090.
}

\lref\ReffertMN{S.~Reffert and E.~Scheidegger,
  ``Moduli stabilization in toroidal type IIB orientifolds,''
  Fortsch. \ Phys. \ {\bf 54}, 462 (2006)
  [arXiv:hep-th/0512227].
}

\lref\Getting{L.E.~Ibanez, F.~Marchesano and R.~Rabadan,
``Getting just the standard model at intersecting branes,''
  JHEP {\bf 0111}, 002 (2001)
  [arXiv:hep-th/0105155].
}

\lref\Conlon{J.P.~Conlon,
``The QCD axion and moduli stabilisation,''
  arXiv:hep-th/0602233.
}

\lref\dealwis{S.~P.~de Alwis,
``Effective potentials for light moduli,''
  Phys.\ Lett.\ B {\bf 626}, 223 (2005)
  [arXiv:hep-th/0506266].
}

\lref\DenefMM{
F.~Denef, M.R.~Douglas, B.~Florea, A.~Grassi and S.~Kachru,
``Fixing All Moduli in a Simple F-Theory Compactification,''
arXiv:hep-th/0503124.
}

\lref\AFIV{G.~Aldazabal, A.~Font, L.E.~Ibanez and G.~Violero,
``D = 4, N = 1, type IIB orientifolds,''
Nucl.\ Phys.\ B {\bf 536}, 29 (1998)
[arXiv:hep-th/9804026].
}

\lref\zwart{G.~Zwart,
``Four-dimensional N = 1 Z(N) x Z(M) orientifolds,''
  Nucl.\ Phys.\ B {\bf 526}, 378 (1998)
  [arXiv:hep-th/9708040].
}

\lref\HORN{R.A. Horn and C.R. Johnson
``Matrix Analysis'', Cambridge University Press, 1990.}

\lref\KT{A.K.~Kashani-Poor and A.~Tomasiello,
``A stringy test of flux-induced isometry gauging,''
  Nucl.\ Phys.\ B {\bf 728}, 135 (2005)
  [arXiv:hep-th/0505208].
}

\lref\FerraraQS{
  S.~Ferrara, L.~Girardello and H.P.~Nilles,
  ``Breakdown Of Local Supersymmetry Through Gauge Fermion Condensates,''
  Phys.\ Lett.\ B {\bf 125}, 457 (1983);\br
M.~Dine, R.~Rohm, N.~Seiberg and E.~Witten,
  ``Gluino Condensation In Superstring Models,''
  Phys.\ Lett.\ B {\bf 156}, 55 (1985).
}

\lref\WittenBN{
  E.~Witten,
  ``Non-Perturbative Superpotentials In String Theory,''
  Nucl.\ Phys.\ B {\bf 474}, 343 (1996)
  [arXiv:hep-th/9604030].
}

\lref\KalloshYU{
  R.~Kallosh and D.~Sorokin,
  ``Dirac action on M5 and M2 branes with bulk fluxes,''
  JHEP {\bf 0505}, 005 (2005)
  [arXiv:hep-th/0501081].
}

\lref\SaulinaVE{
  N.~Saulina,
  ``Topological constraints on stabilized flux vacua,''
  Nucl.\ Phys.\ B {\bf 720}, 203 (2005)
  [arXiv:hep-th/0503125].
}

\lref\KalloshGS{
  R.~Kallosh, A.K.~Kashani-Poor and A.~Tomasiello,
  ``Counting fermionic zero modes on M5 with fluxes,''
  JHEP {\bf 0506}, 069 (2005)
  [arXiv:hep-th/0503138].
}

\lref\BergshoeffYP{
  E.~Bergshoeff, R.~Kallosh, A.K.~Kashani-Poor, D.~Sorokin and A.~Tomasiello,
  ``An index for the Dirac operator on D3 branes with background fluxes,''
  JHEP {\bf 0510}, 102 (2005)
  [arXiv:hep-th/0507069].
}

\lref\ParkHJ{
  J.~Park,
  ``D3 instantons in Calabi-Yau orientifolds with(out) fluxes,''
  arXiv:hep-th/0507091.
}

\lref\LustCU{
  D.~L\"ust, S.~Reffert, W.~Schulgin and P.K.~Tripathy,
  ``Fermion zero modes in the presence of fluxes and a non-perturbative
  superpotential,''
  arXiv:hep-th/0509082.
}

\lref\MartucciRB{
  L.~Martucci, J.~Rosseel, D.~Van den Bleeken and A.~Van Proeyen,
  ``Dirac actions for D-branes on backgrounds with fluxes,''
  Class.\ Quant.\ Grav.\  {\bf 22}, 2745 (2005)
  [arXiv:hep-th/0504041].
}

\lref\CIM{
  D.~Cremades, L.E.~Ibanez and F.~Marchesano,
``SUSY quivers, intersecting branes and the modest hierarchy problem,''
  JHEP {\bf 0207}, 009 (2002)
  [arXiv:hep-th/0201205].
}

\lref\MS{P.~Mayr and S.~Stieberger,
``Threshold corrections to gauge couplings in orbifold compactifications,''
  Nucl.\ Phys.\ B {\bf 407}, 725 (1993)
  [arXiv:hep-th/9303017].
}

\lref\MB{P.~Berglund and P.~Mayr,
``Non-perturbative superpotentials in $F$-theory and string duality,''
  arXiv:hep-th/0504058.
}

\lref\GVW{S.~Gukov, C.~Vafa and E.~Witten,
``CFT's from Calabi-Yau four-folds,''
  Nucl.\ Phys.\ B {\bf 584}, 69 (2000)
  [Erratum-ibid.\ B {\bf 608}, 477 (2001)]
  [arXiv:hep-th/9906070];\br
T.R.~Taylor and C.~Vafa,
``RR flux on Calabi-Yau and partial supersymmetry breaking,''
  Phys.\ Lett.\ B {\bf 474}, 130 (2000)
  [arXiv:hep-th/9912152];\br
  P.~Mayr,
``On supersymmetry breaking in string theory and its realization in brane
worlds,''
Nucl.\ Phys.\ B {\bf 593}, 99 (2001)
  [arXiv:hep-th/0003198].
}

\lref\Ruben{
  M.~Marino, R.~Minasian, G.W.~Moore and A.~Strominger,
``Nonlinear instantons from supersymmetric p-branes,''
  JHEP {\bf 0001}, 005 (2000)
  [arXiv:hep-th/9911206].
}

\lref\BurgessIC{
  C.P.~Burgess, R.~Kallosh and F.~Quevedo,
  ``de Sitter string vacua from supersymmetric D-terms,''
  JHEP {\bf 0310}, 056 (2003)
  [arXiv:hep-th/0309187].
}

\lref\GmeinerVZ{
  F.~Gmeiner, R.~Blumenhagen, G.~Honecker, D.~L\"ust and T.~Weigand,
  ``One in a billion: MSSM-like D-brane statistics,''
  JHEP {\bf 0601}, 004 (2006)
  [arXiv:hep-th/0510170].
}

\lref\GiddingsYU{
  S.~B.~Giddings, S.~Kachru and J.~Polchinski,
  ``Hierarchies from fluxes in string compactifications,''
  Phys.\ Rev.\ D {\bf 66}, 106006 (2002)
  [arXiv:hep-th/0105097].
}

\lref\AnguelovaSJ{
L.~Anguelova and K.~Zoubos,
``Five-brane instantons vs flux-induced gauging of isometries,''
arXiv:hep-th/0606271.
}

\lref\BandosWB{
  I.~Bandos and D.~Sorokin,
``Aspects of D-brane dynamics in supergravity backgrounds with fluxes,
kappa-symmetry and equations of motion. IIB,''
  arXiv:hep-th/0607163.
}


\Title{\vbox{\rightline{LMU--ASC 56/06}
\rightline{MPP--2006--110}
\rightline{DISTA--UPO--06}
\rightline{\tt hep-th/0609013}}}
{\vbox{\centerline{Moduli Stabilization in Type $IIB$ Orientifolds (II)}
}}
\smallskip
\centerline{D. L\"ust$^{a,b}$,\ \ S. Reffert$^b$,\ \ E. Scheidegger$^c$,\ \ 
W. Schulgin$^b$\ \ and\ \ S. Stieberger$^a$}
\bigskip
\centerline{\it $^a$ Arnold--Sommerfeld--Center for Theoretical Physics,}
\centerline{\it  Department f\"ur Physik, Ludwig--Maximilians--Universit\"at M\"unchen,}
\centerline{\it Theresienstra\ss e 37, 80333 M\"unchen, Germany}
\vskip7pt
\centerline{\it $^b$ Max--Planck--Institut f\"ur Physik,}
\centerline{\it F\"ohringer Ring 6, 80805 M\"unchen, Germany}
\vskip7pt
\centerline{\it $^c$ Universita del Piemonte Orientale "A. Avogadro",}
\centerline{\it  Dip. Scienze e Tecnologie Avanzate,}
\centerline{\it Via Bellini, 25/g, 15100 Alessandria, Italy}
\centerline{\it INFN -- Sezione di Torino, Italy}
\bigskip
\centerline{\bf Abstract}
\vskip .2in
\noindent

We discuss general properties of moduli stablization in KKLT scenarios 
in type $IIB$ orientifold compactifications $X_6$.
In particular, we find conditions for the K\"ahler potential 
to allow a KKLT scenario for a manifold $X_6$ 
without complex structure moduli, \ie $h_{(2,1)}^{(-)}(X_6)=0$.
This way, a whole class of \tb orientifolds with $h_{(2,1)}^{(-)}(X_6)=0$ 
is ruled out.
This excludes in particular all $\IZ_N$--and $\IZ_N\times \IZ_M$--orientifolds $X_6$ 
with $h_{(2,1)}(X_6)=0$ for a KKLT scenario. This concerns
$\IZ_3, \IZ_7, \IZ_3\times \IZ_3, \IZ_4\times \IZ_4,
\IZ_6\times \IZ_6$
and $\IZ_2\times \IZ_{6'}$ --both at the orbifold point and away from it.
Furthermore, we propose a mechanism to stabilize the Kaehler moduli accociated 
to the odd cohomology $H^{(1,1)}_-(X_6)$.

In the second part of this work we discuss the moduli stabilization of 
resolved \tb $\IZ_N$-- or $\IZ_N\times \IZ_M$--orbifold/orientifold  
compactifications.
As examples for the resolved $\IZ_6$ and $\IZ_2 \times \IZ_4$ orbifolds we fix all moduli  
through a combination of fluxes and racetrack superpotential.

\Date{}
\noindent

\goodbreak

\listtoc 
\writetoc
\break

\newsec{Introduction}

As it is known already for some time, the landscape of
consistent string compactifications is vast and contains many possible solutions.
In particular, a large class of lower-dimensional
string vacua is formed by four--dimensional, supersymmetric
string compactifications containing massless scalar fields with arbitrary
vacuum expectation values in the low energy
effective field theory. These fields
are in one-to-one correspondence with the continuous
parameters of the underlying background geometry, and
hence they are called moduli fields. 
However, massless moduli are unacceptable from the
phenomenological point of view, since such fields give
rise to a fifth force of about gravitational strength.
Hence mechanisms for stabilizing moduli in string theory
must be explored. Getting rid of moduli means that,
seen from a stringy or geometrical point of view,
we are looking for internal conformal field theories resp. for  background 
spaces which are rigid, \ie do not allow any continuous
deformations. 
On the other hand, within the effective field theory, moduli stabilization is
achieved by creating an effective potential that lifts all flat directions
of the scalar fields.

To arrive at the goal of constructing string vacua with a tiny, positive
de Sitter (dS) cosmological constant, 
$\Lambda_{dS}/M_{\rm Planck}^4\sim 10^{-120}$,
and without massless moduli, the following two-step procedure,
proposed by KKLT in the context of type \tb orientifold
compactifications, seems to be a promising scenario \KKLT:
in the first step one constructs compactifications that lead
to four-dimensional, N=1 supersymmetric anti--de Sitter (AdS) string
vacua with all moduli fixed in the minimum of the $F$-term effective scalar potential.
Here the value of the negative AdS cosmological constant is
of the order of the phenomenologically required scale for supersymmetry breaking
in the supersymmetric Standard Model (SSM), \ie $|\Lambda_{AdS}|\sim {\cal O}(1TeV)^4$.
This method has the advantage that supersymmetric AdS-vacua with stabilized moduli
are much easier to be constructed than supersymmetric Minkowski vacua with stabilized  moduli
fields.
The second step of the KKLT scenario is to break supersymmetry 
by uplifting the AdS minimum to a 
dS vacuum with a tiny positive  cosmological constant.
Possible mechanisms for this uplift are either an explicit supersymmetry breaking
by adding anti $D$-branes or spontaneous supersymmetry breaking
by additional $D$-term contributions \BurgessIC\ to the effective scalar potential.
Although many details of the uplift procedure still have to be worked
out in concrete models, this proposal is attractive since the observed 
supersymmetry breaking scale in the SSM, like
the gravitino mass, is set already by the 
first step of the KKLT procedure, and hence is robust against
the details of the uplift. The second virtue of this scheme is that
the masses of the moduli are also quite insensitive against many details
of the uplift, provided that the (mass)$^2$ matrix $M_{ij}^2$ of the moduli is positive
definite already in the AdS minimum of the effective potential, and that
its eigenvalues are large compared to $|\Lambda_{AdS}|$.
As we will see, the requirement of positive definiteness of $M_{ij}^2$ can
fail in explicit string compactifications.
On the other hand, the KKLT scenario requires a certain amount of fine tuning
of parameters, in particular to achieve the necessary hierarchy of
scales $\Lambda_{dS} \ll |\Lambda_{AdS}| \ll M_{\rm Planck}$.
Taking into account the huge number of points in the string landscape,
this hierarchy can possibly find a statistical or even an anthropic explanation.

In this paper we continue
our previous work \LRSS\ on \tb orientifold compactifications, here
addressing the question for which concrete six--dimensional Calabi-Yau spaces all moduli can be stabilized in the large radius limit.
Concretely we will investigate \tb orientifolds on six--dimensional
toroidal orbifolds away from the singular orbifold point, \ie 
we will consider the case where the orbifold singularities
are resolved by blow--ups. Our choice of background geometries
is dictated by a kind of a compromise: on the one hand, in order to get a concrete
handle on the KKLT scenario, we need to investigate models where calculability of
the effective action after compactification is still possible due to the
orbifold geometry.
On the other hand, these background spaces are in fact smooth
Calabi--Yau manifolds with a lot of non--trivial properties, such as
a relatively large number of moduli fields. A detailed investigation of the properties of these spaces is performed in a companion paper \first. A preview of some of these results has been given in~\ReffertMN.

In addition,  it was independently demonstrated  that \tb orientifolds
on orbifolds have several attractive phenomenological features,
such as a non-vanishing probability to find the SSM in the open string
spectrum on $D$--branes of the theory (for the $\IZ_2\times \IZ_2$ orientifolds this probability
is about one in a billion \GmeinerVZ).
In contrast to the observable gauge and matter sectors of the SSM,
the question of moduli stabilization on the other hand focuses on
the properties of the hidden sector, which is coupled to
the SSM sector through gravitational interactions or also
through new gauge interactions.
Hence the SSM sector and hidden sector can be regarded as two different modules
which eventually have to be implanted at the same time at two different corners of the 
internal CY space.
To analyze whether this is possible is beyond the scope of this paper. In this work
we limit ourselves moreover to the stabilization of the closed string moduli of the background.

The successful implementation of the of the KKLT scenario requires that
the following requirements are met in concrete CY compactifications:

\vskip0.2cm
\noindent
$(i)$ Turning on $3$--form fluxes through $3$--cycles of the CY space must fix
all its complex structure moduli plus the \tb dilaton field at string tree level  \GiddingsYU. 
In the effective action 
non-vanishing $3$--form fluxes generate a tree level N=1 superpotential \GVW\
for the complex structure moduli chiral multiplets and the dilaton, and one has
to check that in the corresponding scalar potential the supersymmetry conditions
(vanishing of all $F$--terms) stabilizes all these scalar fields.
As it turns out, this condition is generically satisfied in \tb orientifold
compactifications.
Without any further input the corresponding supersymmetric flux ground states
are flat Minkowski vacua where all the K\"ahler moduli are still
undetermined flat directions of the effective scalar potential.

\vskip0.2cm
\noindent
$(ii)$ In order to stabilize the K\"ahler moduli of the underlying Calabi-Yau space, 
non-perturbative effects are needed. 
The K\"ahler moduli correspond to the volumes of homologically non-trivial
four--cycles inside the Calabi-Yau manifold, called divisors.
Specifically, a non-vanishing superpotential
for these fields may be generated by Euclidean $D3$--brane instantons which are wrapped
around the corresponding divisors \WittenBN. For the divisor volume to be stabilized 
by Euclidean $D3$--instantons, it must fulfill a certain condition: it must allow for 
precisely two fermionic zero modes
on the $D3$--brane world volume. 
As we will show, these conditions cannot be satisfied for all orientifolds of blown--up
orbifolds geometries, and hence we are left with models, which still possess flat
directions in the K\"ahler moduli space. Another, alternative mechanism to stabilize
K\"ahler moduli fields is provided by gaugino condensation \FerraraQS,  within the
effective Yang-Mills gauge theory on a stack of space-time
filling $D7$--branes, which are also wrapped around a divisor of the CY space.
Again certain conditions have to be met for a non--vanishing gaugino condensate.
First, being space-time filling, the $D7$--branes have to satisfy together with the 
$O7$--orientifold
planes the $RR$--tadpole conditions.
This restricts the number of $D7$--branes and also the choice of divisors around
which they can be wrapped. Second, the matter spectrum of the effective $D7$--brane gauge
theory must be such that a non-perturbative  superpotential is formed by the
gaugino condensates. Specifically, no massless adjoint chiral multiplets should
appear in the effective gauge theory.
This requirement restricts the bosonic zero modes on the $D7$--brane world volume,
\ie the $D7$--branes should not have any open string moduli fields, namely
either Wilson line moduli on its world volume or geometrical moduli that determine
the $D7$--brane positions in the two-dimensional transversal directions inside the CY space.
These conditions essentially mean that the divisor must be rigid inside the CY space.
Furthermore also the spectrum of fundamental chiral fields is restricted, such as
the number of bifundamental representations that originate from $D7$--brane
intersections or from open string $2$--form fluxes on the $D7$--brane world volumes.

\vskip0.2cm
\noindent
$(iii)$ Even if one has succeeded to stabilize all moduli in an AdS vacuum, there
is another condition that is related to the stability of the
orientifold compactification after the uplift to a dS vacuum.
Concretely, it is well known that stable AdS vacua are also possible
for the case that the 
(mass)$^2$ matrix $M_{ij}^2$ of the moduli contains negative eigenvalues as
long as the absolute value of  $M_{ij}^2$ is within the Breitenlohner--Freedman bound \BF.
However if one wants to perform the dS uplift in the way described above,
we have to insist that  $M_{ij}^2$ is positive definite. As we will discuss, this requirement
gives rise to a quite powerful theorem: for CY orientifold compactifications
without complex structure moduli one can show that under very mild
conditions on the K\"ahler potential of the K\"ahler moduli fields, stability after
the uplift cannot be achieved.

\vskip0.2cm
In order to see for which resolved orbifold geometries all moduli can be stabilized
we will essentially follow the following steps:

\vskip0.2cm
\noindent
$\bullet$ We start with some toroidal orbifold and describe the local
patches near the singularity with toric methods. Then we resolve the 
singularities via blow-ups. Next we have to glue together the local patches
in order to get a smooth CY space $Y_6$.
Having done this we can compute
certain geometrical and topological data of $Y_6$.
These are the triple intersection numbers
of the divisors, which enter the K\"ahler potential of the K\"ahler
moduli fields. In addition one also needs the topological Hodge numbers
of the divisors, in order to decide which divisors can contribute to the
non-perturbative superpotential. This first step
is subject of the companion paper \first\ 
on the geometry of blown-up orbifold spaces. In Section 5 we will heavily rely
on these results.

\vskip0.2cm
\noindent
$\bullet$
In the second step we perform an orientifold projection on the smooth
CY, which gives rise to $O3/O7$--planes and also $D3$-- and $D7$--branes
for tadpole cancellation.
Going from N=2 \tb compactifications on a CY covering space $Y_6$, such as
the blown up orbifolds, to its 
N=1 relative on a CY orientifold $X_6$ with
$O3$-- and $O7$--planes leads to several new effects which are 
relevant in the context of the KKLT moduli stabilization program.
First of all, the orientifold projection leads to a field redefinition
of the K\"ahler moduli fields together with the dilaton field. 
Second, 
the $\IZ_2$ orientifold projection will change
some of the triple intersection numbers, which enter
the K\"ahler potential of the K\"aher moduli fields.
Third,
whenever an orbifold group contains subgroups of odd order, some of
the fixed sets under these subgroups will not be invariant under the global
orientifold $\IZ_2$ involution. {\it E.g.} when certain fixed points or lines 
are identified under the orientifold quotient,
the $2$--(co)homology splits into an invariant and an
anti-invariant part under the orientifold action.
This implies that only the geometric divisor moduli of invariant patches
can be fixed by the non-perturbative superpotential.
On the other hand, the moduli associated to non-invariant divisors
are not any more geometrical divisor volumes, but rather
are given by a combination of two-form gauge potentials.
Since the superpotential does not depend on them,
these moduli cannot be fixed by the KKLT mechanism. However, as
we will discuss, they will appear in a $D$--term scalar potential.
Requiring supersymmetric ground-states with vanishing $D$--terms
will allow us to stabilize also these moduli fields.

\vskip0.2cm
\noindent
$\bullet$
Then we determine the K\"ahler potential for the K\"ahler moduli fields using
the triple intersection numbers. In addition one has to compute
also the K\"ahler potential for the complex structure moduli fields.
We will be able to do this explicitly only
for orbifolds with only untwisted complex structure moduli fields.

\vskip0.2cm
\noindent
$\bullet$
Furthermore we compute the non-perturbative superpotential for the K\"ahler moduli fields from
the divisor topologies. In addition we compute the tree level $3$--form flux superpotential
for the dilaton and the complex structure moduli fields.

\vskip0.2cm
\noindent
$\bullet$
Finally we determine the supersymmetric AdS vacua of the scalar potential and
compute the corresponding values of all moduli fields.

\newsec{Moduli stabilization and  stability in \tb orientifolds}

\subsec{Calabi--Yau orientifolds of \tb with $D3/D7$--branes}
\def\n{{h_{(1,1)}^{(+)}(X_6)}}
\def\m{{h_{(1,1)}^{(-)}(X_6)}}

We start with a \tb compactification on a Calabi--Yau (CY) manifold $Y_6$.
This leads to N=2 supersymmetry in $D=4$ dimensions. The geometry
of the manifold $Y_6$ is described by $h_{(1,1)}(Y_6)$ K\"ahler moduli 
and $h_{(2,1)}(Y_6)$ complex structure moduli. These moduli fields 
represent scalar components of N=2 hyper-- and vector multiplets, respectively. 
Together  with the universal hypermultiplet we have 
$h_{(1,1)}(Y_6)+1$ hypermultiplets and $h_{(2,1)}(Y_6)$ vector multiplets.

To arrive at N=1 supersymmetry in $D=4$ we introduce an orientifold projection $\Oc$,
which produces orientifold $O3$-- and $O7$--planes.
To cancel tadpoles and to construct models of phenomenological interest we add
$D3$-- and $D7$--branes. 
The orientifold projection $\Oc$ \doubref\ORI\HB
\eqn\Action{
\Oc=(-1)^{F_L}\ \Om\ \sigma^\star}
acting on the closed \tb string states is given 
by a combination of world--sheet parity transformation $\Om$ and a reflection $\si$
in the internal CY space.
The CY geometry $Y_6$ modded out by the additional involution $\sigma$ is labeled by $X_6$.
To obtain $O3/O7$--planes the action $\si$ must act holomorphically and satisfy
\eqn\satis{
\si^\star\ \Om_{(3,0)} =-\Om_{(3,0)}\ ,}
with $\Om_{(3,0)}$ the holomorphic $3$--form of the CYM $X_6$. 

Due to the holomorphic action of $\si$, the latter splits the cohomology groups
$H^{(p,q)}(Y_6)$ into a direct sum of an even eigenspace $H^{(p,q)}_+(X_6)$
and an odd eigenspace $H^{(p,q)}_-(X_6)$ \HB.
Since the K\"ahler form $J$ is invariant under the orientifold action, 
it is expanded w.r.t. 
a basis of $H^{(1,1)}_+(X_6)$. 
On the other hand, because of \satis, the holomorphic $3$--form
$\Om_{(3,0)}$ may be 
expanded w.r.t.  a real symplectic basis $(\alpha_\lambda,\beta^\lambda)$ of 
$H^{(3)}_-(X_6)$ , \ie
\eqn\expandJ{
J=\sum\limits_{k=1}^\n t^k\ \omega_k\ \ \ ,\ \ \ 
\Om_{(3,0)}=\sum_{\lambda=0}^{h_{(2,1)}^{(-)}(X_6)} X^\lambda \alpha_\lambda-
F_\lambda\beta^\lambda\ ,}
with $(X^\lambda,F_\lambda)$ the periods of the original CYM $Y_6$.
Furthermore, in 
\tb orientifolds with $D3$-- and $D7$--branes, the $NSNS$
two--form $B_2$ and the $RR$ two--form
$C_2$ are odd under the orientifold action $(-1)^{F_L}\Om$. Hence, they are expanded w.r.t. a basis 
of the cohomology $H^{(1,1)}_-(X_6)$, \ie
\eqn\expandB{
B_2=\sum\limits_{a=1}^\m b^a\ \omega_a\ \ \ ,\ \ \ C_2=\sum\limits_{a=1}^\m c^a\ \omega_a\ .}
In \tb orientifolds the fields $b^a$ and $c^a$ give rise to $\m$ complex scalars 
\eqn\combined{
G^a=i\ c^a-S\ b^a\ \ \ ,\ \ \ a=1,\ldots,\m}
of N=1 chiral multiplets \GL, whose vevs eventually should be fixed.
Clearly, $D3$--  and $D7$--branes may be wrapped only around $4$--cycles whose Poincar\'e dual
$2$--form is an element of $H^2_{+}(X_6)$.
To summarize, in addition to the dilaton field 
\eqn\DILA{
S=e^{-\phi_{10}}+i\ C_0}
a CY orientifold compactification $X_6$ has $\n$ K\"ahler moduli $t^k$, 
$\m$ scalars $G^a$ and $h_{(2,1)}^-(X_6)$ complex structure moduli
$u^\lambda$.
As shown\foot{Here and in the following, where no confusion with the orientifold action 
$\Om$ may occur we shall use $\Om$ for $\Om_{(3,0)}$.} 
in Table 1, under the orientifold action $\Oc$ the original set of $h_{(1,1)}(Y_6)+1$ 
N=2 hypermultiplets  and $h_{(2,1)}(Y_6)$ N=2
vectormultiplets is split into a set of N=1 chiral and
vectormultiplets. 
\vskip0.1cm\def\ss#1{{\scriptstyle{#1}}}
{\vbox{\ninepoint{$$
\vbox{\offinterlineskip\tabskip=0pt
\halign{\strut\vrule#
&~$#$~\hfil 
&\vrule$#$ 
&~$#$~\hfil 
&\vrule$#$ 
&~$#$~\hfil 
&\vrule$#$
&~$#$~\hfil 
&\vrule$#$ 
&~$#$~\hfil 
&\vrule$#$
&~$#$~\hfil 
&\vrule$#$\cr
\noalign{\hrule}
&\ &&\ &&\ &&\ &&\   &&\   &\cr
&\ 1\      &&\ {\rm dilaton} && S   &&  {\rm chiral\ multiplet}&& 
\int_{X_6} \Omega\wedge G_3 &&\ \ss{ISD\ 3-form\ flux\ G_3} &\cr
&\ &&\ &&\ &&\ &&\  &&\  &\cr
&\ h^{(-)}_{(2,1)}(X_6)\ &&\ {\rm CS\ moduli}\ && 
\ u^\lambda\ &&{\rm  chiral\ multiplets}&& 
\int_{X_6} \Omega\wedge G_3 &&\ \ss{ISD\ 3-form\ flux\ G_3} &\cr
&\ &&\ &&\ &&\ &&\   &&\  &\cr
&\ h^{(+)}_{(1,1)}(X_6)\   &&\ {\rm K\ddot{a}hler\ moduli}\  
&&\ t^k,\rho^k\ &&{\rm chiral/linear\ multiplets}\ &&\ e^{-T}\  &&\ {D3\ {\rm instanton}\atop {\rm
gaugino\ condensation}}  &\cr
&\ &&\ &&\ &&\ &&\  &&\  &\cr
&\ h^{(-)}_{(1,1)}(X_6)\ &&\ {\rm add.\ moduli}\ &&\ b^a,c^a\ 
&&{\rm chiral\ multiplets}&&\ \int_{C_4} J\wedge B_2&&
\  {\rm \ss{calibration}}\  &\cr
&\ &&\ \ &&\ \ 
&&\ &&\ (D_\mu G^a)^2&&
\  {\rm \ss{massive\ vector}}\  &\cr
\noalign{\hrule}
&\ h^{(+)}_{(2,1)}(X_6)\ &&\ {\rm add.\ vectors}\ 
&& V_\mu^{\tilde j}&&{\rm vector\ multiplets}&& \ \  -\ \  && \ \ -\ \  &\cr
\noalign{\hrule}}}$$
\vskip-6pt
\centerline{\noindent{\bf Table 1:}
{\sl Moduli of Calabi--Yau orientifold $X_6$ and their stabilization mechanism.}}
\vskip10pt}}}
\vskip-0.5cm
\br
The additional $h^{(+)}_{(2,1)}(X_6)$ vectors (and their magnetic duals)
arise from the Ramond $4$--form $C_4$ reduced w.r.t. the cohomology $H^{(3)}_+(X_6)$.
Besides the dilaton field $S$ in the K\"ahler potential for the moduli fields 
\eqn\FULL{
K=-\ln(S+\ov S)-2\ln\lf(\fc{1}{6}\int_{X_6} J\wedge J\wedge J\ri)-\ln\lf(-i\int_{X_6} \Omega \wedge
\ov \Omega\ri)}
only the $\n$ invariant K\"ahler moduli  $t^k$ and the $h_{(2,1)}^-(X_6)$ invariant
complex structure moduli enter explicitly.
However, the string theoretical K\"ahler moduli $t^j$ are not yet scalars 
of an N=1 chiral multiplet.
After defining the proper holomorphic moduli fields $T^j$ (in the string frame\foot{
In the Einstein frame the K\"ahler moduli $t^k$ are multiplied with $e^{-\h\phi_{10}}$.
In the Einstein frame the CY volume reads  $Vol(X_6)=\fc{1}{6}e^{-\fc{3}{2}\phi_{10}}\ 
\Kc_{ijk} t^it^jt^k$.}) \GL
\eqn\defmod{
T^j=\fc{3}{4}\ \ \Kc_{jkl}\ t^k\ t^l-\fc{3}{8}\ e^{\phi_{10}}\ 
\Kc_{jbc}\ \ov G^b\ (G+\ov G)^c+\fc{3}{2}\ i\ \lf(\ \rho^j-\h\ \Kc_{jbc}\ c^b\ b^c\ \ri)\ ,}
the second term $K_{KM}=-2\ln Vol(X_6)=-2\ln\fc{1}{6}\Kc_{ijk} t^it^jt^k$ 
in \FULL\ may be expressed in terms of the N=1 fields $T^j$. This way, 
in the low--energy effective action of \tb CY orientifolds, the fields $G^a$ do enter the
K\"ahler potential 
for the K\"ahler moduli $t^k$ 
through eliminating the moduli $t^k$ via the definition \defmod.
By that the K\"ahler potential $K_{KM}$ for the $\n$ K\"ahler moduli $T^j$  
becomes a complicated function $K_{KM}(S,T^j,G^a)$ depending on the dilaton $S$, 
the $\n$ moduli 
$T^j$ and the $\m$ moduli $G^a$ \GL. 
In \defmod\ the axion $\rho^j$ originates from integrating the $RR$ $4$--form 
along the $4$--cycle $C_j$. The full K\"ahler potential 
\eqn\FULLK{
K=-\ln(S+\ov S)-2\ln Vol(X_6)+K_{CS}}
for the 
dilaton $S$, $\n$ K\"ahler moduli $T^k$, 
$\m$ scalars $G^a$ and $h_{(2,1)}^{(-)}(X_6)$ complex structure moduli takes the form \GL:
\eqn\fullK{
K=-\ln(S+\ov S)+K_{KM}(S,T^j,G^a)+K_{CS}\ .}

To illustrate the structure of the modified K\"ahler potential, let us 
briefly discuss the case $\n=1=\m$. The K\"ahler potential for the single 
K\"ahler modulus $t$ is: $K_{KM}(t)=-2\ln t^3$.
With the intersection numbers $\Kc_{ttt}=6,\ \Kc_{t}=6t^2$ and $\Kc_{tbb}=1$ we derive from 
\defmod\ 
$$T=\fc{9}{2}\ t^2-\fc{3}{8}\ e^{\phi_{10}}\ \ov G\ (G+\ov G)+\fc{3}{2}\ i\ 
\lf(\rho-\h\ c\ b\ri)\ ,$$
and the full K\"ahler potential \fullK\ becomes:
\eqn\fullk{
K=-\ln(S+\ov S)-3\ \ln\fc{1}{9}\lf[\ T+\ov T+\fc{3}{4}\ \fc{(G+\ov G)^2}{S+\ov S}\ \ri]+
K_{CS}\ .}

Before adding background fluxes, in the effective $D=4$ action the 
fields $S,u^\lambda, t^j, b^a$ and $c^a$ have flat directions,
\ie no potential is generated for them and their vevs may assume arbitrary values in
their moduli spaces. Fixing these moduli through some $F$-- or $D$--term potential 
is the main topic of this article.
In the two last columns of Table 1 we have shown the different mechanisms how to stabilize 
these moduli. These mechanisms shall be discussed in Subsections 2.2, 2.3 and Section 3.

\subsec{Stabilization of K\"ahler moduli associated to the cohomology $H^{(1,1)}_{+}(X_6)$}

In the following we shall assume a flux compactification of a \tb CY orientifold 
with $\m=0$ and a general K\"ahler 
potential $K_{KM}$ for the $n:=h_{(1,1)}^{(+)}(X_6)$ K\"ahler moduli:
\eqn\kaehler{
K_{KM}(T^1,\ldots,T^n,\ov T^1,\ldots,\ov T^n)=K(T^1+\ov T^1,\ldots,T^n+\ov T^n)\ .}
We shall adress the case  $\m\neq 0$ in Section 3.
We consider the racetrack superpotential \Race:
\eqn\SUP{
W=W_0(S,U)+\lambda\ \sum_{j=1}^n\gamma_j(S,U)\ e^{a_j\ T^j}\ .}
The first term $W_0$ of \SUP\ represents the tree--level flux superpotential 
\GVW
\eqn\TV{
W_0(S,U)=\int_{X_6} G_3\wedge \Omega}
depending on the
dilaton $S$ and the $h_{(2,1)}^{(-)}(X_6)$ complex structure moduli $U^\lambda$.
Since $\Om\in H^{(3)}_-(X_6)$, we also must have  $G_3\in H^{(3)}_-(X_6)$ with 
$2h_{2,1}^{(-)}(X_6)+2$ complex flux components.
In \eqq \SUP\ we assume $W_0\in \IC$, $\gamma_j\in \IC$, and $a_j \in \IR_{-}$.
In addition, $\lambda\in \IR$ is a real parameter accounting for a possible so--called
K\"ahler gauge (\cf Subsection 5.4). 
The latter may be used to adjust a certain
flux value given by $W_0$ to a given minimum in the K\"ahler moduli space 
$(T^1,\ldots,T^n)$, \cf {\it Ref.} \DenefMM. 
We do not consider a possible open string moduli dependence of the superpotential 
\doubref\Lars\LMRS.

The work of KKLT \KKLT\ proposes a mechanism to stabilize all moduli at a small
positive cosmological constant. This procedure is accomplished through three steps.
One first dynamically fixes the dilaton $S$ and the complex structure moduli $U^\lambda$ 
through the tree--level piece $W_0$ (given in \eqq \TV) of the superpotential. 
This is accomplished with a generic $3$--form flux $G_3$ with both $ISD$-- and $IASD$--flux
components. At the minimum of the scalar potential
in the complex structure and dilaton directions, 
the flux becomes $ISD$ and the potential assumes the value $V_0(S,U^\lambda)=-3e^K |W_0|^2$.
The soft masses $m_S,m_{U^\lambda}$ 
for the dilaton and complex structure scalars are generically 
of the order $\ap/R^3$
\LRSii. In the large radius approximation $\re(T)\gg 1$,  
the non--perturbative terms in \SUP\ only amount to 
a small exponentially suppressed additional contribution to $m_S,m_{U^\lambda}$. According to
\HPNii\ the latter is negligible.
The second step is the addition of the non--perturbative
piece to the superpotential \SUP, which allows the stabilization of the K\"ahler moduli 
$T^j$ at a supersymmetric AdS minimum. The soft masses for the K\"ahler moduli
are much smaller than soft masses $m_S$ and $m_{U^\lambda}$. This property allows us to
separate the first and second step, \ie to effectively first integrate out the dilaton
and complex structure moduli. 
Nonetheless, strictly speaking these two steps
should be treated at the same time and that is what we shall do in Subsection 2.3.
The stability of AdS vacua in gravity coupled to scalar fields has been investigated 
in {\it Ref.}
\BF. Stability is guaranteed, if all scalar masses fulfill the Breitenlohner--Freedman
(BF) bound \BF, \ie their mass eigenvalues do not fall below a certain minimal bound.
The latter is a negative number related to the scalar potential at the minimum.
It can be shown in a completely model independent way that 
all scalars have masses above this bound at any N=1 supersymmetric AdS minimum in 
supergravity theories (\cf \eg \Duff\ and Appendix C of {\it Ref.} \Lukas).
However, the third and final step in the KKLT scenario 
consists in the addition of one anti $D3$--brane, 
\ie a positive contribution to the scalar 
potential, which lifts the AdS minimum to a dS minimum. 
The masses for the moduli fields do not change significantly during this process. 
However stable dS vacua require positive mass eigenvalues. Hence, any negative mass
eigenvalue before the uplift is unacceptable since the effect of the anti $D3$--branes 
on the mass eigenvalues is too small to change a  negative mass to positive.

In \SUP, the sum of exponentials accounts for $D3$--brane instantons and gaugino 
condensation on stacks of $D7$--branes.
The $D3$--instantons come from wrapping (Euclidean)  $D3$--branes on internal
$4$--cycles $C^j_4$ of the CY orientifold $X_6$. The latter have the volume $\re(T^j)$ 
and lead to the instanton effect $e^{-2\pi T^j}$ in the superpotential, \ie $a_j=-2\pi$.
The gauge coupling on a $D7$--brane which is wrapped on the $4$--cycle $C^j_4$
is given by $\re(T^j)$, \cf \eqq \defmod. 
Hence, gaugino condensation on this $D7$--brane yields the effect
$e^{-T^j/b_a}$ in the superpotential. {\it E.g.} for the gauge group $SU(M)$ we have
$b_{SU(M)}=\fc{M}{2\pi}$, \ie $a_j=-\fc{2\pi}{M}$.
On the $D7$--brane,  $\gamma_j(S,U)$ may comprise one--loop effects and further instanton 
effects from $D(-1)$--branes: One loop corrections to the gauge coupling
give rise to \LS
\eqn\ext{\gamma_j\sim \eta(U^\lambda)^{-2/b_a}\ ,}
while additional instantons in the $D7$--gauge theory amount to:
\eqn\exti{\gamma_j\sim e^{-S/b_a\int_{C_4^j} F\wedge F}\ .}

Supersymmetric vacuum solutions are found  by finding the zeros of the $F$--terms:
$\ov F^{\ov M}=K^{\ov M J}\ (\p_J W+W\ K_J)$. Solutions to the equations $F^M=0$ give rise
to extremal points of the scalar potential.
In addition, it has to be verified whether
those zeros lead to a stable minimum.
Since the matrix $K^{\ov M J}$ is positive definite, the zeros $(T^1_0,\ldots,T^n_0)$ 
in the K\"ahler moduli space are determined by the $n$ equations:
\eqn\adsm{
\p_{T^j} W+W\ K_{T^j}=0\ \ \ ,\ \ \ j=1,\ldots,n}
following from the requirement of vanishing $F$--terms. 
These equations turn into the $\n$ relations
\eqn\Relations{
\lambda\ \gamma_i=\lf.-\lf(\ \prod_{j\neq i}^n |a_j|\ \ri)\ e^{-a_i\ T^i}\ W_0
\ \ \fc{K_{T^i}}{\sum\limits_{j=1}^n K_{T^j}\ \prod\limits_{k=1\atop k\neq j}^n |a_k|-
\prod\limits_{k=1}^n |a_k|}\ \ \ \ri|_{T^l=T^l_0}\ \ \ ,\ \ \ i=1,\ldots,n}
to be satisfied at this extremum.
Since $K_{T^i}$ and $a_i$ are real, from \eqq \Relations\ we may easily deduce the 
vevs $t_2^j$ of the axions at the $AdS$--minimum. 
After introducing the phases 
$W_0=|W_0|\ e^{i \varphi}$and $\gamma_i=|\gamma_i|\ e^{i\phi_i}$ 
we obtain ($T^j=t_1^j+it_2^j$)
\eqn\vevaxion{
t_2^i=\fc{1}{a_i}\ \lf[\ \varphi +\pi\ (1+\rho^i)-\phi_i\ \ri]+\fc{2\pi}{a_i}
\ \IZ\ \ \ , \ \ \ 
i=1,\ldots,n}
as vevs for  the axion fields. Above we have introduced the numbers
$$\rho^i=\fc{1}{\pi}\ arg\lf(\fc{\lambda\ K_{T^i}}{\sum\limits_{j=1}^n K_{T^j}\ 
\prod\limits_{k=1\atop k\neq j}^n |a_k|-
\prod\limits_{k=1}^n |a_k|}\ri)\in\{0,1\}\ .$$ 
For the case that an exponential $e^{a_jT^j}$ accounts for gaugino 
condensation in an $SU(M)$ gauge group, we have $a_j=-\fc{2\pi}{M}$ and in \eqq \vevaxion\ 
the additional shift  $\fc{2\pi}{a_j}\ \IZ$ becomes $M\ \IZ$. The latter becomes trivial,  
if the K\"ahler modulus $T^j$ enjoys a discrete shift symmetry, \eg $T^j\ra T^j+1$.
On the other hand, if no such symmetry exists, in \eqq \vevaxion\ the 
additional shifts $\fc{2\pi}{a_j}\ \IZ$ give rise to an infinite number of extrema obtained 
from one another by shifts in the axionic directions $t_2^j$.
A useful relation to be satisfied at the extremum is the ratio
\eqn\ratio{
\fc{\gamma_i}{\gamma_j}=e^{a_j\ T^j-a_i\ T^i}\ \fc{a_j}{a_i}\ \fc{K_{T^i}}{K_{T^j}}\ \ \ ,
\ \ \ i,j=1,\ldots,n\ .}
The latter equation may be written as
\eqn\Ratio{
\fc{|\gamma_i|}{|\gamma_j|}=e^{i\phi_{ji}}\ e^{a_j\ t_1^j-a_i\ t_1^i}\ 
\fc{a_2}{a_1}\ \fc{K_{T^1}}{K_{T^2}}\ \ \ ,\ \ \ \phi_{ji}=\phi_j-\phi_i+a_j\ t_2^j-a_i\ t_2^i\ .}
Since $\phi_{ji}\in \{0, \pi\}$, the directions of the axions $t_2^j$ 
strongly depend on the signs of the first K\"ahler derivatives $K_{T^j}$ and the 
phases $\phi_j$ of the coefficients $\gamma_j$. 

Moreover, from the relation \vevaxion\ we see that any complex phase of $W_0$ and $\gamma_i$ 
may be absorbed into a redefinition of the axion vev at the minimum.
Hence, in the following we may assume without any restriction: 
$$W_0\in \IR^+\ \ \ ,\ \ \ \gamma_j \in \IR^+\ .$$

Finally, at the extremum $(T^1_0,\ldots,T^n_0)$, 
the scalar potential $\tilde V(T^1,\ov T^1,\ldots,T^n,\ov T^n)$ 
assumes the negative value
\eqn\VMINN{
\tilde V_{min}=\lf.-3\ \ e^{K}\ \lf(\ \prod_{k=1}^\n a_k^2\ \ri)\ 
\fc{|W_0|^2}{\lf(\ \sum\limits_{j=1}^n K_{T^j}\ \prod\limits_{k=1\atop k\neq j}^n 
|a_k|-
\prod\limits_{k=1}^n |a_k|\ \ri)^2}\ \ri|_{T^l=T^l_0}}

\subsec{(In)stability of flux compactifications without complex structure moduli}

In this subsection we shall continue and generalize the discussion initiated by the research 
\doubref\ChoiSX\LRSS\  
(see also {\it Refs.} \doubref\dealwis\Japan) on the stability criteria 
in a KKLT setup \KKLT.  In these  references, the conditions under which 
a stable vacuum may be found have been investigated for some selected
orbifold examples. This means in particular that a special K\"ahler potential
for the underlying coset spaces parametrizing the K\"ahler moduli space
has been used. 
The general property for those examples is that without any
complex structure modulus or with fixed complex structures, as it is the case
for many toroidal orbifold examples at the orbifold point, no stable vacuum in a KKLT setup with a 
non--perturbative superpotential exists.
The question we shall address here is, whether this is a general feature, independent
of the form of the K\"ahler potential
of compactifications without complex structure moduli, in particular also holding
for the resolved orbifolds.
Note, that in those cases, the K\"ahler potential assumes a quite different
form than for the coset spaces.

For the case of $h_{(2,1)}^{(-)}(X_6)=0$ and $h_{(1,1)}^{(+)}(X_6)=:n$ 
the superpotential \SUP\ assumes the form:
\eqn\SUPP{
W=B+A\ S+\lambda\ \sum_{i=1}^{\n} \gamma_i\ e^{a_i\ T^i}\ .}
Generically, we have $A,B \in \IC$, $\gamma_j\in \IC$, and $a_i \in \IR_{-}$.
However, we have argued in the previous subsection that without any restriction we 
might assume $A,B, \gamma_j\in \IR^+$. Nevertheless, 
we shall treat them in the following  as arbitrary complex parameters. 
In addition, $\lambda\in \IR$ is a real parameter accounting for a possible so--called
K\"ahler gauge, as discussed in \DenefMM. 

The supersymmetric vacuum solutions are given by solving the equations
$F^M=0$ for $M=S,T^1,\ldots,T^n$. With the K\"ahler potential
$$-\ln(S+\ov S)+K_{KM}(T^1,\ldots,T^n,\ov T^1,\ldots,\ov T^n)\ $$
for the moduli space $(S,T^1,\ldots,T^n)$ we obtain the $F$--term
\eqn\FS{
\ov F_S=e^{\h K}\ (S+\ov S)^{1/2}\ \lf(\ -B+A\ \ov S-\lambda\ \sum_{i=1}^\n
\gamma_i\ e^{a_i\ T^i}\ \ri)}
for the dilaton field $S$.
The solutions $(S_0,T^1_0,\ldots, T^n_0)$ to the vacuum equations
\eqn\VAC{
F^M=0\ \ \ ,\ \ \ M=S,T^1,\ldots,T^n}
are given by the dilaton vev $S_0$
\eqn\fixedto{
S_0=-\fc{\prod\limits_{k=1}^\n |a_i|}{2\ |A|^2}\ 
\fc{\ov A\ B+A\ \ov B}{2\ \sum\limits_{j=1}^\n K_{T^j}\ 
\prod\limits_{k=1\atop k\neq j}^\n |a_k|-
\prod\limits_{k=1}^\n |a_k|}-\fc{1}{2\ |A|^2}\ (\ov A\ B-A\ \ov B)\ ,}
and the $n$ relations:
\eqn\relations{\eqalign{
\lambda\ \gamma_i&=\lf.-\fc{\prod\limits_{j\neq i}^\n |a_j|}{\ov A}\ e^{-a_i\ T^i}\ 
\fc{K_{T^i}\ (\ov A\ B+A\ \ov B)}{2\ \sum\limits_{j=1}^\n K_{T^j}\ 
\prod\limits_{k=1\atop k\neq j}^\n |a_k|-
\prod\limits_{k=1}^\n |a_k|}\ \ \ \ri|_{T^l=T^l_0}\cr
&=\lf.-\fc{2\ A\ \re(S)}{a_i}\ e^{-a_i\ T^i}\ K_{T^i}\ \ \ \ri|_{T^l=T^l_0}\ \ \ ,\ \ \ 
i=1,\ldots,n\ .}}
Similarly as in the previous subsection (\cf \eqq \vevaxion) we may determine
the vevs for the axion fields. With $A=|A|\ e^{i\varphi}$ and $\gamma_i=|\gamma_i|e^{i\phi_i}$
we obtain:
\eqn\Vevaxion{
t_2^i=\fc{1}{a_i}\ \lf[\ \varphi+\pi-\phi_i\ \ri]+\fc{2\pi}{a_i}\ \IZ\ \ \ ,\ \ \ i=1,\ldots,n\ .}
Note, that the above expression is somewhat simpler than \vevaxion, since we have 
solved for $n+1$ equations $F^M=0$ thus allowing to eliminate some more terms.
In fact, the additional relation, valid at the point $(S_0,T^1_0,\ldots,T^n_0)$, is:
\eqn\wehave{
\lf(-2\ \sum\limits_{j=1}^\n K_{T^j}\ \prod\limits_{k=1\atop k\neq j}^\n |a_k|+
\prod\limits_{k=1}^\n |a_k|\ \ri)=
\fc{(\ov A\ B+A\ \ov B)\ \prod\limits_{k=1}^\n |a_i|}{2\ |A|^2\ \re(S)} \ .}
The scalar potential becomes
\eqn\ComPot{\eqalign{
V(S,\ov S,T^1,\ov T^1,\ldots, T^n,\ov T^n)&=
\fc{\tilde V(T^1,\ov T^1,\ldots,T^n,\ov T^n)}{S+\ov S}\cr
&+\fc{e^{K}}{S+\ov S}\ 
\lf|\ B-A\ \ov S+\lambda\ \sum_{k=1}^\n\gamma_k\ e^{a_k\ T^k}\ \ri|^2\ ,}}
with $\tilde V(T^1,\ov T^1,\ldots,T^n,\ov T^n)$  the scalar potential computed in the 
K\"ahler moduli sector $(T^1,\ldots,T^n)$ with K\"ahler potential $K_{KM}$.
At the vacuum solution  $(S_0,T_0^1,\ldots,T_0^n)$, 
the second piece of \ComPot, which is proportional to $|F_S|^2$, vanishes
and the scalar potential assumes the value
\eqn\MIN{\eqalign{
V_{min}&=\fc{\tilde V_{min}}{S_0+\ov S_0}=
\lf.-6\ |A|^2\ \re(S)\ e^{K}\ri|_{S=S_0\ ,\atop T^l=T^l_0}\cr
&=\lf.-3\ \ \fc{e^{K}}{S+\ov S}\ \fc{\prod\limits_{k=1}^\n a_k^2}{|A|^2}\ 
\fc{(\ov A\ B+A\ \ov B)^2}{\lf(\ 2\ \sum\limits_{j=1}^\n K_{T^j}\ 
\prod\limits_{k=1\atop k\neq j}^\n |a_k|-
\prod\limits_{k=1}^\n |a_k|\ \ri)^2}\ri|_{S=S_0\ ,\atop T^l=T^l_0}\ ,}}
with $\tilde V_{min}$ given in \VMINN.

To analyze the stability of the vacuum solution, we have to calculate the second derivatives 
of the scalar potential 
$V(S,\ov S,T^1,\ov T^1,\ldots,T^n,\ov T^n)$ at the supersymmetric point.
After introducing $S=s_1+is_2$, we find that the mixed derivatives 
\eqn\mixed{
\fc{\p^2 V}{\p t_1^i\p t_2^j}=\fc{\p^2 V}{\p t_1^i\p s_2}=
\fc{\p^2 V}{\p t_2^i\p s_1}=\fc{\p^2 V}{\p s_1\p s_2}=0\ \ \ ,\ \ \ i,j=1,\ldots,n}
vanish at the extremum $(S_0,T^1_0,\ldots,T^n_0)$.
Now we derive the necessary criteria for the matrix encoding the scalar masses  
to have positive eigenvalues at the point $(S_0,T^1_0,\ldots,T^n_0)$.
We have to determine the second derivatives of the potential \ComPot\
and check, whether the mass matrix is positive definite.
In particular, due to \mixed,
the mass matrix $M$ 
\eqn\dets{
M=\pmatrix{\fc{\p^2 V}{\p (t^{i_1}_1)^2}&\ldots&
\fc{\p^2 V}{\p t^{i_1}_1\p t^n_1}&\fc{\p^2V}{\p t^n_1\p s_1}\cr
\vdots&\vdots&\vdots&\vdots\cr
\fc{\p^2 V}{\p t_1^n\p t_1^{i_1}}&\ldots&\fc{\p^2V}{\p (t_1^n)^2}
&\fc{\p^2V}{\p t^n_1\p s_1}\cr
\fc{\p^2 V}{\p s_1 \p t_1^{i_1}}&\ldots&\fc{\p^2V}{\p s_1 \p t_1^n}
&\fc{\p^2 V}{\p s_1^2}   }}
of the axions $s_2,t_2^1,\ldots,t_2^n$ has to be positive definite.
For the moment we shall concentrate on the properties of this axionic mass matrix $M$.
We shall derive a general expression for
the determinant $\det(M)$ of the axion mass matrix $M$ and 
express it in terms of its principal minors.
To tackle this project, we first introduce the $2^n-2$ principal submatrices of $M$:
\eqn\Submatrices{
M_{i_1\ldots i_p}=\pmatrix{\fc{\p^2 V}{\p (t^{i_1}_1)^2}&\ldots&
\fc{\p^2 V}{\p t^{i_1}_1\p t^{i_p}_1}&\fc{\p^2V}{\p t^{i_1}_1\p s_1}\cr
\vdots&\vdots&\vdots&\vdots\cr
\fc{\p^2 V}{\p t_1^{i_p}\p t_1^{i_1}}&\ldots&\fc{\p^2V}{\p (t_1^{i_p})^2}
&\fc{\p^2V}{\p t^{i_p}_1\p s_1}\cr
\fc{\p^2 V}{\p s_1 \p t_1^{i_1}}&\ldots&\fc{\p^2V}{\p s_1 \p t_1^{i_p}}
&\fc{\p^2 V}{\p s_1^2}   }\ \ \ ,\ \ \ {p=1,\ldots, n-1\ ,\atop 1\leq i_1<\ldots <i_p\leq n}\ .}
Any principal submatrix of a positive definite matrix must be positive definite \HORN.
Therefore, all the  submatrices $M_{i_1},M_{i_1i_2},\ldots,M_{i_1\ldots i_{n-1}}$
must have positive determinant, \ie those principal minors of $M$ must satisfy
$$\det(M_{i_1\ldots i_p})>0$$ 
in order that $M$ may be positive definite. 
After some tedious algebra the determinant of the matrix $M$ may be 
expressed\foot{We introduce
the $n\times n$ matrix $\Gc=\lf\{\p_{T^i}\p_{\ov T^j} K\ri\}_{i,j,=1,\ldots,n}$, which 
is positive definite.}
in terms of its $2^n-2$ submatrices $M_i,M_{ij},M_{ijk},\ldots$:
\eqn\Find{\eqalign{
\det(M)&=-\lf[\ \re(S)\ |A|^2\ e^{K}\ \ri]^{n}\ \lf(\prod_{j=1}^n a_j \ K_{T^j}\ri)\cr 
&\times \lf\{\ (3\cdot 4)^n\ \fc{\p^2V}{\p s_1^2}
+2\cdot 4^n\ \re(S)^{-1}\ |A|^{2}\ \lf(\prod_{j=1}^n a_j \ K_{T^j}\ri)\  
\fc{e^{K}}{\det(\Gc)}\ri.\cr
&\ \ \ \lf.+\sum_{s=1}^{n-1}\ (3\cdot 4)^{n-s}\ \lf[\ \re(S)\ |A|^2\ e^{K}\ \ri]^{-s}\ 
\sum_{i_1<\ldots<i_s}^n
\fc{\det(M_{i_1\ldots i_s})}{\prod\limits_{r=1}^s a_{i_r} \ K_{T^{i_r}}}\ \ri\}\ \ ,\ \ 
n\geq 1\ .}}
Since $a_i<0$ and $\det(\Gc)>0$ 
from this relation we conclude that the axionic mass matrix $M$ fails to be 
positive definite, if
\eqn\BoxI{
\vbox{\offinterlineskip
\halign{\strut\vrule#
&~$#$~\hfil
&\vrule# \cr
\noalign{\hrule}
&\ \ \       \  \ \ \ \ \ \ \ \ \ \ &\cr
&\ \ \forall\ \ {i=1,\ldots,\n}:\ \ \ \ \ K_{T^i}<0\ \ \   &\cr
&\ &\cr
\noalign{\hrule}}}  }
since in that case \eqq \Find\ yields $\det(M)<0$.
Hence, no stable uplift is possible, if \BoxI\ is fulfilled.

We may also derive a compact expression for the minors $\det(M_i)$ themselves.
For that, we introduce the  $n \times n$ matrix
\eqn\introMatr{
\Oc=\lf\{-K_{T^{i} \ov T^{j}}+\h K_{T^{i}}K_{\ov T^{j}}\ri\}_{i,j=1,\ldots,n}}
and its $n$ principal $n-1 \times n-1$ submatrices
\eqn\introMatri{
\Oc_k=\lf\{-K_{T^{i} \ov T^{j}}+\h K_{T^{i}}K_{\ov T^{j}}\ri\}_{
i,j=1,\ldots,n \atop i,j\neq k}\ \ \ ,\ \ \ 
k=1,\ldots,\n\ .}
Again, after some tedious work we arrive at the following relation:
\eqn\findalso{\eqalign{
\det(M_i)&=-12\ |A|^2\ \re(S)\ e^{K}\ a_i\ K_{T^i}\ \fc{\p^2V}{\p s_1^2}\cr
&+8\ |A|^4\ e^{2K}\ a_i^2\ K_{T^i}^2\ \fc{(-1)^{n}\
\det(\Oc_i)}{\det(\Gc)}
\ \ \ \ ,
\ \ \ i=1,\ldots,\n\ .}}
Since $a_i<0$ and $\det(\Gc)>0$ from this equation we conclude: If 
\eqn\BoxII{
\vbox{\offinterlineskip
\halign{\strut\vrule#
&~$#$~\hfil
&\vrule# \cr
\noalign{\hrule}
&\ \ \       \  \ \ \ \ \ \ \ \ \ \ &\cr
&\ \ \exists\ \ {i\in \{\ 1,\ldots,\n\ \}}: \ \ \ \ \ K_{T^i}<0\ \ \ \bigwedge
 \ \ (-1)^n\ \det(\Oc_i)\leq 0\ \ \ &\cr
&\ &\cr
\noalign{\hrule}}}  }
holds, no stable uplift is possible, either.

Note, that the two (independent) criteria \BoxI\ and \BoxII\ do not depend 
on the flux parameters $A$ and $B$ and the complex
coefficients $\gamma_j$ in \SUPP. The K\"ahler potential \kaehler\ is
enough to check the stability criteria of the KKLT setup \SUPP. According to 
\BoxI\ and \BoxII\ the condition of a positive definite axion mass matrix 
boils down to properties of the first derivatives $K_{T^j}$ at the 
AdS vacuum solution \VAC.
The conditions \BoxI\ and \BoxII\ are two {\it independent} criteria for the AdS solution 
$(S_0,T^1_0,\ldots,T^n_0)$ {\it not} yielding positive masses for the axions.
However, if the conditions are not met, for a KKLT scenario there are further criteria  
to be checked.
The latter arise from calculating the masses for the real parts of $S,T^1,\ldots,T^n$ 
and verifying their positivity.
To summarize, for a KKLT setup with $n$ K\"ahler moduli and no complex structure modulus,
a controlled uplift is {\it not possible} if at least 
one of the two conditions \BoxI\ and \BoxII\
on the K\"ahler potential \kaehler\ is met.

For CY orientifolds $X_6$ with N=1 supersymmetry we may rewrite the inequalities \BoxI\ 
such, that they become conditions on the original Calabi--Yau moduli $t^j$ relevant
for \tb compactifications without orientifold projection. 
As discussed in Subsection 2.1, the orientifold \Action\ splits the forms $\omega_i$
into $\n$ elements $\omega_i$ of $H^{(1,1)}_+(X_6)$ and $\m$ elements of $H^{(1,1)}_-(X_6)$.
The correct K\"ahler moduli $T^j$ 
arise through a Legendre transform on the original Calabi--Yau volume $Vol(t^i)$  \Mario:
With the K\"ahler form 
\eqn\JJ{
J=\sum\limits_{j=1}^{\n} t^j\ \omega_j}
the following intersection numbers may be introduced: 
\eqn\Intersections{\eqalign{
\Kc&=\int_{X_6} J \wedge J \wedge  J=\Kc_{ijk}\ t^i\ t^j\ t^k\ ,\  
\Kc_i=\int_{X_6} \omega_i \wedge J \wedge  J=\Kc_{ijk}\ t^j\ t^k,\cr
\Kc_{ij}&=\int_{X_6} \omega_i\wedge\omega_j\wedge J=\Kc_{ijk}\ t^k\ \ \ ,\ \ \  
\Kc_{ijk}=\int_{X_6} \omega_i\wedge\omega_j\wedge\omega_k\ ,}}
with $\Kc=6\ Vol(X_6)$.
After introducing the correct K\"ahler moduli \defmod\ for our orientifold compactification
$X_6$, the (inverse) metric $G^{ij}$ for the $\n$ moduli of the $H^{(1,1)}_+(X_6)$ cohomology 
becomes \doubref\Mario\GL:
\eqn\newmetric{
G^{ij}=-\fc{2}{3}\ \Kc\ \Kc^{ij}+2\ t^i\ t^j\ \ \ ,\ \ \ i,j=1,\ldots,\n\ ,}
and the following useful  
relations hold for the derivatives on the K\"ahler potential \fullK:
\eqn\Useful{
K_{T^i}=-2\ \fc{t^i}{\Kc}\ \ \ ,\ \ \ K_{T^i\ov T^j}=\fc{G^{ij}}{\Kc^2}\ .}
Equipped with these relations we may rewrite our no--go criteria \BoxI\ 
for a KKLT setup. Since $\Kc>0$, we reformulate \BoxI\ into the condition: 
\eqn\Boxi{
\vbox{\offinterlineskip
\halign{\strut\vrule#
&~$#$~\hfil
&\vrule# \cr
\noalign{\hrule}
&\ \ \       \  \ \ \ \ \ \ \ \ \ \ &\cr
&\ \ \forall\ \ {i=1,\ldots,\n}:\ \ \ \ \ t^i>0\ \ &\cr
&\ &\cr
\noalign{\hrule}}}  }

As we have converted the condition \BoxI\ into \Boxi, we shall rewrite the criterion 
\BoxII\ in terms of the 
quantities introduced in \JJ\ and \Intersections.
First, with \newmetric\ and \Useful\ we obtain for the matrix $\Oc$:
$$\Oc_{ij}=-K_{T^i\ov T^j}+\h\ K_{T^i}\ K_{\ov T^j}=\fc{2}{3}\ \fc{\Kc^{ij}}{\Kc}\ .$$ 
Hence, to obtain the principal minor $\det\Oc_k$, we essentially have to erase the 
$k$--th row and column from the matrix $\Kc^{ij}$ and compute the resulting determinant.
A minor of an inverse matrix $A$ may be computed with the relation \HORN: 
$$\det A^{-1}(\al')=\fc{\det A(\al)}{\det A}\ .$$
Here, $A(\alpha)$ denotes the principal submatrix of $A$, built from the rows and columns 
$\alpha$, 
while $A(\alpha')$ denotes the principal submatrix of $A$, built from $A$ by deleting 
the rows and columns $\alpha$. 
With this information, we obtain:
$$\det(\Oc_k)=\lf(\fc{2}{3\Kc}\ri)^{n-1}\ \fc{\Kc_{kk}}{\det(\Kc)}\ .$$
The intersection form $\Kc_{ij}$ has signature $(1,\n-1)$ \Cand, \ie 
$sign(\det\Kc)=(-1)^{n-1}$. Therefore, the condition $(-1)^n\ \det\Oc_k\leq 0$ of \BoxII\ 
is equivalent to $\Kc_{kk}\geq 0$ and the criterion \BoxII\ is translated into:
\eqn\Boxii{
\vbox{\offinterlineskip
\halign{\strut\vrule#
&~$#$~\hfil
&\vrule# \cr
\noalign{\hrule}
&\ \ \       \  \ \ \ \ \ \ \ \ \ \ &\cr
&\ \ \exists\ \ {i\in\{\ 1,\ldots,\n\ \}}:\ \ \ \ t^i>0\ \ \ \bigwedge\ \ \ \Kc_{ii} \geq 0\ \ \ &\cr
&\ &\cr
\noalign{\hrule}}}  }
Here, $\Kc_{ii}$ is the $i$--th diagonal element of the intersection form $\Kc_{ij}$.
To summarize \Boxii: if we find one positive K\"ahler modulus $t^i>0$
and if in addition we have  $\Kc_{ii}\geq 0$,  a stable uplift is not possible with 
the superpotential \SUPP. The latter consists of a sum over contributing divisors $\Dc_i$ 
whose Poincar\'e
dual classes $\omega_i$ correspond to the K\"ahler moduli $t^i$ in \JJ.
{From} the relation  $\Kc_i\ t^i=\fc{4}{3}\ Vol(\Dc_i)\ t^i=\Kc=6Vol(X_6)>0$ and $Vol(\Dc_i)>0$, we 
see, that there {\it always} exists at least one positive K\"ahler modulus $t^k>0$ in 
the expansion \JJ. Therefore, according to \Boxii\ we only have to verify whether for this
K\"ahler modulus $t^k$, we have $\Kc_{kk} \geq 0$.

The condition \Boxi\ is the inequality \BoxI\ written in terms of the K\"ahler coordinates $t^j$
of $J$ (\cf \eqq \JJ). If all of them are positive, \ie $t^j>0$, 
the mass matrix $M$ for the axions is negative definite.
For instance this is the case, when the K\"ahler class $J$ of $X_6$, defined in \JJ, 
lies in the K\"ahler cone, \ie when we have \Cand
\eqn\CYcond{
\int_{C_2} J>0\ \ \ ,\ \ \ \int_{C_4} J\wedge J>0\ \ \ ,\ \ \ \int_{X_6}
J\wedge J\wedge J>0}
for all curves $C_2$ and $4$--cycles of the CYM $X_6$.
Then we may define the K\"ahler coordinates $t^j$ through $t^j=\int_{C_2^j} J>0$
and meet the criteria \Boxi.

$\underline{\rm To\ summarize:}$ If $(i)$ the K\"ahler class $J$ lies in the 
K\"ahler cone and $(ii)$ each K\"ahler modulus $t^j$ corresponds to a 
$4$--cycle of volume $\re(T^j)$ contributing to \SUPP, 
{\it no} stable uplift is possible for $h^+_{(2,1)}(X_6)=0$.

To this  statement we have to add two refinements for the case that 
 the above two assumptions $(i)$ and $(ii)$ do not hold.
Let us first discuss point $(i)$. The K\"ahler moduli $t^j$ may not be equal to the 
real coordinates $\tilde t^j$ on the space of K\"ahler classes 
of the CY manifold $X_6$. The latter fulfill the positivity conditions \CYcond.
Generically, there are linear relations 
between the supergravity moduli $t^j$ we are working with and
the CY moduli $\tilde t^j$. The latter describe deformations of the K\"ahler form 
$\Jc=\sum\limits_{i=1}^{h_{(1,1)}(X_6)} \tilde t^i\ \tilde\omega_i$, which 
satisfy the positivity conditions \CYcond. One may choose
K\"ahler moduli $\tilde t^j:=\int_{C_2^j}\Jc>0$, while the supergravity moduli 
$t^j$ may not  always be positive. 
Furthermore, the intersection form $\tilde\Kc_{ij}=\tilde\Kc_{ijk} \tilde t^k$ has signature
$(1,h_{(1,1)}(X_6)-1)$.

It may be useful to present an example \LRSS. Let us consider the K\"ahler
moduli space $\fc{SU(1,1)}{U(1)}\otimes \fc{SU(2,2)}{SU(2)\times SU(2)\times U(1)}$
of the $\IZ_4$--orbifold with the K\"ahler potential:
\eqn\examp{
K(t^i,\ov t^i)=-\ln\lf[\ t^1\ t^2\ t^3-\h\ t^3\ (t^4)^2-\h\ t^3\ (t^5)^2\ \ri]\ .}
We have $\Kc_{123}=1, \Kc_{344}=-1, \Kc_{355}=-1$
and the intersection form $\Kc_{ij}$ is:
\eqn\exinter{
\Kc_{ij}=\pmatrix{0  & t^3 & t^2&0&0\cr
                t^3& 0   & t^1&0&0\cr
                t^2&t^1  &0   &-t^4&-t^5\cr
                0&0&-t^4&-t^3&0\cr
                0&0&-t^5&0&-t^3}\ .}
It may be verified, that the signature of $\Kc_{ij}$ is indeed $(1,4)$.
According \defmod, \ie $\re T^j=\fc{3}{4} \Kc_{jkl}\ t^k\ t^l$ 
the holomorphic coordinates $T^i$ or $4$--cycle volumes $\re T^i$ become:
\eqn\fourcyclevol{\eqalign{
\re(T^1)&=\fc{3}{2}\ t^2\ t^3\ \ \ ,\ \ \ \re(T^4)=-\fc{3}{2}\ t^3\ t^4\ ,\cr
\re(T^2)&=\fc{3}{2}\ t^1\ t^3\ \ \ ,\ \ \ \re(T^5)=-\fc{3}{2}\ t^3\ t^5\ ,\cr
\re(T^3)&=\fc{3}{2}\ \lf[\ t^1\ t^2-\h\ (t^4)^2-\h\ (t^5)^2\ \ri]\ .}}
In terms of these moduli, the K\"ahler potential $-2 K(t^i,\ov t^i)$
for the orientifold becomes:
\eqn\becomes{
K(T^i,\ov T^i)=-\ln\fc{1}{6\sqrt 6}\ (T^3+\ov T^3)
\lf[\ (T^1+\ov T^1)\ (T^2+\ov T^2)\ 
-\h\ (T^4+\ov T^4)^2-\h\ (T^5+\ov T^5)^2\ \ri]\ .}
With \exinter\ the equations \newmetric\ and \Useful\ may be verified.
The important point here is, that we must have $t^4,t^5<0$ 
for positive $4$--cycle volumes $\re(T^j)$. Hence, in that example
the supergravity moduli $t^i$ appearing in \JJ\ do not define a K\"ahler class 
which lies in the K\"ahler cone. Nonetheless, though \Boxi\ is not satisfied,
with the K\"ahler metric \becomes, \ie $K_{KM}=K$, we do not obtain a stable uplift \LRSS.

Let us now come to point $(ii)$.
The sum of exponentials in the superpotential \SUPP\ 
consists of  a sum over 
a basis of $\n$ independent divisor classes $\Dc_i$ corresponding to the $\n$ 
$4$--cycles the Euclidean $D3$--branes or $D7$--branes are wrapped on.
The divisors must have arithmetic genus $\chi(\Dc_i)=1$ (\cf Section 4) to guarantee that
$\gamma_i\neq 0$ in the superpotential \SUPP.
Let $\omega_i$ denote the Poincar\'e dual class of $\Dc_i$, which is an element
of $H^{(2)}_+(X_6)$. If those classes $\omega_i$
are identified with the basis of $(1,1)$--forms
in the K\"ahler class \JJ, the moduli $t^j$ may not be identical to the 
K\"ahler deformations $\tilde t^i$ one introduces for the CY orientifold $X_6$ to begin with.
This happens, if the contributing divisor classes $\Dc_i$ are constructed\foot{Note, that 
this change of basis does not alter the positivity properties of the axionic mass matrix $M$, 
introduced in \dets. In the mass matrix $M$, the second derivatives are calculated w.r.t. the
contributing divisor volumes ${\rm Vol}(\Dc_i)$. The mass matrix w.r.t. the original
divisor volumes ${\rm Vol}(\tilde \Dc_i)$ is given by $N^{-1}MN^\star$. Since $N$ has rank $n$
the mass matrix $N^{-1}MN^\star$ is positive definite if $M$ is positive definite \HORN.} 
through linear combinations from the original divisors $\tilde \Dc_i$, 
which are Poincar\'e dual to the classes $\tilde\omega_i$.
However, since the basis $\{\omega_i\}$ may be expressed as linear
combination of the basis $\{\tilde\omega_i\}$ we obtain linear relations 
between the two moduli $t^j$ and $\tilde t^j$.
When one starts with a Calabi--Yau orientifold $X_6$ with the K\"ahler form 
$\Jc=\sum\limits_{A=1}^{h_{(1,1)}(X_6)} \tilde t^i\ \tilde\omega_i$ and divisor classes
$\tilde \Dc_i\in H^{(4)}(X_6,\IZ)$, which are Poincar\'e dual to the classes 
$\tilde\omega_i\in H^{(2)}(X_6,\IZ)$,
one first has to find a set of independent divisors 
$\Dc_i$ with $\chi(\Dc_i)=1$, \ie a linear combination:
\eqn\divrel{
\Dc_i=\sum_{j=1}^\n n_{ij}\ \tilde \Dc_j\ \ \ ,\ \ \ i=1,\ldots,\n\ .}
In other words, there exists are an invertible $n\times n$ matrix $N=(n)_{ij}$ 
such that $\Dc=N\tilde \Dc$.
The new K\"ahler coordinates $t^i$ entering the K\"ahler form \JJ\ are given by:
\eqn\givennewcoordinates{
\pmatrix{t^1 \cr
\vdots\cr
t^n}=(N^t)^{-1}\pmatrix{\tilde t^1 \cr\vdots\cr \tilde t^n}\ .}
In terms of the these K\"ahler coordinates $t^j$ and the basis $\omega_i$,  
the K\"ahler form \JJ\ becomes a linear combination of contributing 
divisors $\Dc_i$ (Poincar\'e dual to the $\omega_i$).
The change of coordinates \givennewcoordinates\ implies the following new
intersection form $\Kc_{ij}$:
\eqn\newinter{
\Kc_{ij}=N\tilde\Kc_{ij} N^t\ .}
Since $N$ is a non--singular matrix, after Sylvester's law of inertia \HORN,
both intersection forms $\Kc_{ij}$ and $\tilde \Kc_{ij}$ have the same inertia, \ie
they have the signature $(1,\n-1)$.

As an instructive example let us discuss the orientifold of the $X_8$ CYM, \ie
$\IP^4[1,1,1,6,9]$ \Race.
We may introduce the K\"ahler form $\Jc=\tilde t^1 \tilde\omega_1+\tilde t^2 \tilde\omega_2$,
with the intersection numbers $\tilde\Kc_{112}=1, \tilde\Kc_{122}=6$ and $\tilde\Kc_{222}=36$
and the intersection form $\tilde\Kc_{ij}=\pmatrix{\tilde t^2&\tilde t^1+6\tilde t^2\cr
                                                \tilde t^1+6\tilde t^2 &\tilde t^1+36\tilde t^2}$.
With this information, the volume of $X_6$ becomes $Vol(X_6)=\fc{1}{6} [\ 3 (\tilde t^1)^2
\tilde t^2+18\tilde t^1 (\tilde t^2)^2+36 (\tilde t^2)^3\ ]$. 
According to the formula $V_{\tilde \Dc_i}=\fc{3}{4}\ 
\tilde\Kc_{ijk} \tilde t^j\tilde t^k$ the divisor volumes of the two $4$--cycles, 
whose divisors $\tilde \Dc_i$ are Poincar\'e dual to $\tilde \omega_i$ are:
\eqn\twodivisors{\eqalign{
V_{\tilde \Dc_1}&=\fc{3}{2}\ \tilde t^2\ (\tilde t^1+3\ \tilde t^2)\ ,\cr
V_{\tilde \Dc_2}&=\fc{3}{4}\ (\tilde t^1+6\ \tilde t^2)^2\ .}}
However, the two divisors $\Dc_i$ contributing to the superpotential \SUPP\ are linear combinations
\divrel\ of the two divisors $\tilde \Dc_i$ \Race, \ie
\eqn\linearDIV{
V_{\Dc_1}=\fc{2}{3}\ (V_{\tilde \Dc_2}-6\ V_{\tilde \Dc_1})\ \ \ ,\ \ \ 
V_{\Dc_2}=\fc{2}{3}\ V_{\tilde \Dc_2}\ ,}
with the divisor volumes:
\eqn\linearDIVV{
\re(T^1):=V_{\Dc_1}=\h\  (\tilde t^1)^2\ \ \ ,\ \ \ 
\re(T^2):=V_{\Dc_2}=\h\ (\tilde t^1+6\ \tilde t^2)^2\ .}
In terms of the latter, the K\"ahler potential $K_{KM}=-2\ln Vol(X_6)$ for the K\"ahler moduli 
$T^1,T^2$ becomes:
\eqn\exiii{
K_{KM}(T^j,\ov T^j)=\ln 1296-2\ln\lf[\ (T^2+\ov T^2)^{3/2}-(T^1+\ov T^1)^{3/2}\ \ri]\ .}
The divisors $\Dc_i$, introduced in \linearDIV,  
correspond to the $2$--forms $\omega_1=-4\ \tilde\omega_1+\fc{2}{3}\ \tilde\omega_2$
and $\omega_2=\fc{2}{3}\ \tilde\omega_2$, respectively. The latter serve as a basis for 
the K\"ahler form \JJ, \ie $J=t^1\omega_1+t^2\omega_2$, 
with the new\foot{According to \Useful\ these coordinates may be read off from the 
first derivatives of the K\"ahler potential $K_{KM}=K$:
\eqn\sincei{
t^1=-3V\ K_{T^1}=-\fc{1}{4}\ \tilde t^1\ \ \ ,\ \ \   
t^2=-3V\ K_{T^2}=\fc{1}{4}\ (\tilde t^1+6\ \tilde t^2)\ .}}
K\"ahler coordinates $t^i$:
\eqn\newcoord{
t^1=-\fc{1}{4}\ \tilde t^1<0\ \ \ ,\ \ \ t^2=\fc{1}{4}\ (\tilde t^1+6\ \tilde t^2)>0\ .}
The volume of $X_6$ may be written $Vol(X_6)=\fc{16}{9}\ [(t^1)^3+(t^2)^3]$, which 
gives the intersection numbers $\Kc_{111}=\Kc_{222}=\fc{32}{3}$ and
the intersection form $\Kc_{ij}=\fc{32}{3}\ \pmatrix{t^1&0\cr
                                                     0&t^2}$, with signature 
$(1,1)$. Clearly, we have $\Kc_{ij}=N\tilde\Kc_{ij} N^t$, with $N=\fc{2}{3}\ \pmatrix{-6&1\cr
                                                                                       0&1}$.
Due to \newcoord, we see that \Boxi\ is not fulfilled, \ie this criterion does not give us
an answer whether an uplift is possible. However, for the $X_8$ CYM from above we find:
$$t^2>0\ \ \ ,\ \ \ \Kc_{22}=\fc{32}{3}\ t^2>0\ .$$
According to the second criterion \Boxii,
no stable uplift would be  possible for the $X_8$ CYM if we assumed the ansatz \SUPP\
for the superpotential with the two contributing divisors \linearDIVV.
{From} this example we learn, that in practice the inequality \Boxi\
may not always  be satisfied and the condition \Boxii\ needs to be borrowed.


As we shall see in Subsection 5.3, 
the condition  \Boxi\ has serious consequences for the resolved orientifolds of orbifolds
without complex structure moduli. In fact, for those orbifolds, it is generically not
possible to find stable minima in a KKLT setup.
This rules out the following orbifolds: $\IZ_3, \IZ_7, \IZ_3\times \IZ_3, \IZ_4\times \IZ_4,
\IZ_6\times \IZ_6, \IZ_2\times \IZ_{6'}$, and $\IZ_{8-I}$ with
$SU(4)^2$ lattice. It would affect even more orbifolds if we would not take into
account the twisted complex structure moduli.

\subsec{Examples}

In the previous subsection we have derived conditions 
on the K\"ahler potential
$K_{KM}=K$ for the K\"ahler moduli in \tb orientifold compactifications without  
complex structure moduli. These conditions are summarized in \eqqs \Boxi,\Boxii\ 
or alternatively in \eqqs \BoxI,\BoxII. If these criteria are 
met, a stable uplift in a KKLT setup is not possible, since otherwise 
the axion mass matrix $M$ would not be positive definite.
This way, it may immediately be  decided whether a given CY compactification
manifold $X_6$ allows moduli stabilization in the framework of KKLT.
In this subsection, we shall apply these conditions to some prominent 
K\"ahler potentials $K$.

For the K\"ahler potential of the unresolved $\IZ_7$ \tb orientifold \LRSS
\eqn\exio{
K=-\ln(T^1+\ov T^1)-\ln(T^2+\ov T^2)-\ln(T^3+\ov T^3)} 
we find: 
\eqn\failio{
K_{T^1}<0\ \ \ ,\ \ \ K_{T^2}<0\ \ \ ,\ \ \ K_{T^3}<0\ .}
According to \BoxI\ the properties \failio\
result in a negative definite axionic mass matrix $M$. 

Furthermore, for 
the K\"ahler potential \becomes\ of the unresolved 
$\IZ_4$ \tb orientifold we have:
\eqn\failii{
t^1>0\ \ \ ,\ \ \ \Kc_{11}=0\ .}
The last relation follows from the intersection form \exinter.
Hence due to the condition \Boxii, no stable uplift is possible.

Finally, the K\"ahler potential for the $\IZ_3$ \tb orientifold, given in \eqq (2.54)
of {\it Ref.} \LRSS\ (with the triple intersection numbers (2.56) from there), shares again the property \failii.
Hence, due to the condition \Boxii, for the unresolved
$\IZ_3$--orientifold is no stable uplift possible either.

Note, that the above three examples, which do not allow a stable uplift
in the KKLT setup,  represent toroidal orbifold/orientifold compactifications,
whose stability criteria have been investigated in great detail in \LRSS\ (\cf also \ChoiSX).
However, it is intriguing, that we may quickly verify these findings by simply
analyzing the conditions \BoxI\ or \BoxII. 

A different example is the K\"ahler potential for the $X_8$ CYM \exiii, 
with $(T^2+\ov T^2)^{3/2}-(T^1+\ov T^1)^{3/2}>0$.
Moduli stabilization of the latter has been recently discussed in {\it Ref.} \Linde,
with the dilaton $S$ and the complex structure moduli being integrated out, \ie
$W_0(S,U)=const$ in their case. For that case all the equations of Subsection 2.1
apply. In particular, since  (\cf also \eqq \sincei)
\eqn\sincei{
K_{T^1}=\fc{3}{2}\ \fc{(T^1+\ov T^1)^{1/2}}
{(T^2+\ov T^2)^{3/2}-(T^1+\ov T^1)^{3/2}}>0\ ,\  
K_{T^2}=-\fc{3}{2}\ \fc{(T^2+\ov T^2)^{1/2}}
{(T^2+\ov T^2)^{3/2}-(T^1+\ov T^1)^{3/2}}<0}
from \Ratio\ it follows, $\phi_{12}=a_1t_2^1-a_2t_2^2=\pi\ ,$
which indeed has been verified in \Linde.
On the other hand, if we used the K\"ahler potential  \exiii\ together with the 
superpotential \SUPP\ we would immediately see, that in this case no stable
uplift would be possible. This may be easily seen from condition \BoxII.
We find:
\eqn\failiii{
K_{T^2}<0\ \ \ ,\ \ \ -K_{T^1\ov T^1}+\h\ K_{T^1}^2=-\fc{3}{8}\ 
\fc{(T^1+\ov T^1)^{3/2}+2\ (T^2+\ov T^2)^{3/2}}
{(T^1+\ov T^1)^{1/2}\ [(T^2+\ov T^2)^{3/2}-(T^1+\ov T^1)^{3/2}]^2}<0\ .}

\newsec{Stabilization of K\"ahler moduli associated to the cohomology $H^{(1,1)}_{-}(X_6)$}

Because the superpotential $W$ does not depend on the $\m$ 
moduli, \ie $\p_{G^a}W~=~0$, the whole KKLT stabilization procedure only works for the $\n$ 
K\"ahler moduli $T^k$, while we have to find a different mechanism to stabilize the 
$2\ \m$ real fields $b^a,c^a$, which are
combined into the $\m$ N=1 chiral fields \combined.
Since the resolved $\IZ_N$ and $\IZ_N\times \IZ_M$ orbifolds, whose vacuum structure
will be the main part of Section 4, generically have those moduli $G^a$ 
(\cf Table~4), fixing them becomes an important question for these vacua.

A first look at the $F$--term potential
\eqn\FPot{
V_F=
K_{I\ov{J}}\ F^I\ov{F}^{\ov{J}}-3\ e^{\kappa_4^2 K}\ 
\kappa_4^2\ |W|^2\ ,}
with $I,\ov J$ running over all fields $S,T^j,G^a,U^\lambda$ and 
the vacuum  conditions 
$$F^{T^j}=0\ \ ,\ \ F^{G^a}=0\ \ ,\ \ F^S=0\ \ ,\ \ F^{U^\lambda}=0$$
in the full moduli space yields ($\p_{G^a} W=0$):
\eqn\firstlook{\eqalign{
\p_S W+W\ \p_S K&=0\ ,\cr
\p_{T^j} W+W\ \p_{T^j} K&=0\ \ \ ,\ \ \ j=1,\ldots,\n\ ,\cr
W\ K_{G^a}&=0\ \ \ ,\ \ \ a=1,\ldots,\m\ ,\cr
\p_{U^\lambda} W+W\ \p_{U^\lambda} K&=0\ \ \ ,\ \ \ \lambda=1,\ldots,h_{(2,1)}^{(-)}(X_6)\ .}}
As demonstrated  in {\it Refs.} \GL, the K\"ahler derivatives $\p_{T^j} K,\p_S K$ and 
$\p_{U^\lambda} K$ 
become non--trivial functions depending on all moduli $T^j,G^a,S$.
Since at the AdS minimum $W\neq 0$, it follows that 
\eqn\itfollows{
K_{G^a}=0\ .}
According to \GL, there is the relation 
$$K_{G^a}=-\fc{3}{2}\ K_{T^j}\ \Kc_{j ac} b^c=3\ \Kc^{-1}\  \Kc_{jac} t^j\ b^c=3\ \Kc^{-1}\  
\Kc_{ac}\ b^c\ .$$ 
Therefore, the solution to \firstlook\ yields:
\eqn\WEhave{
\Kc_{ac}\  b^c=0\ \ \ ,\ \ \ a=1,\ldots,\m\ .}
Thanks to this condition, to be satisfied at the AdS minimum, all the single derivatives 
$\p_S K$ and $\p_{U^\lambda}K$ become the usual ones, and the $F$--term conditions
for the dilaton and complex structure moduli become the usual ones.
Hence we may proceed with the stabilization procedure as outlined before.
The condition \WEhave\ may be solved with $b^a=0$, \ie half of the degrees of freedom
of the fields $G^a=g_1^a+ig_2^a$ may be fixed.
However, after calculating the scalar potential and the masses for the moduli fields
it may be seen, that the masses for the $g^a_1$ fields become negative
\eqn\negb{
m_{g^a_1g^a_1}=-e^{K}\ |W|^2\ \fc{\p^2 K}{\p (g^a_1)^2}\ ,}
with $\fc{\p^2 K}{\p (g^a_1)^2}=4 K_{G^a\ov G^a}$, 
while the masses for the fields $g_2^a$ are zero:
\eqn\zeroc{
m_{g^a_2g^a_2}=0\ .}
The latter observation is clear, since neither
the K\"ahler potential nor the superpotential do depend on them. Therefore, the scalar 
potential
does not depend on them either. Hence, two problems have occurred:
First, the fields $g_2^a$ cannot be stabilized through the $F$--term equations \firstlook.
Second, though the fields $g_1^a$ are stabilized, their masses are negative.
In fact, it has been recently shown in {\it Ref.} \Conlon,  
that the second problem follows from the first problem.  However, this situation is improved in the
presence of additional $D$--terms, which we shall discuss in the following.

In the case of $\m\neq 0$ in the geometry $Y_6$ of the covering space of the orientifold theory,
some divisors $E$ (or divisor orbits under the orbifold group) 
are not invariant under the orientifold action $\sigma$, but
are mapped to other divisors $\tilde E$ (or orbits under the orbifold group), \ie:
\eqn\happen{
\sigma E =\tilde E\ .}
In that case, invariant combinations have to be constructed\foot{As an illustrative
example see the $\IZ_{6-II}$ orbifold in Subsection 5.4.}. Let
\eqn\invDIV{
E_{i}:=\h\ (E+\tilde E)\ \ \ ,\ \ \ E_{a}:=\h\ (E-\tilde E)\ }
such that: 
\eqn\eigenstates{
\sigma E_i=E_i\ \ \ ,\ \ \ \sigma E_a=-E_a\ .}
Then we have $\omega_i\in H^{(1,1)}_+(X_6)$ and $\omega_a\in H^{(1,1)}_-(X_6)$ for their 
corresponding Hodge dual $2$--forms.
This way, the $\m$ odd forms $\omega_a$ are paired with $\m$ even forms $\omega_i$.
After the discussion in Subsection 2.1, a non--vanishing $\m\neq 0$ implies 
a non--trivial bulk $NSNS$ $2$--form $B_2$
\eqn\bulkB{
B_2=\sum\limits_{\tilde a=1}^\m b^{\tilde a}\ \omega_{\tilde a}\ ,}
which (more precisely its pullback on $C_i$) enters the calibration
conditions of a holomorphic 
$4$--cycle $C_i$ on which a $D3$ or $D7$--brane is wrapped upon \Ruben. 
If one supersymmetry should be preserved, those conditions are solved by demanding  \JL
\eqn\demanding{
B_2 \wedge J=0\ .}
Let us assume the $4$--cycle $C_i$ to correspond to the even divisor $E_i$ and the $4$--cycle 
$C_a$ to correspond to the odd divisor $E_a$.
Then, integrating \demanding\ over the odd cycle $C_a$ yields:
\eqn\Yields{
\int_{C_a} B_2\wedge J= t^j\ b^{\tilde a} \int_{X_6} \omega_{\tilde a}
\wedge \omega_j\wedge\omega_a=
\Kc_{j a\tilde a}\  t^j\ b^{\tilde a}=  \Kc_{\tilde a a}\  b^{\tilde a}=0\ .}
Hence, we conclude:
\eqn\conclude{
\Kc_{\tilde a a}\  b^{\tilde a}=0\ .}
The calibration condition \demanding\ 
has to be imposed on all even $4$--cycles $C_i$.
Therefore the condition \demanding\ gives rise to a system 
of $\m$ linear equations for the 
$\m$ variables $b^a$. The intersection form $\Kc_{ab}$ depends also on the
$\n$ K\"ahler moduli $T^j$. However, generically this matrix has non--vanishing
determinant and the only solution to \conclude\ is the trivial solution\foot{As an illustrative
example see the $\IZ_{6-II}$ orbifold in Subsection 5.4.}:
\eqn\mini{
\re(G^a)=0 \ \ \ \Longrightarrow \ \ \ b^a=0\ \ \ ,\ \ \ a=1,\ldots\m\ .}
Interestingly, the requirement \Yields\ is similar to the condition \WEhave\ 
following from the $F$--terms.

For a $D3$--brane wrapped on the $4$--cycle $C_i$, 
the requirement \demanding\ allows to rewrite the real part of the bosonic instanton action
\eqn\realInstanton{
e^{-\phi_{10}}\ \int_{C_i} d^4\xi\ \sqrt{\det(G+B)}
=\h\ \int_{C_i}J\wedge J-e^{-\phi_{10}}\ B\wedge B=\fc{2}{3}\ \re(T^i)\ ,}
with $T^i$ given in \defmod\ (and the internal CY metric $G$).

On the other hand, for a stack of $D7$--branes wrapped on the $4$--cycle $C_i$, 
the requirement \conclude\ may be also anticipated from the $D$--term potential \JL
\eqn\DPot{
V^i_D\sim \fc{1}{\re (T^i)}\  (\ \Kc_{a\tilde a}\  b^{\tilde a}\ )^2\ ,}
associated to a $U(1)$ gauge group\foot{In the case of additional
matter charged under this $U(1)$, the potential \DPot\ receives
an additional term proportional to the matter fields $C_i$. Such matter fields originate 
from open strings stretched between different $D7$ or/and $D3$--branes. For simplicity 
we shall assume that these matter fields are minimized at $C_i=0$.} 
whose gauge coupling is given by $g^{-2}=\re(T^i)$
(the $D7$--brane associated to this $U(1)$ is assumed to wrap the $4$--cycle $C_i$).
For a stack of $D7$--branes with $U(N)$ gauge group, the remaining $SU(N)$ gauge group may 
furnish gaugino condensation. The potential \DPot\  is minimized if condition \Yields\ holds. 

Let us now come to the axionic moduli $c^a$ in  \combined.
The case $\m\neq 0$ implies the non--trivial bulk $RR$ $2$--form $C_2$:
\eqn\bulkC{
C_2=\sum\limits_{a=1}^\m c^a\ \omega_a\ .}
If the stack of $D7$--branes wrapping the $4$--cycle $C_i$ 
has a gauge group with an $U(1)$--factor,
the fields $c^a$ and $\rho^i$ are gauged under this $U(1)$ gauge field $A_\mu$
as a result of the $D7$ world--volume couplings $C_6\wedge F_2$ and $C_4\wedge B_2\wedge F_2$, 
respectively, producing Green--Schwarz terms. 
As a consequence of the definition \combined\ the field $G^a$ 
is charged under this $U(1)$, 
while the K\"ahler modulus $T^i$ remains neutral due to a non--trivial cancellation 
of the gauge transformations of the axionic fields $c^a$ and $\rho^i$  entering the definition
of $T^i$ \defmod, \ie
\eqn\covDER{
D_\mu G^a=\p_\mu G^a-4\ i\ \kappa_4^2\ \mu_7\ (2\pi\ap)\ A_\mu\ \ \ ,\ \ \ D_\mu T^j=\p_\mu T^j\ ,}
\cf {\it Ref.} \JL\ for more details. Hence, the $D$--term potential \DPot\ 
stabilizes the fields $b^a$,
while the covariant derivative $D_\mu G^a$ gives rise to a massive $U(1)$ vector $A_\mu$ 
thus fixing the imaginary part of $G^a$, \ie the field $c^a$ \DSW. The latter is absorbed
into the mass of the gauge field. Since only $\im(T^j)$, but not $\im(G^a)$ couples to 
any $F\wedge F$ piece, the mass term may even occur without
an anomaly as a  St\"uckelberg mass term \doubref\Getting\AM.
To summarize so far, the calibration condition \demanding\ and the gauging \covDER\ of the 
modulus $G^a$ allow to stabilize the K\"ahler moduli associated to the cohomology 
$H_-^{(1,1)}(X_6)$.

\def\fa{^Y\hskip-0.15cm f^a}
\def\f{^Y\hskip-0.15cm f}
The mechanism to stabilize the axionic fields $c^a$ relies on the existence of a 
$U(1)$. If the stack of $D7$--branes sits on an orientifold
plane we encounter an $SO$-- or $Sp$ gauge group and the previous mechanism has to be modified.
In that case, we turn on additional world--volume $2$--form fluxes on that stack of 
$D7$--branes to break off one $U(1)$ factor.
In the orientifold on a stack of $D7$--branes,
there are two kinds of world--volume $2$--form fluxes $\f$ and $\tilde f$. 
The flux $\f$ is inherited from the ambient CY space $Y$, while the flux $\tilde f$
is a harmonic $2$--form of  the $4$--cycle $C_i$. We refer the reader to {\it Ref.} \JL\ 
for a description of the orthogonal  splitting of a general $2$--form flux $f$ into
$f=\f+\tilde f$, with $\f\in im(\iota^\star),\ \tilde f\in coker(\iota^\star)$,
with the map $\iota^\star: H^{(2)}_-(Y_6)\ra  H^{(2)}_-(C_i)$.
In the following, we do not want to consider the class of fluxes $\tilde f$, 
since they would gauge the K\"ahler moduli $T^j$, \ie $\tilde f=0$.
On the other hand, turning on $2$--form fluxes $\f_2$ from the ambient space $Y_6$, 
\eqn\bulkF{
\f_2=\sum\limits_{a=1}^{h_{(1,1)}^{(-)}(Y_6)} f^a\ \iota^\star \omega_a,}
does not gauge the K\"ahler moduli $T^j$ \JL.
These fluxes only modify the previous equations with the effect, 
that the $2$--form $B_2$ is replaced by the combination $B_2-2\pi\ap\ \f_2$, \ie we 
supplement \eqqs \combined,\demanding, \conclude, \realInstanton\ and \DPot\ with the 
following substitution \JL:
\eqn\effect{
B_2\longrightarrow B_2-2\pi\ap\ \f_2\ \ \ ,\ \ \ b^a\longrightarrow b^a-2\pi\ap\ \fa\ .}
Hence the calibration condition \demanding\ is changed into
\eqn\Yieldsi{
\Kc_{\tilde a a}\ \lf(\ b^{\tilde a}-2\pi\ap\ \f^{\tilde a}\ \ri)=0\ ,}
instead of \Yields. Therefore, the solution \mini\ becomes:
\eqn\Mini{
b^a=2\pi\ap\ \fa\ \ \ ,\ \ \ a=1,\ldots\m\ .}
Since the gaugings \covDER\ now hold for the redefined modulus 
$$G^a=i\ c^a-S\ (b^a-2\pi\ap\ \fa),$$
the previous discussion about the stabilization of the axion $c^a$ by making one $U(1)$ gauge
field $A_\mu$ massive may be directly applied.

To conclude, the calibration condition \demanding\ together with the gauging \covDER\ of the 
modulus $G^a$ allow to stabilize this modulus by making one $U(1)$ gauge
field $A_\mu$ massive. If the latter does not exist, it may be generated by turning on 
$2$--form fluxes $\f$ from the ambient space $Y$. It is crucial, that under 
the background fluxes $b^a$ and $\fa$ the K\"ahler modulus $T^i$ remains uncharged, \ie
\covDER\ holds.
 
In N=2 supergravity, background fluxes 
may be associated to some gaugings of continuous $PQ$--symmetries acting on axions of the 
hypermultiplet scalars.  
In the \tb case, the $3$--form fluxes $G_3$ correspond to gaugings of
scalars of the universal hypermultiplet, while the fields $c^a$ and $\rho^i$ are 
gauged by turning on non--trivial $b^a$--fields and the world 
volume $2$--form fluxes $\fa,\tilde f^a$ on the $D$--brane world volume.
Fluxes may protect those isometries, which are gauged in supergravity, from quantum corrections. 
A kind of complementarity arises between turning on fluxes, \ie gauging some continuous 
$PQ$ symmetries of hypermultiplet scalars and instanton corrections, which
break continuous $PQ$ symmetries. 
The compatibility of both effects has been recently studied in \KT. 
After turning on world volume $2$--form fluxes,
the  axions of the K\"ahler moduli $T^i$ and $G^a$ are the potential candidates for being gauged.
Remember, that the K\"ahler moduli $T^i$ correspond  to contributing divisors in the 
non--perturbative superpotential \SUPP\ accounting for $D3$--instanton effects
and gaugino condensation. Hence, their continuous $PQ$ symmetries are broken by instanton effects
rather than by fluxes.
However, in the case that only the flux $\f$ from the ambient space $Y$ is turned on, while
the $2$--form flux $\tilde f$ is absent, 
the K\"ahler modulus $T^i$ indeed remains neutral \JL. 
It is that case we have focused on in the above discussion. 
Hence the instanton effects in the KKLT superpotential \SUPP\ are compatible with 
turning on only $2$--form fluxes $\fa$ from the ambient space $Y$.
We have a complementary situation: one class of K\"ahler moduli $T^i$ is stabilized through
instanton effects in the superpotential, while the other class of K\"ahler moduli $G^a$ is
fixed through gauging isometries \covDER\ and turning on the background fluxes $b^a,\fa$.

\break
\newsec{Instanton effects}

In this section, we  discuss the question for which cases a
non--perturbative superpotential 
from brane instantons is produced. 
As shown in \WittenBN , if a divisor, wrapped by a $M5$-brane in the dual 
 $M$-theory picture (dual to type $IIB$ which one is considering), has holomorphic Euler
characteristic 
$$\chi=h_{(0,0)}-h_{(0,1)}+h_{(0,2)}-h_{(0,3)}=1,$$  
the necessary two 
fermionic zero modes for the instanton contribution 
will be present. To discuss our models we  use the results of 
\multrefv\KalloshYU\SaulinaVE\KalloshGS\BergshoeffYP\ParkHJ,
where it was shown how the zero modes counting is changed in the presence of background 
fluxes. The advantage of the counting procedure in \BergshoeffYP\ is that it is not 
necessary to do an $F$-theory lift, the calculations can be done directly in the type $IIB$ picture. 
Additionally, it was shown in \LustCU\ that only the $(2,1)$--component of the $G_3$-flux may  
lift zero modes. 

To calculate the number of the zero modes, we have to realize what are the possible 
4-cycles wrapped by the $D3$-branes in the compact space. The Hodge numbers 
$h_{(0,0)},h_{(0,1)},h_{(0,2)}$ of the 4-cycle give the number of the zero modes 
with positive ($N_+$) and negative ($N_-$) chirality with respect to the normal bundle 
of the $D3$--brane. If one takes into account background fluxes, orientifold action and 
fixing of the $\kappa$-symmetry, some of the zero modes could be lifted and the index 
$$\chi_{D3}={1\over 2}\left(N_+-N_-\right)$$ will change. $\chi_{D3}$ is not anymore of purely geometrical 
nature. In the case  of type $IIB$, Bergshoeff et al. \BergshoeffYP\ showed that only 
$h_{(0,1)}$ and $h_{(0,2)}$ of $N_+$ can be lifted by fluxes. Thus, if the topology of the divisor 
has vanishing $h_{(0,1)}$, $h_{(0,2)}$, we can neglect the effect of the fluxes altogether and concentrate only on the action of the $O$--planes on the zero mode counting.


The correspondence between zero modes of the Dirac operator on the worldvolume of the 4-cycles and Hodge numbers $h_{(0,0)},h_{(0,1)},h_{(0,2)}$ of these cycles becomes apparent by mapping the spinors to  $(0,p)$-differential forms.{\foot{$\phi_{a_1\ldots a_N}\gamma^{a_1\ldots a_N}|\Omega>\longleftrightarrow\phi_{a_1\ldots a_N}dz^{a_1}\ldots dz^{a_N} $}} Then fermionic zero modes of the Dirac operator correspond  to the harmonic forms  by above mapping.  Locally we can write the world volume spinors on the $D3$--brane as
%
\eqn\spinorenB{\eqalign{
\epsilon_+&=\phi |\Omega> +\phi_{\bar a} \gamma^{\bar a} |\Omega>+\phi_{\overline{ab}}\gamma^{\overline{ab}}|\Omega> \ , \cr
\epsilon_-&=\phi_{\bar z}\gamma^{\bar z} |\Omega> +\phi_{\overline {az}} \gamma^{\overline {az}} |\Omega>+\phi_{\overline{abz}}\gamma^{\overline{abz}}|\Omega> \ ,
}}
where $\epsilon_+$ and $\epsilon_-$ are states with positive and negative chirality 
with respect to the normal bundle $SO(2)$ of the $D3$--brane inside the compact space. $a,b$ are the $D3$--brane worldvolume directions, $z$ is the normal direction to
the worldvolume. 

Note that $\epsilon_+$ and $\epsilon_-$ transform under $SO(4)\times SO(2)\times SO(1,3)$ and the modes $\phi$ have an additional spinor index which transforms in the ${\bf{2}}\oplus {\bf\bar 2}$ under $SO(1,3)$. Thus, the number of the zero modes given by the Hodge numbers of the 4-cycle has  to be doubled.

All modes of $\epsilon_-$ have legs in the normal direction to the
$D3$--brane. By use of Serre's generalization  of the Poincar\'e
duality, these modes can be mapped to those taking values in the
worldvolume of $D3$--brane. This duality maps $(0,p)$-forms with
values in the  bundle $\Omega^{0,p}(X)$ of  the 4--cycle $X$ to
$(0,2-p)$--forms with values in $\Omega^{(0,2-p)}\otimes K$, where $K$
is the canonical bundle of the 4--cycle. In the case of the wrapped
$D3$--brane, the canonical bundle is equal to the normal bundle, so
this duality is realized by multiplying by the covariantly constant
3--form $\Omega_{abc}$ and building the Hodge\foot{\ 
\vskip-1cm\eqn\hodge{\eqalign{
*\Big( \phi_{\overline{a}_1\ldots \overline{a}_{N-p}}\Omega_{a_1\ldots a_N}dz^{\overline{a}_1}\dots &
dz^{\overline{a}_{N-p}}dz^{a_1} \dots dz^{a_{N}}\Big)  \cr &=\epsilon_{{\overline{a}_1}\ldots 
\overline{a}_N} \epsilon_{{a}_1 \ldots {a_N}}\phi^{\overline{a}_1\ldots\overline{a}_{N-p}}
\Omega^{a_1\ldots a_N}\ dz^{\overline{N-p+1}}\ldots dz^{\overline{N}}
}} Note, that in our convention the form is complex conjugated by applying the Hodge star.}
dual.
\eqn\serrduality{\eqalign{
g^{z\bar z}\Omega_{\overline{abz}}\phi_{ z}=\tilde\phi_{\overline{ab}}\ ,\cr
g^{z\bar z}g^{a\bar a}\Omega_{\overline{abz}}\phi_{az}=\tilde\phi_{\overline a} \ , \cr
g^{z\bar z}g^{a\bar a}g^{b\bar b}\Omega_{\overline{abz}}\phi_{abz}=\tilde\phi \ . \cr
}}

%

This means that the numbers of
the modes with positive and negative chirality match. If all 
zero modes are present, the corresponding index
\eqn\indexeq{
\chi_{D3}=\h\ \left(N_+-N_-\right)=\lf(\ h_{(0,0)}+h_{(0,1)}+h_{(0,2)}\
\ri)-\lf(\ h_{(0,0)}+h_{(0,1)}+h_{(0,2)}\ \ri)  }
will be $0$.



\subsec{Calculation of $\chi_{D3}$ for divisors with 
$h_{(0,0)}=1$, $h_{(0,1)}=h_{(0,2)}=0$}

As we shall see in the next section, many of the divisors arising in resolved toroidal orbifold models have the Hodge numbers $h_{(0,0)}=1$, $h_{(0,1)}=h_{(0,2)}=0$.
We will therefore start by calculating the number of zero modes for this especially simple case.
We choose $a,\bar a,b, \bar b$ as holomorphic and antiholomorphic coordinates on the $D3$--brane, 
which take the values $1,\bar 1, 2, \bar2$. $z$ and $\bar z$ should correspond to the transverse directions 
with values $3,\bar 3$. 

The fermionic states on the $D3$--brane corresponding to $h_{(0,0)}$ are
\eqn\spinoren{
\epsilon_+=\phi|\Omega>\ , \qquad \epsilon_- = \phi_{\overline{abz}}\gamma ^{\overline{abz}}|\Omega> \ .
}

On the brane, some of the modes are pure gauge due to the $\kappa$-symmetry. These are the modes which are annihilated by the $\kappa$-symmetry projector $(1-\Gamma_{D3})\theta=0$, where $\Gamma_{D3}=\sigma_2\otimes\gamma_5$ with $\gamma_5$ four ten dimensional $\gamma$-matrices pulled back on the brane. $\theta$ corresponds to two 32--component spinors written in the double spinor formalism \MartucciRB. Additionally, some of the modes can be projected out by the orientifold action. We have to choose $\kappa$-symmetry fixing in such a way that it commutes with the orientifold action \BergshoeffYP. 

There are three different cases to distinguish for the position of the $O7$--plane: it can be on top of  the D3--brane, can intersect it along one direction, or can be parallel to it. We assume that the $O7$--plane fills the non-compact directions.
\vskip 12pt
$\bullet$  {Case 1: an $O$-plane lies on top of  a D3--brane}

\noindent
It is convenient to do the calculations in the local coordinate patch. The
$\kappa$-symmetry fixing condition and the projection through the orientifold action are given by

\eqn\fixingconditions{\eqalign{
(1-\sigma_2\gamma^{1\bar 1 2 \bar 2})\theta =0 \ , \cr
(1-\sigma_2\gamma^{3 \bar 3})\theta =0  \ .
}}
Both conditions written together yield

\eqn\both{
(1-\gamma ^{1\bar 12\bar 23\bar 3})\theta=0 \ .
}
Inserting $\theta=\epsilon_++\epsilon_-$ shows that $\phi$ survives this projection and $\phi_{\overline{abz}}$ not. The index is $\chi_{D3} = h _{(0,0)}=1$.
\vskip 12pt
$\bullet$ {Case 2: Intersection with an $O$--plane along one complex dimension}

The $O$--plane intersects the $D$--brane along complex direction $1$. Then,
$\kappa$-symmetry fixing condition and the projection through the orientifold action are given by
\eqn\conditionsB{\eqalign{
(1-\sigma_2\gamma^{1\bar 1 2 \bar 2})\theta =0 \ , \cr
(1-\sigma_2\gamma^{2 \bar 2})\theta =0   \ .
}}
Both conditions written together give
\eqn\bothB{
(1-\gamma ^{1\bar 1})\theta=0 \ .
}
$\phi$ survives this projection, $\phi_{\overline{abz}}$ not.
From this it follows $\chi_{D3} = h _{(0,0)}=1$.
\vskip 12pt
$\bullet$ {Case 3: No intersection with an $O$--plane}

\noindent
It can be the case, when the $O$--plane is parallel to the $D3$--brane. The orientifold action maps fermions of the brane to the fermions in the mirror brane, so no modes are projected out. There is only the $\kappa$-symmetry fixing condition, by which no modes are cut. The modes $\phi|\Omega>$, $\phi_{\overline{abc}}\gamma^{\overline{abc}}|\Omega>$ corresponding to $h_{(0,0)}$  are present and the index is  $ \chi_{D3} = h _{(0,0)}-h_{(0,0)}=0$.
 
By investigating all configurations of the $O7$--plane we obtain a general statement: 
\vskip 12pt
{\it Divisors with Hodge numbers $h_{(0,0)}=1$, $h_{(0,1)}=h_{(0,2)}=0$
will have the index $\chi_{D3}=1$ if
an $O7$--plane lies on top of them or if it
intersects the divisor along one complex dimension. Otherwise, $\chi_{D3}=0$.
}
\subsec{General case: $h_{(0,1)},\,h_{(0,2)}\neq 0$}

As discussed in the last subsection, locally, there are always only three different configurations of the $O7$--plane relative to the divisor in question. It can be on top of it, intersect it in one complex direction, or be parallel to it. $a,b$ are again the coordinates on the $D3$--brane. The projector equations from the fixing of the $\kappa$-symmetry and the orientifold action will be as in the previous subsection. The only difference is that the modes $\phi_{\overline a}|\Omega>$ and $\phi_{\overline{ab}}\gamma^{\overline{ab}}|\Omega>$ are now present. They can be lifted by fluxes. When turning on fluxes, we assume that they will be of the the most unfavorable form for the presence of zero modes. This would correspond to a general form for the fluxes.
We summarize  the results of the action of the projector equations  in all three cases in the following table:

\vskip0.5cm
{\vbox{\ninepoint{$$
\vbox{\offinterlineskip\tabskip=0pt
\halign{\strut\vrule#
&~$#$~\hfil 
&\vrule$#$ 
&~$#$~\hfil 
&\vrule$#$ 
&~$#$~\hfil 
&\vrule$#$
&~$#$~\hfil 
&\vrule # \cr
\noalign{\hrule}
&  &&O-{\rm plane} &&O-{\rm plane} &&O-{\rm plane}&\cr
&  &&{\rm  on\ top\ of}\ D3 &&{\rm intersects}\ D3 &&{\rm does \ not \ intersect}\ D3 &\cr
\noalign{\hrule}
\noalign{\hrule}
&\hfil \rm chirality              &&\ \ \ +\hfill-\ \ \  \ &&\ \ \ \ \ +\hfill-\ \ \ \ \ \ \   &&\ \ \ \ \ \ +\hfill- \ \ \ \ \ \ \ \  &\cr
\noalign{\hrule}
&\hfil h_{(0,0)}     &&\ \  \ \phi       &&\ \ \ \ \ \phi    &&\ \ \ \ \ \  \phi \hfill \phi_{\overline{abz}} \ \ \ \ \ \   &\cr
&\hfil h_{(0,1)}     &&\ \hfill \phi_{\overline{az}} \ \  \ &&\ \ \ \  [\phi_{\overline a}]\hfill \phi_{\overline{az}} \ \ \ \ \    &&\ \ \ \ \  [\phi_{\overline a}]\hfill \phi_{\overline{az}} \ \ \ \ \ \ \   &\cr
&\hfil h_{(0,2)}     &&\ \  [\phi_{\overline{ab}}]  &&\ \ \ \ \ \hfill \phi_{\overline z} \ \ \ \ \ \   &&\ \ \ \ \   [\phi_{\overline{ab}}]\hfill \phi_{\overline z} \ \ \ \ \ \  \ \ &\cr
\noalign{\hrule}
&\ &&\ &&\ &&\ &\cr
&  \# \rm \ of \ zero \ modes
&&2-2\ h_{(0,1)}^{(-)}+2\ [h_{(0,2)}^{(+)}] &&\hfil 2-2\ h_{(0,1)}^{(-)}
&&2\ [h_{(0,1)}^{(+)}]+2\ [h_{(0,2)}^{(-)}]&\cr
&&&&& -2\ h_{(0,2)}^{(-)}+2\ [h_{(0,1)}^{(+)}] &&-2\
h_{(0,1)}^{(-)}-2\ h_{(0,2)}^{(+)}  &\cr
\noalign{\hrule}
}}$$ 
\vskip-6pt
\centerline{\noindent{\bf Table 2:}
{\sl Zero modes after fixing $\kappa$-symmetry and orientifold projection}}
\vskip10pt}}}
\vskip-0.5cm
\br

In the horizontal line we give the zero modes  associated to the Hodge numbers 
$h_{(0,0)},h_{(0,1)},h_{(0,2)}$. '+' and '-' denote the chirality with respect 
to the normal bundle of the $D3$--brane. In brackets we put the modes which are in 
general lifted in the presence of fluxes, and in the last line we give the number of 
zero modes which are left.
 
Let us discuss this result first before turning on flux. The first column represents 
the case where the influence of the orientifold projection is fully felt by the divisor 
in question. As in the $M$--theory case discussed by Witten, only one chirality survives 
for each Hodge number and the index reduces again to the holomorphic Euler characteristic 
$h_{(0,0)}-h_{(0,1)}+h_{(0,2)}$. The third column corresponds to the case where the influence 
of the orientifold is not felt at all by the wrapped divisor. Both chiralities survive and 
cancel each other out. This agrees with the observation that for a compactification on a CY 
manifold (without orientifold projection), no non--perturbative superpotential is generated. 
The second column represents an intermediate case. It is obvious that the knowledge of the 
Hodge numbers is of prime importance to be able to decide whether a divisor contributes to 
the non--perturbative superpotential.
We can see that without turning on flux, we can get a contribution in the first column for 
$h_{(0,1)}=h_{(0,2)}$. With flux, a contribution is only possible for $h_{(0,1)}=0$. We get 
a contribution from the second column for $h_{(0,2)}=0$ if no flux is turned on, with flux only 
for $h_{(0,1)}=h_{(0,2)}=0$. With or without flux, column three never gives a contribution 
since the number of zero modes is always less than or equal to zero.

In the present work we do not discuss the counting of zero modes for  the
case with 
non--vanishing $2$--form flux $f$ on the $D$--brane world--volume, as
it has been discussed in Section 3. 
Recently, work towards this direction has been accomplished in {\it Ref.} \AnguelovaSJ\ 
for the case of heterotic $M$-theory.
The authors have found that world--volume flux does not change the
zero mode counting for the case of some particular background fluxes.
Those fluxes were chosen such that they do not lift any zero
modes. For the case of $IIB$, Bandos and Sorokin 
derived the Dirac equation for the $D3$--brane in the presence of worldvolume flux \BandosWB. 
Its implication for the zero mode counting has to be analyzed \Waldemar.
Compared to the case without $2$--form flux, there are more complicated 
conditions on the gauge fixing of the $\kappa$-symmetry
and an additional field equation for the $2$--form flux, which  depend
on the topology of the Calabi--Yau manifold. 

We finish this section 
by the remark  that the formalism described above requires the wrapped
4--cycle to be K\"ahler. 
In our models it is the case because we have to deal with divisors
which are hyperplanes in the 
Calabi--Yau manifolds and hyperplanes of a  K\"ahler manifold are
K\"ahler. This can be also 
directly observed from the topology of the divisors, which are either
products or fibrations 
of tori and  $\IP^1$s.

\newsec{Fixing  all moduli  in resolved 
\tb $\IZ_N$-- and $\IZ_N\times \IZ_M$--orientifolds}

\subsec{Type $IIB$ orientifolds of $\IZ_N$-- and $\IZ_N\times \IZ_M$--orbifolds}

In the following we shall investigate moduli stabilization for 
a class of \tb orientifold compactifications $X_6$.
We shall discuss orientifolds $X_6$ of the resolved toroidal orbifolds $Y_6$
\eqn\orbi{
Y_6={T^6}/{\Gamma}\ \ \ \ \ ,\ \ \ \ \ \Gamma=\IZ_N\ \ \ ,\ \ \ \IZ_N\times \IZ_M\ ,}
with orbifold group $\Gamma$.

\vskip0.5cm
{\vbox{\ninepoint{$$
\vbox{\offinterlineskip\tabskip=0pt
\halign{\strut\vrule#
&~$#$~\hfil 
&\vrule$#$
&~$#$~\hfil 
&\vrule$#$ 
&~$#$~\hfil 
&\vrule$#$
&~$#$~\hfil 
&\vrule$#$ 
&~$#$~\hfil 
&\vrule$#$
&~$#$~\hfil 
&\vrule$#$&\vrule$#$\cr
\noalign{\hrule}
&\ \IZ_N &&{\rm lattice}\ T^6 && h^{\rm untw.}_{(1,1)} &&  h^{\rm untw.}_{(2,1)} &&
h_{(1,1)}^{\rm twist.}&& h^{\rm twist.}_{(2,1)} &\cr
\noalign{\hrule}\noalign{\hrule}
& \ \IZ_3      &&\   SU(3)^3            &&9 &&0&&27 &&0&\cr
& \  \IZ_4      &&\    SU(4)^2         &&5 &&  1&&20&&0&\cr
& \  \IZ_4      &&\     SU(2)\times SU(4)\times SO(5)  &&5&&  1&&22&&2&\cr
& \    \IZ_4       &&\     SU(2)^2\times SO(5)^2   &&5  &&  1&&26&&6&\cr
& \  \IZ_{6-I}     &&  (G_2\times SU(3)^{2})^{\flat}   &&5 &&0&&20&&1&\cr
& \  \IZ_{6-I}     &&    SU(3)\times G_2^2  &&5 && 0&&24&&5&\cr
& \  \IZ_{6-II}    && SU(2)\times SU(6)     && 3  && 1&&22&&0&\cr
& \  \IZ_{6-II}    && SU(3)\times SO(8)  &&3 &&1&& 26&&4&\cr
& \  \IZ_{6-II}    && (SU(2)^2\times SU(3)\times SU(3))^{\sharp}   &&3 &&1&&28&&6&\cr
& \  \IZ_{6-II}    &&  SU(2)^2\times SU(3)\times G_2   &&3 &&1&&32&&10&\cr
& \  \IZ_7         &&  SU(7)                               &&3 &&0&&21&&0&\cr
& \    \IZ_{8-I}     &&    (SU(4)\times SU(4))^*    &&3 && 0&&21&&0&\cr
& \    \IZ_{8-I}     &&   SO(5)\times SO(9)          &&3& &  0 &&24&&3&\cr
& \    \IZ_{8-II}    &&    SU(2)\times SO(10)        &&3 &&  1&&24&&2&\cr
& \    \IZ_{8-II}      &&    SO(4)\times SO(9)    &&3&&  1&&28&&6&\cr
& \  \IZ_{12-I}    &&  E_6    &&3&&0&&22&&1&\cr
& \  \IZ_{12-I}    &&  SU(3)\times F_4   &&3 &&0&&26&&5&\cr
& \    \IZ_{12-II}    &&   SO(4)\times F_4   &&3&&  1&&28&&6&\cr
& \    \IZ_2 \times\IZ_2      &&   SU(2)^6   &&3 && 3&&48&&0&\cr
& \    \IZ_2 \times\IZ_4     &&   SU(2)^2\times SO(5)^2  &&3&&  1&&58&&0&\cr
& \    \IZ_2 \times\IZ_6      &&    SU(2)^2\times SU(3)\times G_2 &&3&&  1&&48&&2&\cr
& \    \IZ_2 \times\IZ_{6'} &&    SU(3)\times G_2^2 &&3 &&  0&&33&&0&\cr
& \  \IZ_3 \times\IZ_3      && SU(3)^3    &&3&&0&&81&&0&\cr
& \  \IZ_3 \times\IZ_6      && SU(3)\times G_2^2  &&3 &&0&&70&&1&\cr
& \    \IZ_4 \times\IZ_4      &&   SO(5)^3   &&3&&  0&&87&&0&\cr
& \    \IZ_6 \times\IZ_6      &&   G_2^3  &&3&&  0&&81&&0&\cr
\noalign{\hrule}}} $$
\vskip-6pt
\centerline{\noindent{\bf Table 3:}
{\sl Orbifold groups, lattices and Hodge numbers for $\IZ_N$ and $\IZ_N\times\IZ_M$ orbifolds.}}
\vskip10pt}}}
\vskip-0.5cm
\br
The group generator $\theta\in\Gamma$ acts as follows on the complex coordinates of $T^6$
$$\theta:\ (z^1,z^2,z^3)\longrightarrow (e^{2\pi i\  {v^1}}\ z^1\ ,\ 
e^{2\pi i \ {v^2}}\ z^2\ ,\ e^{2\pi i \ {v^3}}\ z^3)\ ,$$
with $\pm v^1\pm v^2\pm v^3=0$ to furnish  $SU(3)$ holonomy ($\theta\ \Omega=\Omega$) \DHVW.
In Table 3, we give a list of possible $\IZ_N$ and $\IZ_N\times\IZ_M$ orbifolds, together 
with the torus lattices they live on and their Hodge numbers.
The lattices marked with $\flat$, $\sharp$, and $*$ are realized as generalized Coxeter twists, 
the automorphism being in the first and second case
$S_1S_2S_3S_4P_{36}P_{45}$ and in the third  
$S_1S_2S_3P_{16}P_{25}P_{34}$.

The twist elements $\theta,\ldots,\theta^{N-1}$ produce conical singularities.
In a small neighborhood around them, the space locally looks like 
$\IC^3/\Gamma$ (isolated singularity) 
or $\IC^2/\Gamma^{(2)}\times \IC$  (non--isolated singularity).
In {\it Ref.} \first\ 
these singularities are resolved using the methods of toric geometry resulting in
a smooth Calabi--Yau space $Y_6$. Afterwards a consistent orientifold action $\Oc$ is 
introduced, resulting in the Calabi--Yau orientifold $X_6$.
After resolving the orbifold, three kinds of divisors $\Dc$ appear, 
namely $E_\alpha$, $D_{i\alpha}$, and $R_i$. 
The divisors $E_\alpha$ are the exceptional divisors arising from
the resolution of an orbifold singularity $f_\alpha$ (or an orbit under the orbifold group), 
while the divisors $D_{i\alpha}$ denote hyperplanes passsing through fixed points: 
$D_{i\alpha}=\{z^i=z^i_{fixed,\alpha}\}$. The divisors $R_i=\{z^i=c\}$
for $c\not =z^i_{fixed,\alpha}$ are hyperplanes not passing through a
fixed point \first. As opposed to the $D_{i\alpha}$ they are allowed to move.

As already anticipated in Section 3, some divisors $E$ (or divisor orbits
under the orbifold group on the $T^6$) in the geometry of the covering space $Y_6$
may not be invariant under the orientifold action $\sigma$, \cf \eqq \happen.
In this case, a pair of divisors $(E_i,E_a)$, which are eigenstates (with eigenvalues 
$\pm 1$) under $\sigma$ may be constructed, \cf \eqq \eigenstates.
To this end, the original number of divisors $h_{(1,1)}(Y_6)$ is 
split into $h^{(+)}_{(1,1)}(X_6)$ even  and $h^{(-)}_{(1,1)}(X_6)$ odd  divisors.
These numbers are determined for the orientifolds of the resolved orbifolds 
\orbi\ in {\it Ref.} \first\ and are displayed in Table 4.
In the previous section we have discussed the different stabilization mechanisms 
for these two kinds of K\"ahler moduli.

\vskip0.5cm
{\vbox{\ninepoint{$$
\vbox{\offinterlineskip\tabskip=0pt
\halign{\strut\vrule#
&~$#$~\hfil 
&\vrule$#$
&~$#$~\hfil 
&\vrule$#$ 
&~$#$~\hfil 
&~$#$~\hfil 
&\vrule$#$
&~$#$~\hfil 
&\vrule$#$ 
&~$#$~\hfil 
&\vrule$#$&\vrule$#$\cr
\noalign{\hrule}
&\ \Gamma && h^{(+)}_{(1,1)} && h_{(1,1)}^{(-)}&&\ \Gamma &&\ h^{(+)}_{(1,1)} 
&&\ h_{(1,1)}^{(-)}&\cr
\noalign{\hrule}\noalign{\hrule}
& \ \IZ_3             &&23 &&13&& \    \IZ_{8-II}      &&\ 27 && \  4\ &\cr
& \  \IZ_4            &&25 &&  6&& \    \IZ_{8-II}       &&\ 31 &&\   0\ &\cr
& \  \IZ_4       &&27&&  4&& \  \IZ_{12-I}    &&\ 18&&\ 6\ &\cr
& \    \IZ_4        &&31  &&  0&& \  \IZ_{12-I}       &&\ 22 &&\ 6\ &\cr
& \  \IZ_{6-I}       &&19 &&6&& \    \IZ_{12-II}       &&\ 31&& \  0\ &\cr
& \  \IZ_{6-I}   && 23 && 6&& \    \IZ_2 \times\IZ_2         &&\ 51 && \  0\ &\cr
& \  \IZ_{6-II}       && 19  && 6&& \    \IZ_2 \times\IZ_4      &&\ 61&&\   0\ &\cr
& \  \IZ_{6-II}   &&23 &&6&& \    \IZ_2 \times\IZ_6      &&\ 51&& \  0\ &\cr
& \  \IZ_{6-II}    &&21 &&8&& \    \IZ_2 \times\IZ_{6'}  &&\ 36 && \  0\ &\cr
& \  \IZ_{6-II}     &&25 &&8&& \  \IZ_3 \times\IZ_3      &&\ 47 &&\ 37\ &\cr
& \  \IZ_7            &&15 &&9&& \  \IZ_3 \times\IZ_6      &&\ 51 &&\ 22\ &\cr
& \    \IZ_{8-I}         &&24 &&  5&& \    \IZ_4 \times\IZ_4       &&\ 90&& \  0\ &\cr
& \    \IZ_{8-I}          &&27& &  0&& \    \IZ_6 \times\IZ_6      &&\ 84&&\   0\ &\cr
\noalign{\hrule}}} $$
\vskip-6pt
\centerline{\noindent{\bf Table 4:}
{\sl Hodge numbers $h_{(1,1)}(X_6)$ after the orientifold action}}
\vskip10pt}}}
\vskip-0.5cm
\br

We choose the orientifold action such that it gives rise to $O3$--planes and $O7$--planes. 
On the local $\IC^3/\Gamma$ patches, an involution, possibly involving the new coordinates 
associated to the exceptional divisors is chosen, see Section 5 of \first.

Since each $O7$--plane induces $-8$ units of $D7$--brane charge, we choose to cancel this 
tadpole locally by placing a stack of 8 coincident $D7$--branes on top of each divisor fixed 
under the combination of the involution and the scaling action. Each such stack therefore 
carries an SO(8) gauge group.
For the $D3$--brane charge, the case is a bit more involved. 
The contribution from the $O3$--planes is (in the orientifold quotient $X_6$ of $Y_6$)
$$ Q_3(O3)=-{1\over 4}\  n_{O3}\ ,$$
where $n_{O3}$ denotes the number of $O3$--planes. 
The $D7$--branes also contribute to the $D3$--tadpole (in the orientifold quotient $X_6$)
$$ Q_3(D7)=-\h\ \sum_a\,{n_{D7,a}\,\chi(\Dc_a)\over 24}\ ,$$
where $n_{D7,a}$ denotes the number of $D7$--branes in the stack located on the 
divisor $\Dc_a$. As we have seen, the $\Dc_a$ can be local $D$--divisors as well as 
exceptional divisors $E$.
The last contribution to the $D3$--brane tadpole comes from the $O7$--planes (in the orientifold quotient $X_6$):
$$ Q_3(O7)=-\h\ \sum_a\,{\chi(\Dc_a)\over 6}\ .$$
So the total $D3$--brane charge that must be cancelled is:
\eqn\totaltadpole{Q_{3,tot}=-{n_{O3}\over4}-\h\ \sum_a\,{(n_{D7,a}+4)\,\chi(\Dc_a)\over 24}\ .}
These are the values for the orientifold quotient $X_6$, 
in the double cover $Y_6$ this value must be multiplied by two (\cf Subsection 5.3).
Because we would like to avoid mobile $D3$--branes, this tadpole will
be saturated by 3--form flux
$G_3$.

The formula \totaltadpole\ for the total $D3$--brane charge $Q_{3,tot}$ differs from the 
known  tadpole equation for the singular orbifold case by the second term. 
The latter is induced by the curvature of the $D7$--branes which is absent in the singular 
case. 
In that case, the number of orientifold $O3$--planes is always $64$, \ie $n_{O3}=64$,
and \totaltadpole\ boils down to $Q_{3,tot}=-16$ \LRSi.
In the CFT description, this tadpole originates from the total leading divergent contribution 
of the Klein bottle amplitude $Z_\Kc(1,1)$ of the untwisted orbifold sector.
However, there are additional tadpole contributions from other orbifold sectors 
to be cancelled.
More precisely, the tadpole arising from the Klein bottle amplitude $Z_\Kc(1,\th^k)$ and in 
addition for even $N$ the $\IZ_2$--twisted tadpole related to 
 $Z_\Kc(\th^{N/2},\th^k)$ have to be cancelled
($k=0,\ldots,N-1$).
The tadpoles from the sector $(1,1)$ and for even $N$ also from the sector $(1,\th^{N/2})$
may be cancelled by introducing the right amount of $D3$--brane (or/and $3$--form flux) 
and $D7$--branes, respectively. On the other hand,
the divergences of the Klein bottle amplitude  $Z_\Kc(1,\th^k),\ k\neq 0$ 
or for even $N$ from the combination 
$Z_\Kc(1,\th^k)+Z_\Kc(\th^{N/2},\th^k),\ k\neq0,N/2$ 
can only be cancelled against any of the annulus and M\"obius strip contributions 
in the case that the orbifold group $\Gamma$ is 
$\IZ_3,\,\IZ_{6-I},\,\IZ_{6-II},\,\IZ_7$ or $\IZ_{12-I}$ \AFIV\ or $\IZ_2\times\IZ_2,
\IZ_3\times\IZ_3, \IZ_6\times\IZ_6, \IZ_2\times\IZ_3,\IZ_2\times \IZ_6, \IZ_2\times \IZ_6'$ 
\zwart. 
Hence singular orbifolds have much more constraining tadpole equations, which
are non--trivial to fulfill for all $\IZ_N$-- and $\IZ_N\times \IZ_M$ orbifolds.
However, if one introduces discrete torsion or vector structure tadpoles from all orbifold
sectors may be completely cancelled in all singular orbifold cases~\Rabadan.

Nevertheless, the orientifolds $X_6$ constructed geometrically in {\it Ref.} \first\ 
in the large radius regime from resolved orbifolds $Y_6$ 
need not have a CFT counterpart in their orbifold limit, 
since $D$--branes (in particular stacks of $D7$ and $O7$--branes) wrapping cycles  
which vanish in the orbifold limit, give rise to  extra non--perturbative 
states in the orbifold limit.

\subsec{Resolved toroidal orientifolds as candidate models for a KKLT scenario}

In \DenefMM, a toroidal orbifold model, namely type $IIB$ string theory compactified on the 
orientifold of the resolved $T^6/\IZ_2\times\IZ_2$, was checked for its suitability as a 
compactification manifold for the KKLT proposal. Since the $F$--theory lift of this example 
is known, Witten's criterion could be checked directly and the results of \DenefMM\ strongly 
indicate that in this model, all geometric moduli can be fixed.

The methods to obtain a smooth Calabi--Yau manifold from a toroidal orbifold and to 
subsequently pass to the corresponding orientifold as described in {\it Ref.} \first\ enable 
us to explicitly check other toroidal orbifolds for their suitability as candidate models for 
the KKLT proposal.

The requirement that the scalar mass matrix be positive, discussed in
Section 2.3, places severe constraints on the list of possible
models. Those orbifolds without complex structure moduli do not give
rise to stable vacua after the uplift to dS space. Thus
$\IZ_3,\,\IZ_7,\,\IZ_{8-I}$ on $SU(4)^2$, $\IZ_2\times\IZ_{6'}$,
$\IZ_3\times\IZ_{3}$, $\IZ_4\times\IZ_{4}$ and $\IZ_6\times\IZ_{6}$
are excluded from the list of possible models given in Table 3. For
more details \cf Subsection 5.5.

Since the stabilization of twisted complex structure moduli via 3--form flux is not well 
understood yet, the models with $h_{(2,1)}^{twist.}(X_6)\neq 0$ cannot be checked explicitly. 
Yet considerations regarding the topology of their divisors suggest that they might not 
be suitable candidate models anyway, as will be explained later on. 
The only models which are not yet excluded and are directly amenable to our methods are 
thus $T^6/\IZ_4  $  on $SU(4)^2 $, $T^6/ \IZ_{6-II} $  on $ SU(2)\times SU(6)$, the above 
mentioned $T^6/\IZ_2 \times\IZ_2$, and $T^6/\IZ_2 \times\IZ_4$. The example $T^6/\IZ_4  $  on $SU(4)^2 $ contains five instead of the usual three untwisted K\"ahler moduli. Since it is not clear how these two extra non--diagonal untwisted K\"ahler moduli contribute to the superpotential, this example will not be discussed explicitly.

The question one would like to answer is: Do enough of the divisors of the above models 
contribute to the non-perturbative superpotential that all K\"ahler moduli can be fixed?
To answer this question, the topologies of the divisors must be studied. 
In Section 4.3 of \first\ it was shown that there are four basic topologies for the divisors 
of the resolved toroidal orbifolds: The divisors $R_i$ inherited from the covering $T^6$ have 
the topology of either {\bf (i)} $K3$ or {\bf (ii)} $T^4$. 
The exceptional divisors $E_i$ which arise in the blowing up process can be birationally 
equivalent to either 
{\bf (iii)} a rational surface (\ie $\IP^2$ or $\IF_n$) or 
{\bf (iv)} $\IP^1\times T^2$. 
The same is true for the $D$--divisors, which correspond to planes fixed at the loci of the 
fixed points and are linear combinations of the $R$s and $E$s. The rational surfaces have 
$h_{(1,0)}=h_{(2,0)}=0$ and therefore $\chi({\cal O}_{S})=1$. Since $h_{(1,0)}$ and $h_{(2,0)}$ 
are birational invariants, the number of blow--ups which depends on the triangulation of the 
resolution is irrelevant here. $\IP^1\times T^2$ has $h_{(1,0)}=1,\ h_{(2,0)}=0$,  
$T^4$ has  $h_{(1,0)}=2,\ h_{(2,0)}=1$, which both results in $\chi({\cal O}_{S})=0$. $K3$ 
has $h_{(1,0)}=0,\ h_{(2,0)}=1$ and therefore $\chi({\cal O}_{S})=2$.

Since except for $T^6/\IZ_2 \times\IZ_2$, the $F$-theory lifts of these models are not known, 
it must be determined directly in type $IIB$ which divisors contribute to the non--perturbative 
superpotential. Here, we make use of the index for the Dirac operator on the world--volume 
of the Euclidean $D3$--brane \indexeq. 
The values of the index for the four divisor topologies arising from resolutions of toroidal orbifolds are given in Table 5.

\vskip0.5cm
{\vbox{\ninepoint{$$
\vbox{\offinterlineskip\tabskip=0pt
\halign{\strut\vrule#
&~$#$~\hfil 
&\vrule$#$
&~$#$~\hfil 
&\vrule$#$ 
&~$#$~\hfil 
&\vrule$#$
&~$#$~\hfil 
&\vrule$#$&\vrule$#$\cr
\noalign{\hrule}
&{\rm Topology} &&O7\ {\rm on\ top}&& {\rm inters.\ in\ 1\ dim.} && {\rm no\ intersection}&\cr
\noalign{\hrule}\noalign{\hrule}
& \ K3      &&\   2/[1]           &&0 &&0/[-1]&\cr
& \ T^4      &&\   0/[-1]         &&0/[-2] && 0/[-3]&\cr
& \  \IP^1\times T^2      &&\  0 &&  1/[0]  &&0/[-1]&\cr
& \  \IP^2,\ \IF_n      &&\     1  &&\ 1  &&  0&\cr
\noalign{\hrule}}} $$
\vskip-6pt
\centerline{\noindent{\bf Table 5:}
{\sl Index $\chi_{D3}$ for the four basic topologies}}
\vskip10pt}}}
\vskip-0.5cm
\br
The numbers in square brackets are the values of the index in the case that the corresponding zero modes have been lifted by flux, \cf Table 2. We see thus that for the case that the $O7$--plane does not intersect the divisor, we never get a contribution, so we better seek an orientifold action which leads to many $O7$--plane solutions. $K3$ can contribute for the case that the $O7$ lies on top of the divisor if the $h_{(2,0)}^{(+)}$ zero modes are lifted by flux. In our set--up, the case that the $O7$ lies on top of the divisor cannot arise, since only the inherited divisors $R_i$ can have the topology of $K3$, and these divisors are never wrapped by $O7$--planes. A divisor with the topology of $T^4$ can likewise never contribute. $\IP^1\times T^2$ can contribute in case of an intersection with the $O7$--plane in one direction if {\it no} zero modes are lifted by flux. The rational surfaces always contribute except if there is no intersection irrespective of the background flux. To summarize: All those models are likely to allow the stabilization of all geometric moduli for which 

$(i)$ the fixed points and fixed lines are all in equivalence classes with only one member, giving rise to $E$ and $D$ divisors which are birationally equivalent to rational surfaces and

$(ii)$ an orientifold action exists which gives rise to enough $O7$--plane solutions that each divisor intersects an $O7$--plane in at least one complex dimension.

When these conditions are met, it is likely that all geometric moduli will be stabilized when the full scalar potential is minimized.

Requirements $(i)$ and $(ii)$ are both met by $T^6/\IZ_4$  on $SU(4)^2$, $T^6/ \IZ_{6-II}$ on $SU(2)\times SU(6)$, $T^6/\IZ_2 \times\IZ_2$ and $T^6/\IZ_2 \times\IZ_4$, therefore we expect that all geometric moduli can be stabilized in these cases.

Models with fixed lines without fixed points on them which lie in orbits of length greater than one do not satisfy criterion (i) since the divisors corresponding to these fixed lines have the topology of $\IP^1\times T^2$. These are exactly the models with $h^{(2,1)}_{twist.}\neq0$. Unless an elaborate configuration of $O$--planes can be chosen such that all these divisors intersect on $O7$--plane along one dimension, these examples in general allow only for a partial stabilization of the geometric moduli via Euclidean $D3$--brane instantons.
It should be stressed that examples like these are still not completely hopeless since additional effects might lead to the complete stabilization of all moduli.
On the other hand, this survey again confirms the old suspicion that manifolds with the right geometrical properties to allow the stabilization of all K\"ahler moduli by Euclidean $D3$--brane instantons or gaugino condensates are not very generic.

So far, we discussed the conditions for a contribution to the non--perturbative superpotential from Euclidean $D3$--instantons. Since we cancel the $O7$--tadpole by placing $D7$--branes on top of the $O7$--planes, a gaugino condensate can arise on the world--volume of the $D7$--branes. As mentioned before, for a contribution to the non--perturbative superpotential to arise from a gaugino condensate, we should have

\noindent $(a)$ no bifundamental matter. This is given when the different divisors on which $D7$--branes are wrapped do not intersect. This condition can be easily checked by inspection of the toric diagram of the resolved patches.

\noindent $(b)$ no adjoint matter. This depends on the Hodge numbers of the divisor which is wrapped by the brane. For rational surfaces, \ie $h_{(1,0)}=h_{(2,0)}=0$, this criterion is fulfilled.

In the following, moduli stabilization will be discussed in detail for
the two examples $T^6/ \IZ_{6-II}$ on $SU(2)\times SU(6)$ and
$T^6/\IZ_2 \times\IZ_4$ on $SU(2)^2\times SO(5)^2$.
In Subsection 5.3 stabilization of the dilaton and complex structure
moduli through $3$--form flux $G_3$ is discussed and in Subsection 5.4, 
the stabilization of the K\"ahler moduli.

\subsec{Complex structure and dilaton stabilization through $3$--form flux}

For the orbifolds $X_6$ with $h^{(-)}_{(2,1)}(X_6)= 1$
the K\"ahler potential for the dilaton and complex structure modulus ($U\equiv U^3$) 
\FULLK\ is:
\eqn\BECOMES{
K_0=-\log(S+\bar S)-\log(U+\bar U)\ ,}
while the tree--level superpotential \TV\ may be written as
\eqn\WandK{
W_0=A+B\ S+U\ (C+D\ S) \ ,}
with $A,B,C,D \in \IC$ to be specified later. With the $F$--terms
\eqn\FSU{\eqalign{
\ov F^{\ov S}&=\lf(\fc{S+\ov S}{U+\ov U}\ri)^{1/2}\ \lf[\ 
-A+B\ \ov S-U\ (C-D\ \ov S)\ \ri]\ ,\cr
\ov F^{\ov U}&=\lf(\fc{U+\ov U}{S+\ov S}\ri)^{1/2}\ \lf[\ 
-A-B\ S+\ov U\ (C+D\ S)\ \ri]\ ,}}
we may cast the scalar potential 
$$V=g_{S\ov S}\ F^S\ov F^{\ov S}+g_{U\ov U}\ F^U\ov F^{\ov U}-3\ e^{K_0}\ |W_0|^2\ $$
into the form:
\eqn\superpotential{\eqalign{
V&= \fc{1}{U+\bar U} \fc{1}{S+\bar S}\ \lf[\ |\ A-B\ \ov S+U\ (C-D\ \ov S)\ |^2
+|\ A+B\ S-\ov U\ (C+D\ S)\ |^2\ri.\cr
&\hskip3cm\lf.-3\ |\ A+B\ S+ U\ (C+D\ S)\ |^2\ \ri]\ .}}
The extremal points in the moduli space $(S,U)$ are determined
by the solutions of the equations $F^S,F^U=0$:
\eqn\AXION{
s_2=\fc{i}{2}\ \fc{\ov B\ C-B\ \ov C-\ov A\ D+A\ \ov D}{\ov B\ D+B\ \ov D}\ \ \ ,
\ \ \ 
u_2=\fc{i}{2}\ \fc{-\ov B\ C+B\ \ov C-\ov A\ D+A\ \ov D}{\ov C\ D+C\ \ov D}\ ,}
and similarly for the real parts $s_1,u_1$. 

The $3$--form flux $G_3=F_3+i\ S\ H_3$
\eqn\flux{
\hskip-0.5cm{1\over{(2\pi)^2\alpha'}}\ G_3=\sum\limits_{i=0}^{3} 
\lf[\ (a^i+i\ S\ c^i)\ 
\alpha_i+(b_i+i\ S\ d_i)\ \beta^i\ \ri]
+\sum\limits_{j=1}^{6} \lf[(e^j+i\ S\ g^j)\ \gamma_j+
(f_j+i\ S\ h_j)\ \delta^j\ri]}
entering \TV\ is given as linear combination w.r.t.
the integer cohomology basis $\{\al_i,\beta^i\}_{i=0,\ldots,3}$
and $\{\gamma_j,\delta^j\}_{j=1,\ldots,6}$ \LRSS.
This gives rise to 20 real flux components to be constrained by the 
respective orbifold group $\IZ_N$. This allows to express the complex parameters
$A,B,C,D$ through the eight integers $a^0,a^1,b_0,b_1,c^0,c^1,d_0,d_1$. For more details 
\cf \LRSS. The $F$--flatness conditions $F^S,F^{U}=0$ force 
the complex structure to align such, that the flux $G_3$ 
\eqn\FLUXG{\eqalign{
\hskip-0.5cm{1\over{(2\pi)^2\alpha'}}\ G_3&=\fc{i}{2\ \re(U)}\ \lf\{\ 
\lf[\ \ov A-\ov B\ S+\ov U\ (\ov C-\ov D\ S)\ \ri]\ dz^1\wedge dz^2\wedge dz^3\ri.\cr
&-\lf[\ A+B\ S+U\ (C+D\ S)\ \ri]\ d\ov z^1\wedge d\ov z^2\wedge d\ov z^3\cr
&+\lf[\ A+B\ S-\ov U\ (C+D\ S)\ \ri]\ d\ov z^1\wedge d\ov z^2\wedge d z^3\cr
&-\lf.\lf[\ \ov A-\ov B\ S-U\ (\ov C-\ov D\ S)\ \ri]\ d z^1\wedge d z^2\wedge d\ov z^3\ 
\ri\}}}
becomes $ISD$, \ie it has only
$(2,1)$ and $(0,3)$--components at the extremum.
The flux $G_3$ induces the contribution of
\eqn\NFLUX{
N_{flux}={1\over{(2\pi)^4\ap^2}}\ \int_{Y_6} F_3\wedge H_3}
to the total $D3$--brane charge \totaltadpole. Generically, this integral is calculated 
in the orientifold cover $Y_6$. Therefore the number $N_{flux}$ has to be twice
the negative value of the total $D3$--brane charge \totaltadpole, \ie
\eqn\idFLUX{
N_{flux}=-2\ Q_{3,tot}}
to cancel the latter by flux only.

\ \br
$\underline{(i)\ \ 
\IZ_{6-II}-{\rm orbifold\ on\ the\ } SU(2)\times SU(6)\ {\rm lattice}:}$
\ \br
\ \br
The $\IZ_{6-II}$--orbifold has the action $(v^1,\,v^2,\,v^3)=({1\over6},\,{1\over3},\,-{1\over2})$.
The $3$--form flux \flux\ constrained by the $\IZ_{6-II}$--orbifold group becomes:
\eqn\GdreiZsechs{\eqalign{
\fc{1}{(2\pi)^2\alpha'}\ G_3=&\fc{1}{3}\ (a_0+iS c_0)\ 
(3\ \alpha_0+2\ \beta_3+\gamma_1-2\gamma_2-
2\ \gamma_3+\gamma_4-\delta_5)\cr
&+(b_0+iSd_0)\ (-\alpha_3+\beta_0+\gamma_5-\gamma_6) \cr
&+\fc{1}{2}\ (b_1+iSd_1)\ (2\ \beta_1+\beta_2+\delta_1-\delta_2-2\ \delta_3-\delta_4)\cr
&+(a_1+iSc_1)\ (\alpha_1+\alpha_2+\beta_3-\gamma_2-\gamma_3-\delta_6)\ .}}
This flux correspond to the flux number:
\eqn\NflussZ{
N_{\rm {flux}}=2\ b_0\ c_0+b_1\ (c_0+3\ c_1)-2\ a_0\ d_0-d_1\ (a_0+3\ a_1)\ .}
For the $\IZ_{6-II}$ orbifold with $SU(2)\times SU(6)$ lattice the coefficients 
$A,B,C,D$ entering \superpotential\ become:
\eqn\Zsechs{\eqalign{
A&=-\fc{\sqrt 3}{2}\ b_1+i b_0 \ \ \ ,\ \ \ B=-d_0-\fc{\sqrt 3\,i}{2}d_1 \ , \cr
C&=a_0+i \lf(\fc{a_0}{\sqrt 3}+\sqrt 3 a_1\ri) \ \ \ ,\ \ \ 
D=-\lf(\fc{c_0}{\sqrt 3}+\sqrt 3 c_1\ri)+ic_0\ .}}
With this information, \eqq \WandK\ yields the superpotential:
\eqn\WflussZ{\eqalign{
W_0=&-\fc{\sqrt{3}}{2}\ b_1+i\ b_0-S\ \lf(d_0+\fc{\sqrt{3}i}{2}\ 
d_1\ri) \cr
&+U\ \lf[\ a_0+i\lf(\fc{a_0}{\sqrt{3}}+\sqrt{3}\ a_1\ri)\ \ri]-
S\ U\ \lf(\fc{c_0}{\sqrt{3}}+\sqrt{3}\ c_1-i\ c_0\ri)\ .}}
Since the total $D3$--brane charge  
in the CY orientifold is $Q_{3,tot}=-22$ (see Section 5.4), we look for  fluxes
\GdreiZsechs\ 
with $N_{flux}=44$ on the covering space $Y_6$. Furthermore, the fields
$S=s_1+is_2$ and $U=u_1+iu_2$ should be fixed (\cf \AXION) to realistic values.
A reasonable  value for $\re S$ is $s_1\sim 3.6$, which corresponds to a string 
coupling constant $g_{\rm string}\sim 0.27$ at the string scale. Besides, the complex structure
modulus $U$ is expected to be around the $\rho$--point in the fundamental region,
with $\rho=\h+\fc{i}{2}\sqrt 3$. An additional constraint may be imposed on the
tuning parameter $e^{K_0}|W_0|^2$, which should be small to avoid higher 
order effects in the full non--perturbative superpotential \SUP.
After a systematic scan in the flux space $(a^0,a^1,b_0,b_1,c^0,c^1,d_0,d_1)\in \IZ^8$
we find hundreds of vacua, which meet these criteria. 
A set of equivalent vacua, differing only in the discrete flux parameters
$(a^0,a^1,b_0,b_1,c^0,c^1,d_0,d_1)$, is given in the following Table~6.
\vskip0.5cm
{\vbox{\ninepoint{$$
\vbox{\offinterlineskip\tabskip=0pt
\halign{\strut\vrule#
&~$#$~\hfil 
&\vrule$#$&\vrule$#$
&~$#$~\hfil 
&\vrule$#$ 
&~$#$~\hfil 
&\vrule$#$
&~$#$~\hfil 
&\vrule$#$ 
&~$#$~\hfil 
&\vrule$#$
&~$#$~\hfil 
&\vrule$#$
&~$#$~\hfil 
&\vrule$#$&\vrule$#$\cr
\noalign{\hrule}
&\ (a^0,\ b_0,\ c^0,\ d_0,\ a^1,\ b_1,\ c^1,\ d_1)\  &&&  s_1 && s_2   && u_1&& u_2 && m_S &&  m_U &\cr
\noalign{\hrule}\noalign{\hrule}
&(-5 ,\  12 ,\  0 ,\  2 ,\  -4 ,\  -8 ,\  -1 ,\  0)\  &&& 3.15788 && 5.83333 && 1.26315 && \
0.0666667 &&   2.18 && 13.68 &\cr
 &(-5 ,\  10 ,\  0 ,\  2 ,\  -3 ,\  -8 ,\  -1 ,\  0)\  &&& 3.15788 && 4.83333 && 1.26315 && \
0.0666667 &&   2.18 && 13.68 &\cr
&(-5 ,\  6 ,\  0 ,\  2 ,\  -1 ,\  -8 ,\  -1 ,\  0)\  &&& 3.15788 && 2.83333 && 1.26315 && \
0.0666667 &&   2.18 && 13.68 &\cr
&(-5 ,\ 0 ,\  0 ,\  2 ,\  2 ,\  -8 ,\  -1 ,\  0)\  &&& 3.15788 && -0.166667 && 1.26315 && \
0.0666667 &&   2.18 && 13.68 &\cr
&(-5 ,\  -4 ,\  0 ,\  2 ,\  4 ,\  -8 ,\  -1 ,\  0)\   &&& 3.15788 && -2.16667 && 1.26315 && \
0.0666667 &&   2.18 && 13.68 &\cr
&(-5 ,\  -8 ,\  0 ,\  2 ,\  6 ,\  -8 ,\  -1 ,\  0)\  &&& 3.15788 && -4.16667 && 1.26315 && \
0.0666667 &&   2.18 && 13.68 &\cr
&(-5 ,\ -12 ,\  0 ,\  2 ,\  8 ,\  -8 ,\  -1 ,\  0)\  &&& 3.15788 && -6.16667 && 1.26315 &&\ 
0.0666667 &&   2.18 && 13.68 &\cr
&(5 ,\  10 ,\  0 ,\  -2 ,\  -7 ,\  8 ,\  1 ,\  0)\  &&& 3.15788 && -5.16667 && 1.26315 && \
0.0666667 && 2.18 && 13.68 &\cr
&(5 ,\  8 ,\  0 ,\  -2 ,\  -6 ,\  8 ,\  1 ,\  0)\  &&& 3.15788 && -4.16667 && 1.26315 && \
0.0666667 &&   2.18 && 13.68 &\cr
&(5 ,\  6 ,\  0 ,\  -2 ,\  -5 ,\  8 ,\  1 ,\  0)\   &&& 3.15788 && -3.16667 && 1.26315 && \
0.0666667 &&    2.18 && 13.68 &\cr
&(5 ,\  2 ,\  0 ,\  -2 ,\  -3 ,\  8 ,\  1 ,\  0)\   &&& 3.15788 && -1.16667 && 1.26315 && \
0.0666667 &&   2.18 && 13.68 &\cr
\noalign{\hrule}}}$$
\vskip-6pt\hskip0cm
\centerline{\noindent{\bf Table 6:}
{\sl Discrete landscape of supersymmetric AdS minima
for $N_{flux}=44$,}}
\centerline{$e^{K_0/2}\ |W_0|=0.34864$ and $V_0=-0.364644$.}
\vskip10pt}}}
\vskip-0.5cm
\ \br
Clearly, the axionic vev $s_2$ may be shifted back into the fundamental region $s_2\equiv~-0.166667$, 
while the flux number $N_{flux}$ in \NFLUX\ and $K_0,W_0$ are preserved \KachruA.
Furthermore, in Table 7 
we present a set of supersymmetric AdS minima in the $(S,U)$--space
with different tuning parameters $e^{K_0}|W_0|^2$.
\vskip0.0cm\hskip-2cm
{\vbox{\ninepoint{$$
\vbox{\offinterlineskip\tabskip=0pt
\halign{\strut\vrule#
&~$#$~\hfil 
&\vrule$#$&\vrule$#$
&~$#$~\hfil 
&\vrule$#$ 
&~$#$~\hfil 
&\vrule$#$
&~$#$~\hfil 
&\vrule$#$ 
&~$#$~\hfil 
&\vrule$#$
&~$#$~\hfil 
&\vrule$#$
&~$#$~\hfil 
&\vrule$#$
&~$#$~\hfil 
&\vrule$#$
&~$#$~\hfil 
&\vrule$#$&\vrule$#$\cr
\noalign{\hrule}
&\ (a^0,\ b_0,\ c^0,\ d_0,\ a^1,\ b_1,\ c^1,\ d_1)\  &&&  s_1 && s_2   && u_1&& 
u_2 && -V_0 && e^{K/2}\ |W_0|&& m_S&&  m_U&\cr
\noalign{\hrule}\noalign{\hrule}
&(\ -2 ,\  -7 ,\  -1 ,\  2 ,\  3 ,\  -8 ,\  0 ,\  -2 )\ &&& 3.8092 && -0.3 && 2.11622 && \
0.722222 &&  0.02273 && 0.087046 && 1.52 && 4.91 &\cr
&(\ 0 ,\  -10 ,\  -1 ,\  -1 ,\  5 ,\  -3 ,\  1 ,\  -2 )\ &&& 3.7944 && 3.95 && 1.2648 && \
0.316667 && 0.278999 && 0.304958 && 1.51 && 13.67 &\cr
&(\ 0 ,\  -10 ,\  -1 ,\  2 ,\  3 ,\  -6 ,\  0 ,\  -2 )\ &&&  3.7934 && -1.9 && 2.10745 && \
0.944444 && 0.296119 && 0.314176 && 1.51 && 4.92 &\cr
&(\ 2 ,\  -10 ,\  -1 ,\  -1 ,\  3 ,\  -1 ,\  1 ,\  -2 )\ &&&  3.7918 &&  2.65 && 1.26392 && \
0.55 && 0.324662 && 0.328969 && 1.51 && 13.67 &\cr
&(\ 0 ,\  -10 ,\  -1 ,\  -1 ,\  6 ,\  -6 ,\  1 ,\  -2 )\ &&&  3.7296 && 4.7 && 1.036 && \
0.305556 && 1.40124 && 0.683432 && 1.51 && 19.86 &\cr
&(\ 0 ,\  -10 ,\  -1 ,\  -1 ,\  4 ,\  0 ,\  1 ,\  -2 )\ &&&  3.7095 && 3.2 && 1.5456 && \
0.333333 && 1.75049 && 0.763869 && 1.51 && 8.87 &\cr
&(\ 5 ,\  -9 ,\  -1 ,\  -1 ,\  0 ,\  3 ,\  1 ,\  -2 )\ &&&  3.6575 && 0.35 && 1.21918 && \
0.783333 && 2.64978 && 0.93982 && 1.51  && 13.95  &\cr
\noalign{\hrule}}}$$
\vskip-6pt\hskip0.2cm
\centerline{\noindent{\bf Table 7:}
{\sl Supersymmetric AdS minima
in the $(S,U)$--space for $N_{flux}=44$ and specific $e^{K_0/2}\ |W_0|$.}}
\vskip10pt}}}
\vskip-0.5cm
\br
\ \br
$\underline{(ii)\ \ 
\IZ_2\times \IZ_4-{\rm orbifold\ on\ the\ } SU(2)^2\times SO(5)^2\ {\rm lattice}:}$
\ \br
\ \br
The $\IZ_2\times\IZ_4$--orbifold has the two actions
$(v^1,v^2,v^3)=\h(1,0,-1)$ and $(w^1,w^2,w^3)=\fc{1}{4}(0,1,-1)$.
The $3$--form flux \flux\ constrained by the $\IZ_2\times\IZ_4$--orbifold group becomes: 
\eqn\GdreiZzweiZvier{\eqalign{
\fc{1}{(2\pi)^2\alpha'}\ G_3=&(a_3+iSc_0)\ (-\alpha_2+\alpha_3)+(a_0+iSc_0)\ 
\lf(\alpha_0-\alpha_2-\fc{1}{2}\beta_1\ri)\cr
&+(b_2+iS d_2)\ \lf(\alpha_1+\fc{1}{2}\beta_0+\beta_2\ri)+(b_3+iS d_3)\ 
\lf(\alpha_1+\fc{1}{2}\beta_0+\beta_3\ri)\ .}}
The coefficients $A,B,C,D$ entering \superpotential\ 
are given in the case of $\IZ_2\times \IZ_4$--orbifold with  $SU(2)^2\times SO(5)^2$ lattice by:
\eqn\Zzweivier{\eqalign{
A&=-\fc{1+2}{2}\left(b_2-i\ b_3\right)\ \ \ ,\ \ \ B=\fc{1-i}{2}\left(d_2-i\ d_3\right)\ , \cr
C&=\fc{1+i}{2}a_0+a_3\ \ \ ,\ \ \ D=\fc{-1+i}{\ 2}c_0+i\ c_3\ .}}
Furthermore, the flux number is:
\eqn\NflussZweiVier{
N_{\rm{flux}}=a_3 d_2-b_2 c_3+b_3\ (c_0+c_3)-(a_0+a_3)\ d_3\ . }
With this information, the superpotential \WandK\ becomes:
\eqn\WflussZweiVier{\eqalign{
W_0=&-\fc{1+i}{2}\ (b_2-i\ b_3)+S\ \fc{(1-i)}{2}\ (d_2-i\ d_3) \cr
&+U\lf(\fc{1+i}{2}\ a_0+a_3\ \ri)+S\ U\ \lf(\fc{-1+i}{2}\ c_0+i\ c_3\ri)\ .}}

We search for fluxes \GdreiZzweiZvier\ with $N_{flux}=52$. We fix the value of the $s_1$ at
$3.24$, which corresponds to a string coupling constant $g_{{\rm string}}=0.30$ at the string scale.
A set of equivalent vacua, differing only in the discrete flux parameters $(a^0,b_2,c^0,d_2,a^3,b_3,c^3,d_3)$,
is given in the following Table 8.

\vskip0.5cm\hskip-1.4cm
{\vbox{\ninepoint{$$
\vbox{\offinterlineskip\tabskip=0pt
\halign{\strut\vrule#
&~$#$~\hfil 
&\vrule$#$&\vrule$#$
&~$#$~\hfil 
&\vrule$#$ 
&~$#$~\hfil 
&\vrule$#$
&~$#$~\hfil 
&\vrule$#$ 
&~$#$~\hfil 
&\vrule$#$
&~$#$~\hfil 
&\vrule$#$
&~$#$~\hfil 
&\vrule$#$&\vrule$#$\cr
\noalign{\hrule}
&\ (a^0,\ b_2,\ c^0,\ d_2,\ a^3,\ b_3,\ c^3,\ d_3)\  &&&  s_1 && s_2   && u_1&& 
u_2 && m_S&&  m_U&\cr
\noalign{\hrule}\noalign{\hrule}
&( 14 ,\  14 ,\  0 ,\  1 ,\  -3 ,\  14 ,\  -2 ,\  -5)\ &&&  3.23796 && -1.875 && 0.925131 \
&& -1.46429 &&  2.46  &&  30.21 &\cr
&( 14 ,\  15 ,\  0 ,\  1 ,\  -3 ,\  15 ,\  -2 ,\  -5)\ &&&   3.23796 && -2.125 && 0.925131 \
&& -1.53571 &&  2.46  &&  30.21 &\cr
&( -8 ,\  8 ,\  4 ,\  1 ,\  11 ,\  8 ,\  -2 ,\  -3)\ &&&   3.23796 && -1.875 && 0.925131 && \
0.535714 &&  2.46  &&  30.21 &\cr
&( -14 ,\  14 ,\  0 ,\  5 ,\  11 ,\  14 ,\  2 ,\  -1)\ &&&   3.23796 && 1.875 && 0.925131 && \
1.46429 &&  2.46  &&  30.21 &\cr
&( -14 ,\  15 ,\  0 ,\  5 ,\  11 ,\  15 ,\  2 ,\  -1)\ &&&   3.23796 && 2.125 && 0.925131 && \
1.53571 && 2.46  &&  30.21 &\cr
&(  8 ,\  8 ,\  4 ,\  3 ,\  3 ,\  8 ,\  -2 ,\  -1)\ &&&   3.23796 &&1.875 && 0.925131 && \
-0.535714 && 2.46  &&  30.21 &\cr
&( 14 ,\  -16 ,\  0 ,\  -3 ,\  -19 ,\  -16 ,\ -2,\  -1)\ &&&   3.23796 && 6.125 && 0.925131 &&\ 
0.535714 && 2.46  &&  30.21 &\cr
&(  14 ,\  -15 ,\  0 ,\  -3 ,\  -19 ,\  -15 ,\  -2 ,\  -1)\ &&&  3.23796 && 5.875 && 0.925131 &&\ 
0.464286 &&  2.46  &&  30.21 &\cr
&(14 ,\  -14,\  0,\ -5,\ -11,\ -14,\ -2,\ 1)\ &&& 3.23796 && 1.875 && 0.925131 &&\ 
1.46429 && 2.46 && 30.21&\cr
\noalign{\hrule}}}$$
\vskip-6pt\hskip0cm
\centerline{\noindent{\bf Table 8:}
{\sl Discrete landscape of supersymmetric AdS minima
for $N_{flux}=52$,}}
\centerline{$e^{K_0/2}\ |W_0|=0.310374$ and $V_0=-0.288997$.}
\vskip10pt}}}
\vskip-0.5cm
\ \br
In the next table, we present a set of supersymmetric AdS minima in the $(S,U)$-space with 
same tuning parameter $e^{K_0}|W_0|^2$, but different choices for $S$ and $U$.

\vskip0.0cm
{\vbox{\ninepoint{$$
\vbox{\offinterlineskip\tabskip=0pt
\halign{\strut\vrule#
&~$#$~\hfil 
&\vrule$#$&\vrule$#$
&~$#$~\hfil 
&\vrule$#$ 
&~$#$~\hfil 
&\vrule$#$
&~$#$~\hfil 
&\vrule$#$ 
&~$#$~\hfil 
&\vrule$#$
&~$#$~\hfil 
&\vrule$#$
&~$#$~\hfil 
&\vrule$#$&\vrule$#$\cr
\noalign{\hrule}
&\ (a^0,\ b_2,\ c^0,\ d_2,\ a^3,\ b_3,\ c^3,\ d_3 )\  &&&  s_1 && s_2   && u_1&& 
u_2 && m_S&&  m_U&\cr
\noalign{\hrule}\noalign{\hrule}
&\ (-13 ,\  17 ,\  3 ,\  1 ,\  11 ,\  17 ,\  -1 ,\  -5)\ &&& 3.70 && -2.85714 && 1.29518 \
&& 1.85 &&     1.88 && 15.42 &\cr
&\ (-10 ,\  16 ,\  3 ,\  1 ,\  9 ,\  16 ,\  -1 ,\  -5)\ &&& 3.70 && -2.14286 && 1.52375 \
&& 1.82353 &&     1.88 && 11.14 &\cr
&\ ( 9 ,\  15 ,\  1 ,\  3 ,\  -1 ,\  15 ,\  -2 ,\  -5)\ &&& 3.70 &&-1.14286 && 1.52375 \
&& -2.17647 &&     1.88 && 11.14 &\cr
&\ ( -14 ,\  20 ,\  3 ,\  -1 ,\  16 ,\  20 ,\  -2 ,\  -4)\ &&& 3.70 && -6.14286 && \
1.29518 &&  1.15 &&     1.88 && 15.41 &\cr
&\ (-9 ,\  15 ,\  1 ,\  5 ,\  8 ,\  15 ,\  1 ,\  -3)\ &&& 3.70 && 1.14286 && 1.52375 && \
2.17647 &&     1.88 && 11.14 &\cr
&\ (-4 ,\  15 ,\  2 ,\  2 ,\  6 ,\  15 ,\  -1 ,\  -5)\ &&& 3.70 && -1.85714 && 3.23796 \
&& 1.625 &&     1.87 && 2.48 &\cr
&\ (2 ,\  20 ,\  1 ,\  -1 ,\  13 ,\  20 ,\  -2 ,\  -3)\ &&& 3.70 && -8.14286 && 1.52375 \
&& -0.176471 &&     1.88 && 11.14 &\cr
&\ (-6 ,\  20 ,\  2 ,\  -1 ,\  12 ,\  20 ,\  -2 ,\  -4)\ &&& 3.24 && -5.875 && 2.15864 \
&& 0.583333 &&     2.46 && 5.56&\cr
&\ (-5,\  10 ,\  3 ,\  2 ,\  7 ,\  10 ,\  -1 ,\  -4)\ &&& 3.24 && -0.875 && 1.61898 && \
1.0625 &&     2.46 && 9.87&\cr
&\ (3 ,\  11 ,\  2 ,\  2 ,\  3 ,\  11 ,\  -2 ,\  -4)\ &&& 3.24 && -1.375 && 2.15864 && \
-0.916667 &&     2.46 && 5.56&\cr
&\ (-2 ,\ 15 ,\  2 ,\  3 ,\  4 ,\  15 ,\  -1 ,\  -5)\ &&& 3.24 && -0.875 && 4.31728 && \
1.16667 &&    1.38 && 2.47 &\cr
&\ (-6 ,\  14 ,\  2 ,\  4 ,\  6 ,\  14 ,\  0 ,\ -5)\ &&& 3.24 && -0.125 && 2.15864 &&\ 
2.41667 && 2.46 && 5.56 &\cr
\noalign{\hrule}}}$$
\vskip-6pt\hskip0.2cm
\centerline{\noindent{\bf Table 9:}
{\sl Supersymmetric AdS minima
in the $(S,U)$--space for $N_{flux}=52$,}}
\centerline{\sl $e^{K_0/2}\ |W_0|=0.310374$ and $V_0=-0.28900$.}
\vskip10pt}}}
\vskip-0.5cm
\br
\ \br
$\underline{(iii)\ \ \IZ_4-{\rm orbifold\ on\ the\ } SU(4)^2\ {\rm lattice}:}$
\ \br
\ \br
The $\IZ_{4}$--orbifold has the action $(v^1,\,v^2,\,v^3)=({1\over4},\,{1\over4},\,-{1\over2})$.
The $3$--form flux \flux\ constrained by the $\IZ_4$--orbifold group becomes 
\eqn\GdreiZvier{\eqalign{
\fc{1}{(2\pi)^2\alpha'}G_3=&(a_0+iS c_0)\ 
(\alpha_0+\alpha_3+\beta_2-\gamma_2-\gamma_3-\delta_5)\cr
&+\fc{1}{2}\ (b_0+iSd_0)\ (-\alpha_2+2\beta_0+\beta_3+\gamma_4+\gamma_5-\gamma_6+\delta_3) \cr
&+\fc{1}{2}\ (b_1+iSd_1)\ (\alpha_2+2\beta_1+\beta_3-\gamma_4+\gamma_5-\gamma_6-2\delta_2-
\delta_3)\cr
&+(a_1+iSc_1)\ (\alpha_1+2\alpha_3+\beta_2-\gamma_2-2\gamma_3+\delta_4-\delta_5-\delta_6)\ .}}
In the case of the $\IZ_4$--orbifold with the lattice $SU(4)^2$ the 
coefficients $A,B,C,D$ entering \superpotential\ are given  by:
\eqn\Zvier{\eqalign{
A&=-b_1+i\ b_0 \ \ \ ,\ \ \ B=-d_0-i\ d_1 \ , \cr
C&=a_0+i\ (a_0+2a_1) \ \ \ ,\ \ \ D=-c_0-2c_1+i\ c_0 \ .}}
Furthermore, the flux number is:
\eqn\NflussZZ{
N_{\rm{flux}}=2\ \lf[\ (b_0+b_1)\ c_0 +b_1 c_1-2 a_1d_1-(d_0+d_1)\ a_0\ \ri]\ .}
With this information the superpotential \WandK\ becomes:
\eqn\WflussZZ{\eqalign{
W_0=&-b_1+i \ b_0-S\ \lf(d_1+i \ d_0\ri) \cr
&+U\ \lf[\ a_0+i\ (a_0+2\ a_1) \ \ri]-S\ U\ \lf(c_0+2\ c_1-i\ c_0\ri)\ .}}

\subsec{K\"ahler moduli stabilization}

We consider the racetrack superpotential \SUP
\eqn\SUP{
W=\tilde W_0+\sum_{j=1}^\n\gamma_j\ e^{a_j\ T^j}\ ,} 
with $\tilde W_0$ related to the tree--level flux superpotential \WandK, by
$\tilde W_0=-e^{K_0/2} |W_0|$. The redefined quantity $\tilde W_0$ makes sure, that 
the minimization procedure w.r.t. the set of K\"ahler moduli $T^1,\ldots,T^n$ in the 
K\"ahler gauge $K_0\equiv 0$ yields the correct 
value $-3e^{K}|\tilde W_0|^2=-3e^{K_0+K} |W_0|^2$ in the scalar potential \VMINN.
This value accounts for the contribution of the 
dilaton and complex structure stabilization procedure, which is  decoupled and performed
in the previous subsection.
Here and in the following $K$ is the K\"ahler potential \kaehler\ 
for the $n=\n$ K\"ahler moduli $T^1,\ldots,T^n$. According to Subsection 2.2, we may 
assume $\gamma_j \in \IR^+$, \ie any complex phase of $\gamma_j$ has been 
put into the axionic vevs \vevaxion\ of the K\"ahler moduli $T^j$.
The supersymmetric vacua are given by the equations \adsm, \ie by the critical
points of $e^{K/2} W$. These equations fix the real part
of the K\"ahler moduli $T^j$, \ie the divisor volumes ${\rm Vol}(\Dc_i)$
of an even  divisor $\Dc_j$:
\eqn\divvol{
{\rm Vol}(\Dc_j)=\re(T^j)=\fc{3}{4}\ \Kc_{ijk}\ t^j\ t^k=
\fc{3}{2}\ \fc{\p}{\p t^j}\ {\rm Vol}(X_6)\ .}

To ignore $\ap$--corrections, the K\"ahler moduli $T^j$ or divisor volumes ${\rm Vol}(\Dc_j)$
should be stabilized at large values, resulting in a large CY volume ${\rm Vol}(X_6)$.
The $F$--flatness conditions \adsm\ roughly give rise to the relations
$\tilde W_0\sim \gamma_j\ e^{a_j\ T^j}$. Hence,
a smaller $\tilde W_0$ or larger coefficients $\gamma_j$ 
yield larger divisor volumes $\re T^j$.
Hence a small $\tilde W_0$ or large divisor volumes guarantee that $\ap$--corrections 
may be neglected.
In \SUP, the exponentials $e^{a_j\ T^j}$ should be small $\sim \Oc(10^{-4})$, such that
multi--instanton processes or multi--wrappings may be neglected.
In principle, this means that $\tilde W_0$ should be also of this order $\sim 10^{-4}$ 
\KKLT.
Furthermore, in {\it Ref.} \KKLT\ it has been argued that due to the smallness of these 
exponentials
a dependence of the coefficients $\gamma_j$ on the dilaton $S$ and complex structure 
moduli $U^j$
does not change the critical points of the dilaton and complex structure moduli much, 
as derived in the previous subsection, as long as 
the relative derivatives $\gamma_j^{-1}\ \p_{S,U^\lambda}\gamma_j$ and 
$\gamma_j^{-1}\ \p^2_{S,U^\lambda}\gamma_j$ are not huge.

\ \br
$\underline{(i)\ \ 
\IZ_{6-II}-{\rm orbifold\ on\ the\ } SU(2)\times SU(6)\ {\rm lattice}:}$
\ \br
\ \br
We consider the resolved $\IZ_{6-II}$ orbifold $Y_6$ on the lattice
$SU(2)\times SU(6)$ which has $h_{(1,1)}(Y_6)=25$.  
\ifig\ffixsixiiaa{Schematic picture of the fixed set configuration of
$\IZ_{6-II}$ on 
$SU(2)\times SU(6)$}{\epsfxsize=0.6\hsize\epsfbox{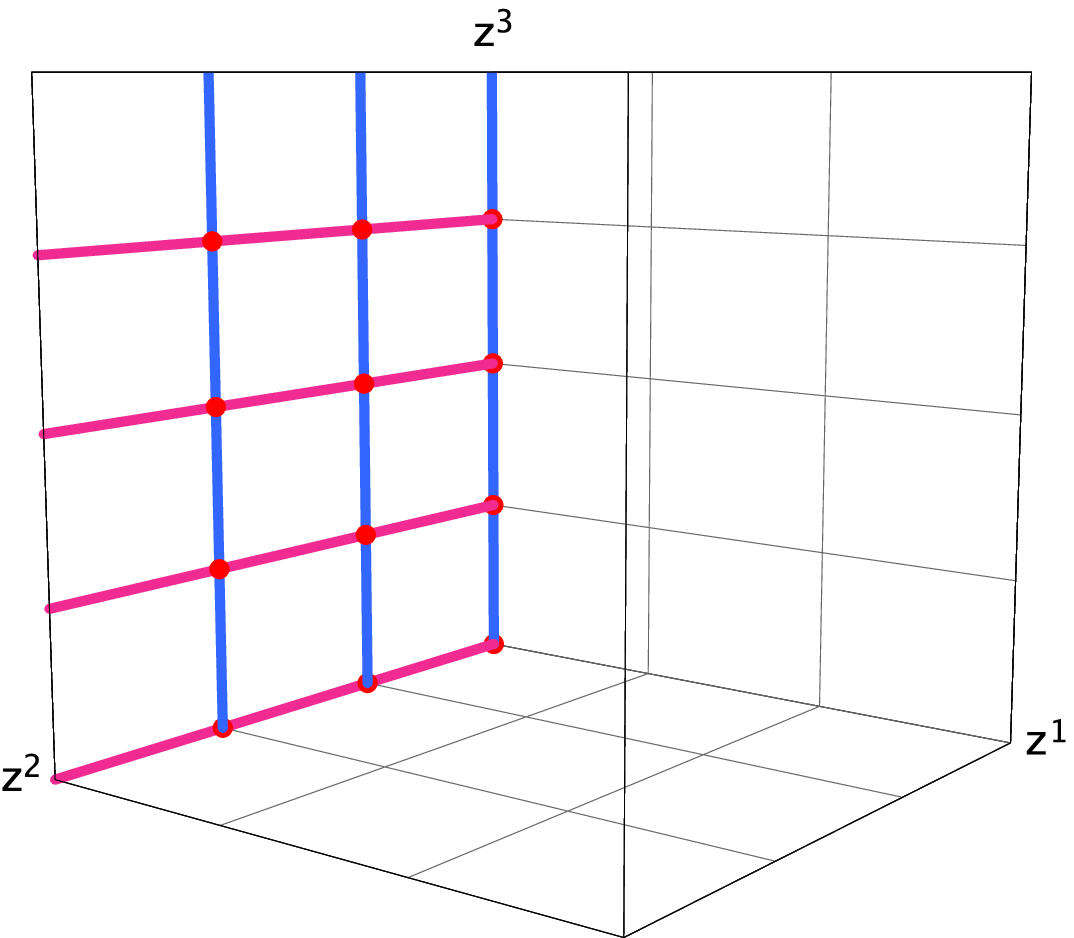}}
\hskip-0.75cm The configuration of the fixed point set is displayed in Figure\ffixsixiiaa\ in a schematic way, 
where each complex coordinate is shown as a coordinate axis and the opposite faces of the 
resulting cube of length 1 are identified.
We see that there are 12 local $\IC^3/\IZ_{6-II}$ patches which each sit at the intersection of 
two fixed lines, 3 $\IC^2/\IZ_3$ fixed lines in the $z^3$ direction originating from the order 
three element $\theta^2$ and 4 $\IC^2/\IZ_2$ fixed lines in the $z^2$ direction originating 
from the order two element $\theta^3$. 
\ifig\fsixii{Toric diagram of the resolution of ${\bf C}^3/\IZ_{6-II}$ and dual graph}{\epsfxsize=1\hsize\epsfbox{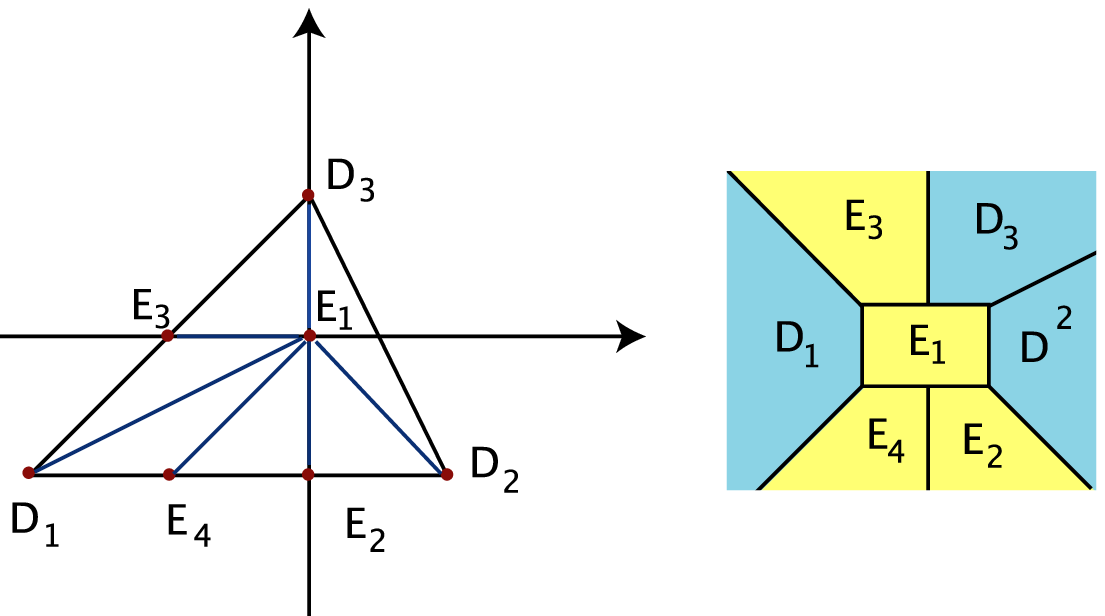}}
The resolution of the $\IC^3/\IZ_{6-II}$ singularity is described by the toric diagram in Figure\fsixii.
From these two figures, we can read off the exceptional divisors \first, which together with the 
inherited divisors $R_i$ form a basis for $H^{(1,1)}(X_6)$:
\eqn\DIVS{
R_1\ \ \ ,\ \ \ R_2\ \ \ ,\ \ \ R_3\ \ \ ,\ \ \ 
E_{1,\beta\gamma}\ \ \ ,\ \ E_{3,\gamma}\ \ \ ,\ \ \ 
E_{2,\beta}\ \ \ ,\ \ \ E_{4,\beta}\ ,}
with $\beta=1,2,3\ ,\ \gamma=1,\ldots,4$.
In addition, the orbifold fixed points give rise to the eight divisors $D_1, D_{2,\beta}$
and $D_{3,\gamma}$.
The topologies of these divisors were determined in \first. $E_{1,\beta\gamma}$ is a blow--up of $\IF_1$ 
in two points, while the remaining exceptional divisors $E_{2,\beta},E_{3,\gamma},E_{4,\beta}$ are 
all $\IP^1\times\IP^1$. $D_1, D_{2\beta}, D_{3\gamma}$ are blow--ups of $\IP^1\times\IP^1$ in 12, 8, and 9 
points, respectively. Finally, $R_1$ is a $T^4$ and $R_2,R_3$ are K3 surfaces.

Any orientifold projection $\Oc$ with $O3$-- and $O7$--planes is such that the twelve divisors 
$E_{1,2\gamma}$, $E_{1,3\gamma}$, $E_{2,2}$, $E_{2,3}$, $E_{4,2}$, and $E_{4,3}$ are not invariant under 
the orientifold action $\sigma$ \first:
\eqn\notinV{\eqalign{
\sigma\  E_{1,2\gamma}&=E_{1,3\gamma}\ \ \ ,\ \ \ 
\sigma\  E_{1,3\gamma}=E_{1,2\gamma}\ ,\cr
\sigma\  E_{2,2}&=E_{2,3}\ \ \ ,\ \ \ \sigma\  E_{2,3}=E_{2,2}\ ,\cr
\sigma\  E_{4,2}&=E_{4,3}\ \ \ ,\ \ \ \sigma\  E_{4,3}=E_{4,2}\ ,}}
and the six pairs of eigenstates $(E,\tilde E)$ 
under $\sigma$ have to be constructed (\cf also Section 3):
\eqn\eigenST{\eqalign{
E_{1,\gamma}&:=\h\ (\ E_{1,2\gamma}+E_{1,3\gamma}\ )\ \ \ ,\ \ \ 
\tilde E_{1,\gamma}:=\h\ (\ E_{1,2\gamma}-E_{1,3\gamma}\ )\ ,\cr
E_{2}&:=\h\ (\ E_{2,2}+E_{2,3}\ )\ \ \ ,\ \ \ 
\tilde E_{2}:=\h\ (\ E_{2,2}-E_{2,3}\ )\ ,\cr
E_{4}&:=\h\ (\ E_{4,2}+E_{4,3}\ )\ \ \ ,\ \ \ 
\tilde E_{4}:=\h\ (\ E_{4,2}-E_{4,3}\ )\ .}}
As a consequence, we have $h_{(1,1)}^{(-)}(X_6)=6$ (\cf also Table 4).
Furthermore, the divisors $D_{2,2}$ and $D_{2,3}$ are mapped to each other under $\sigma$
and only the combination $D_2=\h(D_{2,2}+D_{2,3})$ is invariant.
To summarize, the orientifold action $\Oc$ splits the divisors \DIVS\
into the even divisors 
\eqn\evenDIVS{
H_+^{(4)}(X_6)\ \owns\ 
E_{1,1\gamma}\ ,\ E_{1,\gamma}\ ,\ E_{2,1}\ ,\ E_2\ ,\ 
E_{4,1}\ ,\ E_4\ ,\ E_{3,\gamma}\ ,\ 
D_1\ ,\ D_{2}\ ,\ D_{2,1}\ ,\ 
D_{3,\gamma}\ ,\ R_1,\ ,\ R_2\ ,\ R_3} 
and into the odd divisors
\eqn\oddDIVS{
H_-^{(4)}(X_6)\ \owns\  
\tilde E_{2}\ \ \ ,\ \ \ \tilde E_{4}\ \ \ ,\ \ \ \tilde E_{1,\gamma}\ \ \ ,\ \ \ \tilde D_2}
with $\gamma=1,\ldots,4$.
We choose the orientifold action $\sigma$ such that its fixed point set consists of seven 
$O7$--planes wrapped on the divisors $D_1, D_{3,\gamma},E_{2,1}$ and $E_2$. In addition, 
there are twelve $O3$--planes at $z^2=0\ ,\ z^1\neq 0$. 
Because of $\chi(D_1)=16$, $\chi(D_{3,\gamma})=13$ and $\chi(E_{2,1})=\chi(E_2)=4$, the total
$D3$--brane charge $Q_{3,tot}$ in \eqq \totaltadpole\ is $Q_{3,tot}=-22$.
The Poincar\'e dual $2$--forms $\omega_i$ of the $19$ invariant divisors
represent a basis for the K\"ahler form~$J$:
\eqn\Jj{\eqalign{
J=r&_1\ R_1+r_2\ R_2+r_3\ R_3-t_2\ E_2-t_4\ E_4-t_{2,1}\ E_{2,1}-t_{4,1}\ E_{4,1}
\cr
&-\sum_{\gamma=1}^4 \lf(\ t_{1,\gamma}\ 
E_{1,\gamma}+t_{1,1\gamma}\ E_{1,1\gamma} +t_{3,\gamma}\ E_{3,\gamma}\ \ri)\ ,}}
with the $19$ K\"ahler coordinates $r_1,\ r_2,\ r_3,\ t_{1,\gamma},\ t_{1,1\gamma},\ t_2,\ 
t_{2,1},\ t_{3,\gamma},\ t_4,\ t_{4,1}$.
The intersection numbers computed in \first\ are split (\cf Subsection 2.1)
into the two sets of intersection
numbers $\Kc_{ijk}$
\eqn\interni{\eqalign{
&R_1R_2R_3=3\ \ \ ,\ \ \ R_3E_{2,1}E_{4,1}=1\ \ \ ,\ \ \ 
R_3E_{2}E_{4}=\h\ \ \ ,\ \ \ 
R_2E_{3,\gamma}^2=-1\ ,\cr 
&R_3E_{2,1}^2=-4\ \ \ ,\ \ \ \ R_3E_{4,1}^2=-1\ \ \ ,\ \ \ 
R_3E_{2}^2=-2\ \ \ ,\ \ \ \ R_3E_{4}^2=-\h\ ,\cr
&E_{1,1\gamma}^3=3\ \ \ ,\ \ \ E_{1,\gamma}^3=\fc{3}{4}\ \ \ ,\ \ \ E_{2,1}^2E_{4,1}=-4
\ \ \ ,\ \ \ E_{2}^2E_{4}=-1\ ,\cr
&E_{2,1}^3=32\ \ \ ,\ \ \ E_{4,1}^3=4\ \ \ ,\ \ \ 
E_{2}^3=8\ \ \ ,\ \ \ E_{4}^3=1\ \ \ ,\ \ \ E_{3,\gamma}^3=4\ ,\cr
&E_{1,1\gamma}E_{2,1}^2=-4\ \ \ ,\ \ \ E_{1,1\gamma}E_{4,1}^2=-1\ \ \ ,\ \ \ 
E_{1,1\gamma}E_{3,\gamma}^2=-1\ ,\cr 
&E_{1,\gamma}E_{2}^2=-1\ \ \ ,\ \ \ E_{1,\gamma}E_{4}^2=-\fc{1}{4}\ \ \ ,\ \ \ 
E_{1,\gamma}E_{3,\gamma}^2=-1\ ,\cr
&E_{1,1\gamma}E_{2,1}E_{4,1}=1\ \ \ ,\ \ \ E_{1,\gamma}E_{2}E_{4}=\fc{1}{4}\ ,}}
and $\Kc_{iab}$:
\eqn\intern{\eqalign{
&R_3\tilde E_2\tilde E_4=\h\ \ \ ,\ \ \ R_3\tilde E_2^2=-2\ \ \ ,\ \ \ 
R_3\tilde E_4^2=-\h\ ,\cr
&E_{1,\gamma}\tilde E_{1,\gamma}^2=\fc{3}{4}\ \ \ ,\ \ \ E_4\tilde E_2^2=-1\ \ \ ,\ \ \ 
E_2\tilde E_2\tilde E_4=-1\ ,\cr
&E_2\tilde E_2^2=8\ \ \ ,\ \ \ E_4\tilde E_4^2=1\ ,\cr
&\tilde E_{1,\gamma}\tilde E_2E_2=-1\ \ \ ,\ \ \ E_{1,\gamma}\tilde E_2^2=-1
\ \ \ ,\ \ \ \tilde E_{1,\gamma}\tilde E_4E_4=-\fc{1}{4}\ \ \ ,\ \ \ 
E_{1,\gamma}\tilde E_4^2=-\fc{1}{4}\ ,\cr
&E_{1,\gamma}\tilde E_{2}\tilde E_{4}=\fc{1}{4}\ \ \ ,\ \ \ 
\tilde E_{1,\gamma} E_{2}\tilde E_{4}=\fc{1}{4}\ \ \ ,\ \ 
\tilde E_{1,\gamma}\tilde E_{2} E_{4}=\fc{1}{4}\ .}}
The volume  ${\rm Vol}(X_6)=\fc{1}{6}\ \int_{X_6} J\wedge J\wedge J =\fc{1}{6}\ \Kc_{ijk}\ 
t^i\ t^j\ t^k$ of the CY orientifold $X_6$ becomes:
\eqn\volsix{\eqalign{\hskip-0.75cm
{\rm Vol}(X_6)&=3\ r_1\ r_2\ r_3+
r_3\ (t_{2,1}\ t_{4,1}+\h\ t_{2}\ t_{4})-
\sum\limits_{\gamma=1}^4\ (\ t_{1,1\gamma}\ t_{2,1}\ t_{4,1}+
\fc{1}{4}\ t_{1,\gamma}\ t_2\ t_4\ )\cr
&-\fc{1}{2}\ r_2\ 
\sum\limits_{\gamma=1}^4\ t_{3,\gamma}^2-r_3
\ (\ 2\ t_{2,1}^2+\fc{1}{2}\,t_{4,1}^2+t_{2}^2+\fc{1}{4}\ t_{4}^2\ )\cr
&-\fc{1}{2}\,\sum\limits_{\gamma=1}^4 (\ t_{1,1\gamma}^3+\fc{1}{4}\ t_{1,\gamma}^3\ )+
2\ t_{2,1}^2\ t_{4,1}+\fc{1}{2}\ t_2^2\ t_4
-{4\over 3}\ (\ 4\,t_{2,1}^3+\fc{1}{2}\,t_{4,1}^3+t_{2}^3+\fc{1}{8}\ t_{4}^3\ )-\fc{2}{3}\,
\sum\limits_{\gamma=1}^4 t_{3,\gamma}^3\cr
&+\sum\limits_{\gamma=1}^4\ \lf(\ 2\ t_{1,1\gamma}\ t_{2,1}^2+\fc{1}{2}\ 
t_{1,1\gamma}\ t_{3,\gamma}^2+\fc{1}{2}\,t_{1,1\gamma}\ t_{4,1}^2+\h\ 
t_{1,\gamma}\ t_2^2+\h\  
t_{1,\gamma}\ t_{3,\gamma}^2+\fc{1}{8}\ t_{1,\gamma}\ t_{4}^2\ \ri)\ .}}
\def\Vol{ {\rm Vol} }
According to \eqq \divvol\ from \interni\ or \volsix, the $19$ 
divisor volumes are derived:
\eqn\DIVOLS{\eqalign{
\Vol(R_1)&=\fc{9}{2}\ r_2\ r_3\ \ \ ,\ \ \ \
Vol(R_2)=\fc{9}{2}\ r_1\ r_3-\fc{3}{4}
\ \sum_{\gamma=1}^4 t_{3,\gamma}^2\ ,\cr
\Vol(R_3)&=\fc{9}{2}\ r_1\ r_2+\fc{3}{2}\ t_{2,1}\ t_{4,1}+\fc{3}{4}\ t_2\ t_4-
3\ t_{2,1}^2-\fc{3}{4}\ t_{4,1}^2-\fc{3}{2}\ t_2^2-\fc{3}{8}\ t_4^2\ ,\cr
\Vol(E_{1,\gamma})&=-\fc{9}{16}\ t_{1,\gamma}^2+\fc{3}{4}\ t_2^2+\fc{3}{16}\ t_4^2
+\fc{3}{4}\ t_{3,\gamma}^2-\fc{3}{8}\ t_2\ t_4\ ,\cr
\Vol(E_{1,1\gamma})&=-\fc{9}{4}\ t_{1,1\gamma}^2+3\ t_{2,1}^2+\fc{3}{4}\ t_{4,1}^2
+\fc{3}{4}\ t_{3,\gamma}^2-\fc{3}{2}\ t_{2,1}\ t_{4,1}\ ,\cr
\Vol(E_2)&=\fc{3}{4}\ r_3\ t_4-3\ r_3\ t_2+\fc{3}{2}\ t_2\ t_4-6\ t_2^2-\fc{3}{8}\ 
\sum_{\gamma=1}^4
(\ t_{1,\gamma}\ t_4-4 t_2\ t_{1,\gamma}\ )\ ,\cr
\Vol(E_{2,1})&=\fc{3}{2}\ r_3\ t_{4,1}-6\ r_3\ t_{2,1}+6\ t_{2,1}\ t_{4,1}-
24\ t_{2,1}^2-\fc{3}{2}\ \sum_{\gamma=1}^4
(\ t_{1,1\gamma}\ t_{4,1}-4 t_{2,1}\ t_{1,1\gamma}\ )\ ,\cr
\Vol(E_{3,\gamma})&= -\fc{3}{2}\ r_2\ t_{3,\gamma}-3\ t_{3,\gamma}^2+
\fc{3}{2}\ t_{1,1\gamma}\ t_{3,\gamma}+\fc{3}{2}\ t_{1,\gamma}\ t_{3\gamma}\ ,}}
$$\eqalign{
\Vol(E_4)&=\fc{3}{4} \ r_3\ t_2-\fc{3}{4}\ r_3\ t_4+\fc{3}{4}\ t_2^2-\fc{3}{4}\ t_4^2
+\fc{3}{8}\ \sum_{\gamma=1}^4t_{1,\gamma}\ t_4-\fc{3}{8}\ \sum_{\gamma=1}^4
t_{1,\gamma}\ t_2\ ,\cr
\Vol(E_{4,1})&=\fc{3}{2}\ r_3\ t_{2,1}-\fc{3}{2}\ r_3\ t_{4,1}+3\ t_{2,1}^2-3\ t_{4,1}^2
+\fc{3}{2}\ \sum_{\gamma=1}^4t_{1,1\gamma}\ t_{4,1}-\fc{3}{2}\ \sum_{\gamma=1}^4
t_{1,1\gamma}\ t_{2,1}\ .}$$
The seven (invariant) 
planes $D_{i\alpha}$ localized at the fix points are given through the relations~\first:
\eqn\Planes{\eqalign{
D_1&=\fc{1}{3}\ \lf(R_1-E_{2,1}-4\ E_{4,1}-2\ E_2-8\ E_{4,2}\ \ri)
-\fc{1}{3}\ \sum_{\gamma=1}^4\lf(\ 3\ E_{3,\gamma}+ 2\
E_{1,\gamma}+E_{1,1\gamma}\ \ri)\ ,\cr}}
$$\eqalign{
D_{2}&=\fc{1}{3}\ \lf(\ R_2-E_{2}-E_{4}\ \ri)-\fc{1}{3}\ \sum_{\gamma=1}^4 E_{1,\gamma}\ ,
\ D_{2,1}=\fc{1}{3}\ \lf(\ R_2-E_{2,1}-E_{4,1}\ \ri)
-\fc{1}{3}\ \sum_{\gamma=1}^4 E_{1,1\gamma}\ ,\cr
D_{3,\gamma}&=R_3-E_{1,1\gamma}-2\ E_{1,\gamma}-E_{3,\gamma}\ .}$$
In the superpotential \SUP, we have two sets of contributing divisors $\Dc_i$:
On the seven divisors $\Dc_{D7}=\{D_1, D_{3,\gamma},E_{2,1},E_2\}$, a stack of 
one $O7$--plane and eight $D7$--branes is wrapped. Gaugino condensation takes place in the $SO(8)$ gauge theory. Therefore, we have $a_j=-\fc{2\pi}{6}$ for the set $\Dc_{D7}$
of divisors contributing in \SUP.
On the other hand, divisors in the set 
$\Dc_{D3}=\{D_2,D_{2,1},E_{1,\gamma},E_{1,1\gamma},E_{3\gamma},E_4,E_{4,1}\}$
can be wrapped by Euclidean $D3$--branes. Since all $D$ and $E$
divisors intersect one of the divisors carrying an $O7$--plane in at
least one complex dimension (\cf 
Figures\ffixsixiiaa\ and\fsixii), the condition $\chi_{D3}=1$ for
a non--vanishing instanton contribution in the superpotential \SUP\ is always met for
the set $\Dc_{D3}$ of divisors.
In total we have $23$ contributing divisors and the superpotential \SUP\ reads:
\eqn\SUPreads{
W=W_0(S,U)+\sum_{\Dc_i\in \Dc_{D7}} e^{-2\pi\ \fc{\Vol(\Dc_i)}{6}}+\sum_{\Dc_i\in \Dc_{D3}}
e^{-2\pi\ \Vol(\Dc_i)}\ .}
Now we are ready to stabilize all $19$ K\"ahler moduli $r_i, t_{1,\gamma}, t_{1,1\gamma},
t_2,t_{2,1}, t_{3,\gamma}, t_4$ and $t_{4,1}$.
The $D3$--brane charge $Q_{3,tot}=-22$ is completely cancelled  by the $3$--form flux $G_3$,
given in \eqq \GdreiZsechs.
In fact, in the previous subsection we have presented critical points for
the dilaton and complex structure moduli corresponding to a set of flux solutions, 
with $N_{flux}=44$. 
For a $\tilde W_0=-0.34864$, corresponding to the critical points of Table 6,
we find the following $23$ divisor volumes (measured in string units):
\eqn\findDIV{
\matrix{\Vol(D_1) =4.92087\ , & \Vol(D_{2,1})= 17.1883\ ,&\Vol(D_{2,2}) = 17.9329\ ,\cr
\Vol(D_{3,\gamma}) =35.1656\ ,&\Vol(E_2) = 3.55689\ , & \Vol(E_4)= 0.710518\ ,\cr  
\Vol(E_{2,1}) = 4.84171\ , & \Vol(E_{4,1}) = 0.922548\ , & \Vol(E_{3,\gamma}) = 1.01315\ ,\cr
\Vol(E_{1,1\gamma}) = 1.06872\ ,&\Vol(E_{1,\gamma}) = 0.884484\ ,&}\ \ \ 
\gamma=1,\ldots,4\ ,}
corresponding to the sizes of the nineteen K\"ahler moduli:
\eqn\findDIVV{\eqalign{
r_1&=3.04765 \ \ \ ,\ \ \  r_2 =2.91779 \ \ \ ,\ \ \ r_3 = 4.53928\ ,\cr
t_{1,\gamma}&= 1.52711\ \ \ ,\ \ \ t_{1,1\gamma}= 0.869367\ \ \ ,\ \ \ 
t_{3,\gamma} = 0.46524\ \ \ ,\ \ \ \gamma=1,\ldots,4\ ,\cr
t_{2,1} &=0.443261\ \ \ ,\ \ \ t_2 = 0.663503\ \ \ ,\ \ \  
t_4 = 0.967525\ \ \ ,\ \ \ t_{4,1} = 0.634432\ .}}
The divisor volumes give rise to the total volume $\Vol(X_6)=115.94$. This is large enough,
that one--loop (and higher loop) corrections to the K\"ahler potential are suppressed, 
with a string--coupling
constant $g_{\rm string}\sim 0.3$ (\cf Subsection 5.3).
Furthermore, the divisor volumes \findDIV\ are large enough to suppress higher order
instanton effects (\eg from multi--wrapped instantons) 
in \SUPreads, since $e^{-2\pi \Vol(E_2)/6}\sim 0.02$ and even
smaller for the other divisors.

The six divisors \oddDIVS\ or their corresponding cohomology elements give
rise to the non--vanishing bulk $2$--forms $B_2,C_2$ (\cf \eqqs
\bulkB\ and \bulkC), with the twelve real scalars $b^a,c^a$. According
to \eqq \combined\ the latter are combined into the six complex
scalars $G^a$, defined in \eqq \combined. 
Following Section 3 for stabilizing the moduli $b^a$ we solve the
calibration condition \conclude.
For this we need the intersection form $\Kc_{ab}$ and calculate its
determinant for the values \findDIVV. The matrix $\Kc_{ab}$ is given by:
\def\ss#1{{\scriptstyle{#1}}}
\eqn\INTERSIX{\hskip-1cm
\Kc_{ab}=\pmatrix{\ss{-2\ r_3+8\ t_2-t_4-\sum\limits_\gamma   t_{1,\gamma}}&
\ss{\h\ r_3-t_2+\fc{1}{4}\ \sum\limits_\gamma   t_{1,\gamma}}&
\ss{-t_2+\fc{1}{4}\ t_4}&\ss{-t_2+\fc{1}{4}\ t_4}&\ss{-t_2+\fc{1}{4}\ t_4}&
\ss{-t_2+\fc{1}{4}\ t_4}\cr
\ss{2\ r_3-t_2+2\ \sum\limits_\gamma   t_{1,\gamma}}&
\ss{-\h\ r_3+t_4-\fc{1}{4}\ \sum\limits_{\gamma}t_{1,\gamma}}&
\ss{-\fc{1}{4}\ t_4+\fc{1}{4}\ t_2} &     \ss{-\fc{1}{4}\ t_4+\fc{1}{4}\ t_2}&
\ss{-\fc{1}{4}\ t_4+\fc{1}{4}\ t_2}&\ss{-\fc{1}{4}\ t_4+\fc{1}{4}\ t_2}\cr
\ss{-t_2+\fc{1}{4}\ t_4}&\ss{-\fc{1}{4}\ t_4+\fc{1}{4}\ t_2} & \ss{\fc{3}{4}\ t_{1,1}}
&0 &0&0\cr
\ss{-t_2+\fc{1}{4}\ t_4}&\ss{-\fc{1}{4}\ t_4+\fc{1}{4}\ t_2} & 0&\ss{\fc{3}{4}\ t_{1,2}}
&0&0\cr
\ss{-t_2+\fc{1}{4}\ t_4}&\ss{-\fc{1}{4}\ t_4+\fc{1}{4}\ t_2} & 0&0& \ss{\fc{3}{4}\ t_{1,3}}
&0\cr
\ss{-t_2+\fc{1}{4}\ t_4}&\ss{-\fc{1}{4}\ t_4+\fc{1}{4}\ t_2}&0&0&0&\ss{\fc{3}{4}\ t_{1,4}}}
\ .}
For the values \findDIVV\ we find $\det(\Kc)=-91.25\neq 0$. Hence, the
only non--trivial solution for \conclude\ is:
$$b^a=0\ .$$ Furthermore, in Section 3 a mechanism has been proposed to
also stabilize the fields $c^a$ by turning on the $2$--form flux
$^Yf_2$ from the ambient space $Y_6$ (\cf \eqq \bulkF\ and also the discussion at the end of
section 4).
To this end, for the $\IZ_6$ orbifold we have stabilized all
$27$ moduli fields.


\ \br
$\underline{(ii)\ \ 
\IZ_2\times \IZ_4-{\rm orbifold\ on\ the\ } SU(2)^2\times SO(5)^2\ {\rm lattice}:}$
\ \br
\ \br
As our second example, we consider the resolved $\IZ_2\times \IZ_4$ orbifold $Y_6$
on the lattice $SU(2)^2\times SO(5)^2$. This orbifold has $h_{(1,1)}(Y_6)=61$
K\"ahler moduli. We summarize here the relevant data from  \first.  
\ifig\ffixtwofour{Schematic picture of the fixed set configuration of
$\IZ_2\times \IZ_4$}{\epsfxsize=0.6\hsize\epsfbox{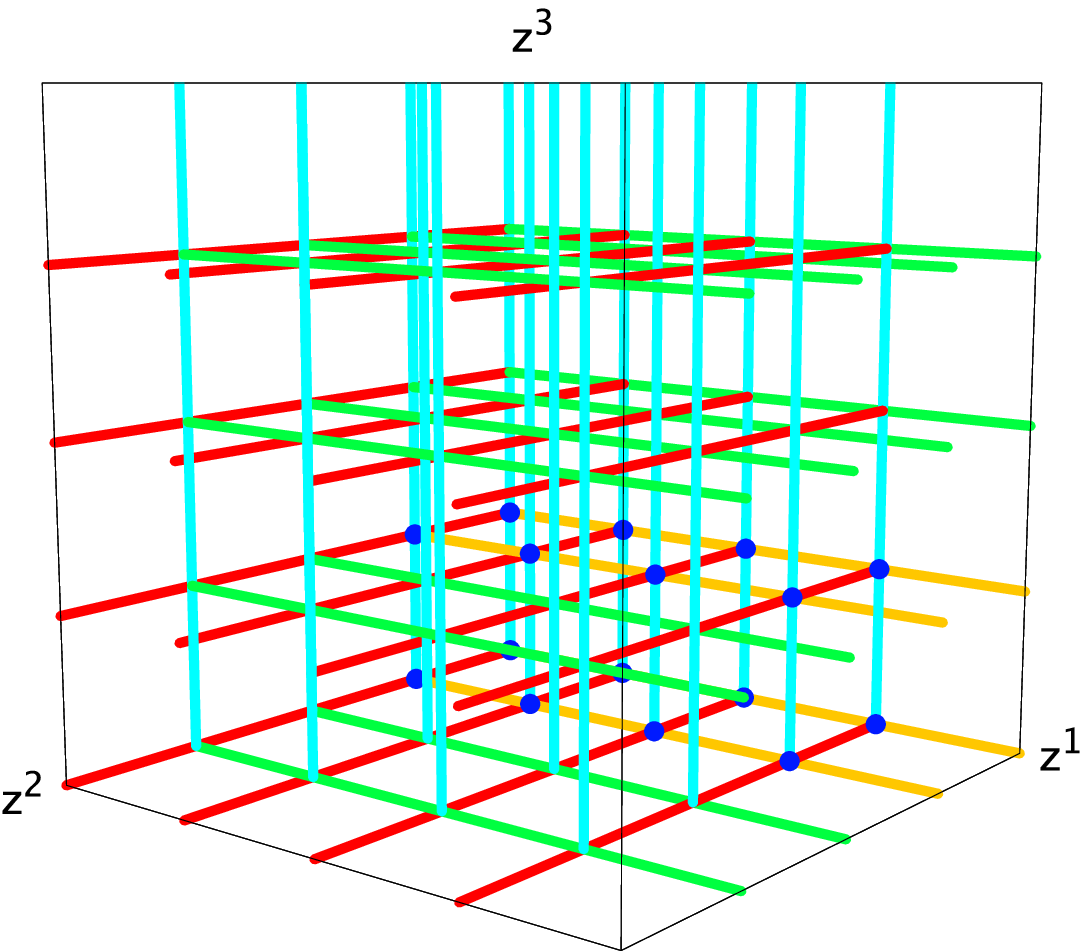}}
\hskip-0.75cm
The configuration of the fixed point set is displayed in Figure\ffixtwofour\ in a schematic way.
\ifig\frtwofour{Toric diagram of two of the resolutions of ${\bf
C}^3/\IZ_{2}\times\IZ_4$ and dual
graphs}{\epsfxsize=0.85\hsize\epsfbox{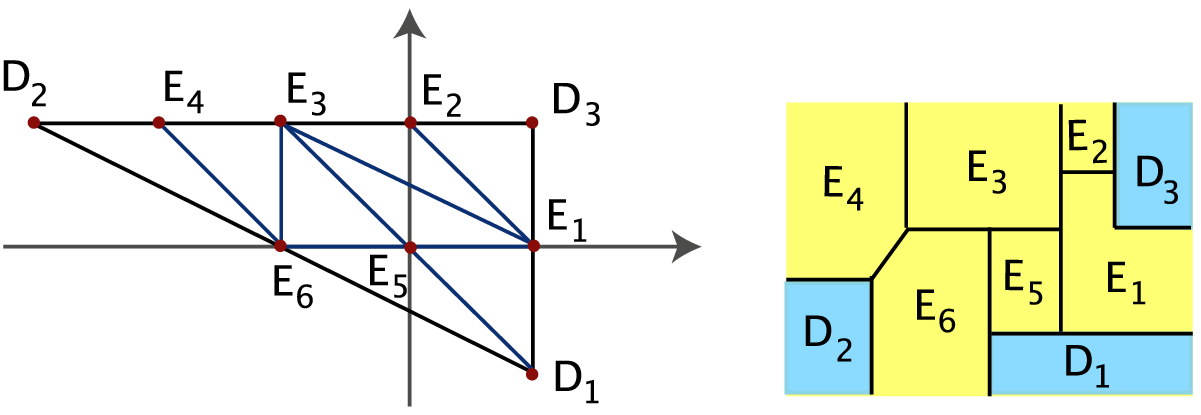}}
\hskip-0.75cm There are 16 local $\IC^3/\IZ_2\times\IZ_4$ patches.
The resolution of the $\IC^3/\IZ_2\times\IZ_4$ singularity is
described by the toric diagram in Figure\frtwofour. There are four
$\IC^2/\IZ_4$ fixed lines in the $z^1$ direction from the order four
element $\theta^2$. Furthermore, there are $12+12+(10-4)=30$
$\IC^2/\IZ_2$ fixed lines from the order two elements: From
$\theta^1$, $\theta^1(\theta^2)^2$, and $(\theta^2)^2$ in the $z^2$,
$z^3$, and $z^1$ direction, respectively. The intersection points of
three $\IZ_2$ fixed lines are locally described by the resolved 
$\IC^3/\IZ_2\times\IZ_2$ patches.
\ifig\frtwotwo{Toric diagram of the resolution of ${\bf
C}^3/\IZ_{2}\times\IZ_2$ and 
its dual graph}{\epsfxsize=0.85\hsize\epsfbox{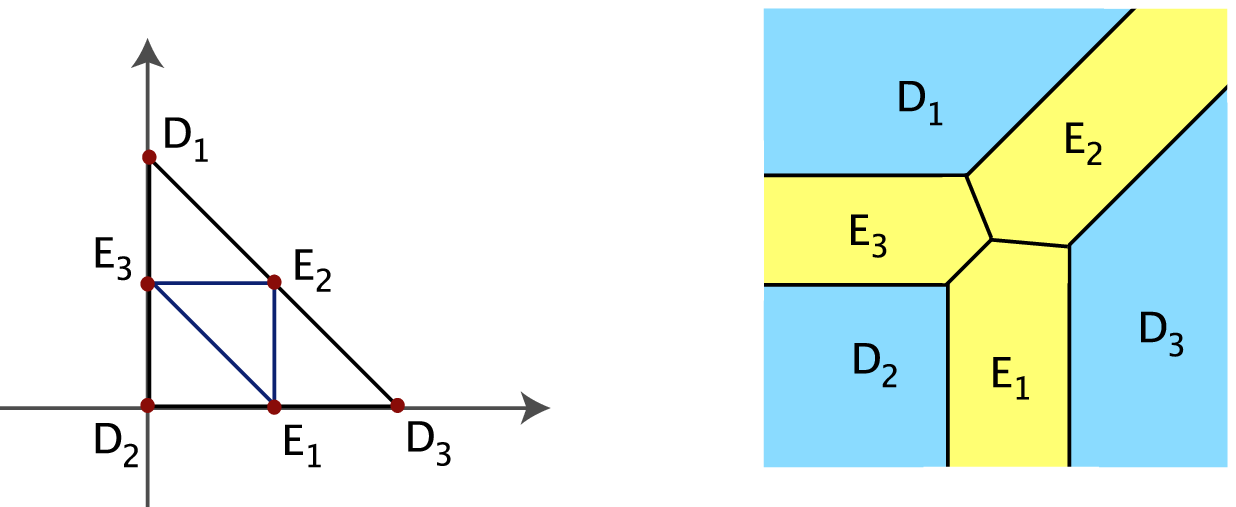}}
\hskip-0.75cm The resolution is described by the toric diagram in
Figure\frtwotwo. From these two figures, we can 
read off the exceptional divisors \first, which together with the
inherited divisors $R_i$ form a 
basis for $H^{(1,1)}(Y_6)$:
\eqn\DIVStwofour{
R_1\ ,\  R_2\ ,\  R_3\ ,\ E_{1,\alpha\gamma'}\ ,\ E_{2,\beta\gamma}\
,\ E_{3,\mu}\ ,\ E_{4,\beta\gamma}\ ,\ E_{5,\alpha\beta\gamma}\ ,\ E_{6,\alpha\beta'},}
with $\alpha=1,\ldots,4$, $\beta=1,2$, $\beta'=1,2,3$, $\gamma=1,2$,
$\gamma'=1,2,3$ 
and $\mu=1,\ldots,10$. The divisors $E_{3,\mu}$, $\mu=1,2,4,5$ will
also be denoted by 
$E_{3,\beta\gamma}$, $\beta,\gamma=1,2$. In addition, the orbifold fixed points give rise to the ten divisors \first:
\eqn\ADDDD{
D_{1,\al}\ \ ,\ \ D_{2,\beta'}\ \ ,\ \ D_{3,\gamma'}\ .}
The topology of these divisors was determined in \first. The divisors
$E_{1,\alpha\gamma}$ and $E_{6,\alpha\beta}$ are blow--ups of
$\IP^1\times\IP^1$ in 5 points, the divisors $E_{1,\alpha3}$,
$E_{3,\mu}$, $\mu=3,6,\ldots,10$, $E_{6,\alpha3}$, and $D_{1\alpha}$
are blow--ups of $\IP^1\times\IP^1$ in 
4 points, the divisors $E_{3,\mu}$, $\mu=1,2,4,5$ are blow--ups of
$\IP^1\times\IP^1$ 
in 8 points. The divisors $E_{2,\beta\gamma}$, $E_{4,\beta\gamma}$,
$D_{2\beta'}$, and $D_{3\gamma'}$ 
are $\IP^1\times\IP^1$, while the divisors $E_{5,\alpha\beta\gamma}$
are $\IF_1$, and the $R_i$ are K3 surfaces. 

The orientifold projection $\Oc$ leaves all the divisors \DIVStwofour\
and \ADDDD\ invariant, hence $h_{(1,1)}^{(-)}(X_6)=0$ 
(\cf also Table 4). We choose an orientifold action $\sigma$ such that
its fixed point set 
consists of 14 $O7$--planes wrapped on the divisors $D_{1,\alpha}$,
$D_{2\beta'}$, $D_{3,\gamma'}$, and $E_{3,\mu}$, $\mu=1,2,4,5$. There are no
$O3$--planes. Because of 
$\chi(D_{1\alpha})=8$, $\chi(D_{2\beta'})=\chi(D_{3,\gamma})=4$, and $\chi(E_{3,\mu})=12$, the total
$D3$--brane charge $Q_{3,tot}$ in \eqq \totaltadpole\ is $Q_{3,tot}=-26$.
The Poincare dual $2$--forms $\omega_i$ of the $61$ invariant divisors
\DIVStwofour\ 
represent a basis for the K\"ahler form $J$:
\eqn\Jjj{\eqalign{
J&=r_1\ R_1+r_2\ R_2+r_3\
R_3-\sum_{\beta,\gamma=1,2}\lf(t_{2,\beta\gamma}\ E_{2,\beta\gamma}+
t_{4,\beta\gamma}\
E_{4,\beta\gamma}+\sum_{\alpha=1}^4t_{5,\alpha\beta\gamma}\ E_{5,\alpha\beta\gamma}\ri)\cr
&-\sum_{\al=1}^4\lf(\sum_{\gamma=1,2,3}t_{1,\alpha\gamma}\ E_{1,\alpha\gamma}+
\sum_{\beta=1,2,3} t_{6,\al\beta}\ E_{6,\al\beta}\ri)-
\sum_{\mu=1}^{10}t_{3,\mu}\  E_{3,\mu}\ ,}}
with the $61$ K\"ahler coordinates
$r_i,\ t_{1,\al\gamma},\ t_{2,\beta\gamma},\ t_{3,\mu},\
t_{4,\beta\gamma},\ 
t_{5,\al\beta\gamma},\ t_{6,\al\beta}$.
The orientifold action changes the triple intersection numbers
$\Kc_{ijk}$. They become \first:
\eqn\ringtwofourA{\eqalign{
  R_1R_2R_3 &=2,\ \ \ \ \ \ R_1E_{2\beta\gamma}^2 =-1,\ \ \ \ \ \ R_1E_{3\mu''}^2 =-1, \cr
   R_1E_{4\beta\gamma}^2 &=-1,\ \ \ \ \ \ R_2E_{1\alpha\gamma}^2 =-1,\ \ \ \ \ \ R_2E_{1\alpha3}^2 = -2, \cr
  R_3E_{6\alpha\beta}^2 &=-1,\ \ \ \ \ \ R_3E_{6\alpha3}^2 = -2,\ \ \ \ \ \ E_{1\alpha\gamma}^3 =3/2, \cr
  E_{1\alpha3}^3 &=2,\ \ \ \ \ \ E_{1\alpha\gamma'}^2E_{3\mu''} = -1/2,\ \ \ \ \ \ E_{1\alpha\gamma}^2E_{6\alpha3} = -1/2,\cr
   E_{1\alpha3}^2E_{6\alpha\beta} &= -1/2,\ \ \ \ \ \ E_{1\alpha3}^2E_{6\alpha3} = -1, \ \ \ \ \ \ 
  E_{1\alpha\gamma}E_{2\beta\gamma}^2 =-1, \cr
  E_{1\alpha\gamma'}E_{3\mu''}^2 &= -1/2, \ \ \ \ \ \ 
  E_{1\alpha3}E_{3\mu_1}E_{6\alpha\beta} =1/2,\ \ \ \ \ \ E_{1\alpha\gamma}E_{3\mu_2}E_{6\alpha3} =1/2,\cr
   E_{1\alpha3}E_{3\mu_3}E_{6\alpha3} &=1/2,\ \ \ \ \ \ E_{1\alpha\gamma}E_{5\alpha\beta\gamma}^2 = -1, \ \ \ \ \ \ 
  E_{1\alpha3}E_{6\alpha\beta}^2 =-1/2,\cr
   E_{1\alpha\gamma}E_{6\alpha3}^2 &=-1/2,\ \ \ \ \ \ E_{1\alpha3}E_{6\alpha3}^2 =-1,\ \ \ \ \ \ E_{2\beta\gamma}^3 = 4, \cr
  E_{3\mu''}^3 &= 2, \ \ \ \ \ \ 
  E_{3\mu''}^2E_{6\alpha\beta'} = -1/2, \ \ \ \ \ \ 
   E_{3\mu''}E_{6\alpha\beta'}^2 = -1/2, \cr
   E_{4\beta\gamma}^3 &= 4,\ \ \ \ \ \ E_{4\beta\gamma}^2E_{6\alpha\beta} = -1,  \ \ \ \ \ \ 
  E_{5\alpha\beta\gamma}^3 = 4, \cr
  E_{5\alpha\beta\gamma}^2E_{6\alpha\beta}&=-1,\ \ \ \ \ \
E_{6\alpha\beta}^3 =3/2,\ \ \ \ \ \ E_{6\alpha3}^3 = 2\ .}}
The intersection numbers involving the $E_{3,\mu}, \ \mu=1,2,4,5$, which are fixed are
\eqn\ringtwofourB{\eqalign{
R_1E_{3\mu}E_{4\beta\gamma} &= 1,\ \ \ \ \ \ R_1E_{2\beta\gamma}E_{3\mu} =1,\ \ \ \ \ \ E_{1\alpha\gamma'}^2E_{3\mu} = -1,\cr
 E_{1\alpha\gamma}E_{2\beta\gamma}E_{3\mu} &= 1,\ \ \ \ \ \ E_{1\alpha\gamma'}E_{3\mu}^2 = -2, \ \ \ \ \ \  E_{1\alpha\gamma}E_{3\mu}E_{5\alpha\beta\gamma} = 1,\cr
  E_{2\beta\gamma}^2E_{3\mu} &= -4,\ \ \ \ \ \ E_{2\beta\gamma}E_{3\mu}^2 = 4,\ \ \ \ \ \ E_{3\mu}^2E_{4\beta\gamma} =4,\cr
  E_{3\mu}^2E_{6\alpha\beta'} &= -2,\ \ \ \ \ \ E_{3\mu}E_{4\beta\gamma}^2 = -4,\ \ \ \ \ \ E_{3\mu}E_{4\beta\gamma}E_{6\alpha\beta} = 1,\cr
  E_{3\mu}E_{5\alpha\beta\gamma}^2 &=-2,\ \ \ \ \ \
E_{3\mu}E_{5\alpha\beta\gamma}
E_{6\alpha\beta} =1,\ \ \ \ \ \ E_{3\mu}E_{6\alpha\beta'}^2 = -1\ .}}
Furthermore, the orientifold action changes 
the linear relations \first\ between the divisors $D_i$ and $R_i$: 
\eqn\globalreltwofour{\eqalign{
  R_1&=2\,D_{1,\alpha}+\h\sum_{\gamma=1}^3
E_{1,\alpha\gamma}+\h\sum_{\beta,\gamma=1,2} 
E_{5,\alpha\beta\gamma}+\h\sum_{\beta=1}^3 E_{6,\alpha\beta}\ \ \ ,\ \ \ \alpha=1,\ldots,4\ ,\cr
  R_2&=4\,D_{2,\beta}+\sum_{\gamma=1,2}\lf(\h\ 
E_{2,\beta\gamma}+2\ E_{3,\beta\gamma}+\fc{3}{2}\
E_{4,\beta\gamma}\ri)+\h\sum_{\alpha=1}^4
\sum_{\gamma=1,2}E_{5,\alpha\beta\gamma}\cr
     &    +\sum_{\alpha=1}^4E_{6,\alpha\beta}+E_{3,\mu}\ \ \ ,\ \ \ 
(\beta,\mu)\in\{\ (1,3)\ ,\ (2,6)\ \}\ ,\cr
  R_2&=2\,D_{2,3}+\h\sum_{\alpha=1}^4E_{6,\alpha3}+\h\sum_{\mu=7}^{10}
E_{3,\mu}\ ,}}
$$\eqalign{
  R_3&=4\,D_{3,\gamma}+\sum_{\alpha=1}^4E_{1,\alpha\gamma}+\sum_{\beta=1,2}
\lf(\fc{3}{2}\ E_{2,\beta\gamma}
+2\ E_{3,\beta\gamma}+\h\ E_{4,\beta\gamma}\ri)+\h
\sum_{\alpha=1}^4\sum_{\beta=1,2}E_{5,\alpha\beta\gamma}\cr
     &+E_{3,\mu}\ \ \ ,\ \ \ (\gamma,\mu)\in\{\ (1,7)\ ,\ (2,8)\ \}\ ,\cr
  R_3&=2\,D_{3,3}+\h\sum_{\alpha=1}^4E_{1,\alpha3}+\h\sum_{\mu=3,6,9,10} E_{3,\mu}\ .}$$
With the intersection numbers \ringtwofourA, \ringtwofourB\ 
and the relation \divvol, the divisor volumes $\Vol(E)$ and $Vol(D)$ 
of the $68$ divisors \DIVStwofour\ and
\ADDDD\ may be calculated. Since the expressions are rather long we do not display them here.

In the superpotential \SUP, the $68$ divisors split into two sets of contributing divisors $\Dc_i$:
On the 14 divisors $\Dc_{D7}=\{D_{1,\alpha},\ D_{2,\beta},\ D_{3,\gamma},\ E_{3,1},\ E_{3,2},\ 
E_{3,4},\ E_{3,5}\}$, a stack of 
one $O7$--plane and eight $D7$--branes is wrapped. Gaugino
condensation takes place in the pure $SO(8)$ gauge theory. Therefore
we have again $a_j=-\fc{2\pi}{6}$ for the set $\Dc_{D7}$ of divisors contributing in \SUP.
On the other hand, Euclidean $D3$--branes can be wrapped on the divisors in the set 
$\Dc_{D3}=\{E_{3,3},E_{3,6},E_{3,7},E_{3,8},E_{3,9},E_{3,10}, E_{1,\al\gamma},E_{6,\al\beta},
E_{2,\beta\gamma},E_{4,\beta\gamma},E_{5,\al\beta\gamma}\}$
of the $54$ remaining divisors. Since all $D$ and $E$ divisors intersect one of the divisors
carrying an $O7$--plane in at least one complex dimension, the condition $\chi_{D3}=1$ for
a non--vanishing instanton contribution in the superpotential \SUP\ is always met for
the set $\Dc_{D3}$ of divisors (\cf Section 4).
In total we obtain for full  the superpotential \SUP:
\eqn\SUPreads{
W=W_0(S,U)+\sum_{\Dc_i\in \Dc_{D7}} e^{-2\pi\ \fc{\Vol(\Dc_i)}{6}}+\sum_{\Dc_i\in \Dc_{D3}}
e^{-2\pi\ \Vol(\Dc_i)}\ .}
Now we are ready to stabilize all $61$ K\"ahler moduli $r_i,\ t_{1,\al\gamma},\ 
t_{2,\beta\gamma},\ t_{3,\mu},\ t_{4,\beta\gamma},\ t_{5,\al\beta\gamma},\ t_{6,\al\beta}$.
The $D3$--brane charge $Q_{3,tot}=-26$ is completely cancelled  by the $3$--form flux $G_3$,
given in \eqq \GdreiZzweiZvier.
In fact, in the previous subsection we have presented critical points for
the dilaton and complex structure moduli corresponding to a set of flux solutions, 
with $N_{flux}=52$. 
For a $\tilde W_0=-0.3104$ corresponding to the critical points of Table 8
we find the following $68$ divisor volumes (measured in string units):
\eqn\findDIVi{\eqalign{
\Vol(D_{1,\alpha})&=14.00 \ \ \ ,\ \ \  \Vol(D_{2,3})=5.543 \ \ \ ,\
\ \ \Vol(D_{3,3})=5.60\ ,\cr
\Vol(D_{2,\gamma})&=10.30\ \ \ ,\ \ \ \Vol(D_{3,\gamma})=11.07   \ \ ,\ \ \gamma=1,2\ ,\cr
\Vol(E_{5,\alpha\beta\gamma})&=1.30  \ \ \ ,\ \ \  \Vol(E_{4,\beta\gamma})=1.59\ ,\cr
\Vol(E_{2,\beta\gamma})&=3.30 \ \ ,\ \ \al=1,\ldots,4\ ,\ \beta,\gamma=1,2\ ,\cr
\Vol(E_{1,\alpha\gamma})&=9.82 \ \ \ ,\ \ \ \Vol(E_{6,\alpha\gamma})=14.72 \ \ ,\ \ 
\al=1,\ldots,4\ ,\ \gamma=1,2\ ,\cr
\Vol(E_{1,\alpha3})&=15.23\ \ \ ,\ \ \ \Vol(E_{6,\alpha3})=21.62 \ \ ,\
\ \al=1,\ldots,4\ ,\cr
\Vol(E_{3,\mu})&=8.06\ \ ,\ \ \mu=1,2,4,5\ \ \ ,\ \ \
\Vol(E_{3,\mu})=27.15\ \ ,\ \ \mu=3,6\ ,\cr
\Vol(E_{3,\mu})&=19.26\ \ ,\ \ \mu=7,8\ \ \ ,\ \ \ \Vol(E_{3,\mu})=27.00\ \ ,\ \ \mu=9,10\ .}}
corresponding to the sizes of the $61$  K\"ahler moduli:
\eqn\findDIVI{\eqalign{
r_1&=7.826 \ \ \ ,\ \ \  r_2 =5.410 \ \ \ ,\ \ \ r_3 =4.593\ ,\cr
t_{5,\alpha\beta\gamma}&=0.770  \ \ \ ,\ \ \  t_{4,\beta\gamma}=0.107\ ,\cr
t_{2,\beta\gamma}&=0.239 \ \ ,\ \ \al=1,\ldots,4\ ,\ \beta,\gamma=1,2\ ,\cr
t_{1,\alpha\gamma}&=0.858 \ \ \ ,\ \ \ t_{6,\alpha\gamma}=1.860  \ \ ,\ \ 
\al=1,\ldots,4\ ,\ \gamma=1,2\ ,\cr
t_{1,\alpha3}&=0.686 \ \ \ ,\ \ \ t_{6,\alpha3}=1.320 \ \ ,\ \ \al=1,\ldots,4\ ,\cr
t_{3,\mu}&=0.037\ \ ,\ \ \mu=1,2,4,5\ \ \ ,\ \ \ t_{3,\mu}=0.569\ \
,\ \ \mu=3,6\ ,\cr
t_{3,\mu}&=0.302\ \ ,\ \ \mu=7,8\ \ \ ,\ \ \ t_{3,\mu}=0.413\ \ ,\ \ \mu=9,10\ .}}
The divisor volumes give rise to the total volume $\Vol(X_6)=229.22$. This is large enough,
that one--loop corrections to the K\"ahler potential are suppressed.
Furthermore, the divisor volumes \findDIVi\ are large enough to suppress higher order
instanton effects in \SUPreads, since \eg $e^{-2\pi
\Vol(E_{5,\al\beta\gamma})}\sim 0.0002$ 
and even smaller for the other divisors.
To this end, for the $\IZ_2\times\IZ_4$ orbifold we have stabilized all
$63$ moduli fields.

\subsec{Orbifolds without complex structure moduli}

It has already been discussed in \doubref\ChoiSX\LRSS\ and thoroughly
in Subsection 2.4. that \tb orientifolds of toroidal 
$\IZ_N$ and $\IZ_N\times \IZ_M$ orbifolds without complex structure are not 
suitable candidates for a KKLT scenario. This investigation has been limited to the singular orbifold case. According to Table 3, this concerns the following orbifolds:
$\IZ_3,\IZ_7,\IZ_3\times \IZ_3$,\ $\IZ_4\times \IZ_4,\ 
\IZ_6\times \IZ_6,\ \IZ_2\times \IZ_{6'}$ and $\IZ_{8-I}$ on the $SU(4)^2$ lattice.
In this subsection we shall extend this discussion to resolved orbifolds.
We want to apply our stability criteria, derived in  
Subsection 2.3, to check whether those orbifolds with the generic superpotential \SUPP\ may
be suitable for a stable uplift.

We consider an orientifold  compactification on a resolved orbifold $X_6$ 
with $n:=\n$ K\"ahler moduli (and $h_{(2,1)}^{(-)}(X_6)=0$). 
There are $n_1$ untwisted moduli $r^i$ and 
$n_2$ blow--up moduli $t^\al$, with $n=n_1+n_2$.
We may introduce the K\"ahler form $\Jc$:
\eqn\orbJ{
\Jc=\sum_{i=1}^{n_1}r^i\ u_i-\sum_{\al=1}^{n_2} t^\al\ v_\al\ ,}
with the K\"ahler coordinates $r^i>0$ and $t^\al>0$. The volume of $X_6$ is given by
$${\rm Vol}(X_6)=\fc{1}{6}\ \int_{X_6}\Jc\wedge\Jc\wedge\Jc\ .$$
We consider the set of $n$ divisors $\tilde \Dc_i\in \lf\{\bigcup_{i=1}^{n_1}R_i\ ,\ 
\bigcup_{\al=1}^{n_2}\tilde E_\alpha\ri\}$, which are
Poincar\'e dual to the classes $\tilde \omega_i \in \lf\{\bigcup_{i=1}^{n_1}u_i\ ,\ 
\bigcup_{\al=1}^{n_2} v_\alpha\ri\}$. Their volumes
are given as usual by:
\eqn\Holdd{\eqalign{
V_{r^i}&=\fc{3}{2}\ \fc{\p}{\p r^i}\ {\rm Vol}(X_6)\ \ \ ,\ \ \ i=1,\ldots,n_1\cr
V_{t^\alpha}&=\fc{3}{2}\ \fc{\p}{\p t^\alpha}\ {\rm Vol}(X_6)\ \ \ ,\ \ \ \al=1,\ldots,n_2\ .}}
Following the discussion in Subsection 2.3, the divisors $\tilde\Dc_i$
do not yet necessarily represent contributing divisors $\Dc_i$ in \SUPP.
However, for resolved orbifolds $X_6$, the latter are generically
related to the $n$ original divisors $\tilde\Dc_i$ through linear combinations of the form
\eqn\Holds{\eqalign{
D_i&=\lambda_i\ R_i+\sum_{\alpha=1}^{n_2} \mu_{i\alpha}\ \tilde E_\alpha\ \ \ ,
\ \ \ i=1,\ldots,n_1\ ,\cr
E_\alpha&=-\rho_\alpha\tilde E_\al\ \ \ ,\ \ \ \al=1,\ldots,n_2\ ,}}
with some positive rational  numbers $\lambda_i,\mu_{i\al},\rho_\al\in \IQ^+$.
Following \eqq \divrel, the invertible $n \times n$--matrix $N$ assumes the form
\eqn\integerN{
N=\pmatrix{\lambda&\mu\cr
0& -\rho}\ ,}
with the submatrices
$$\lambda=\pmatrix{\lambda_1&\     &\  \cr  
                          \ &\ddots &\  \cr  
                          \ &\     &\lambda_{n_1}}\ \ \ ,\ \ \ 
\mu=\pmatrix{\mu_{11}&\ldots &\mu_{1n_2}\cr
             \vdots  &\ &\vdots\cr
             \mu_{n_11}&\ldots&\mu_{n_1n_2}}\ \ \ ,\ \ \ 
\rho=\pmatrix{\rho_1&\     &\  \cr  
                          \ &\ddots &\  \cr  
                          \ &\     &\rho_{n_2}}\ .$$
According to \eqq \givennewcoordinates, the $n$ new K\"ahler coordinates corresponding to the 
contributing divisors \Holds\ are related to the original orbifold coordinates $r^i,t^\al$ 
by the matrix $(N^t)^{-1}$, \ie:
\eqn\integerN{
\pmatrix{\lambda^{-1}&0 \cr
\ &\ \cr
\rho^{-1}\ \mu^t\  \lambda^{-1}& -\rho^{-1}}\ \pmatrix{r^1 \cr
                                                      \vdots \cr
                                                      r^{n_1}\cr
                                                      t^1\cr
                                                      \vdots\cr
                                                      t^{n_2}}\ .}
Note, that the matrices $\lambda^{-1}$ and $\rho^{-1}\ \mu^t\  \lambda^{-1}$ have only 
positive entries and the negative $n_2\times n_2$--matrix $-\rho^{-1}$ only acts on the 
blow--up moduli $t^\al$, which generically are very small.
Hence  all the $n$ new K\"ahler coordinates, corresponding to the divisors \Holds,
are positive, \ie the condition \Boxi\ is fulfilled.

For concreteness, let us discuss an example.
The (symmetrically) resolved $\IZ_2\times\IZ_2$ orientifold has $3$ untwisted K\"ahler moduli
corresponding to the three radii $r^i$ of the six--torus $T^6=(T^2)^3$
and $48$ twisted K\"ahler moduli $t^i$ describing the blow--up modes or exceptional divisors 
of the orbifold singularities. In the limit of $r_i=r$ and $t_i=t$, 
the volume of this orientifold becomes \DenefMM
\eqn\volDD{
Vol(X_6)=r^3-24\ r\ t^2+48\ t^3\ .}
{From} $\Kc=6Vol(X_6)$ the intersection numbers $\tilde\Kc_{ijk}$ may be read off
\eqn\interASD{
\tilde\Kc_{rrr}=6\ \ \ ,\ \ \ \tilde\Kc_{rtt}=-48\ \ \ ,\ \ \ \tilde\Kc_{ttt}=288\ ,}
with the intersection matrix
\eqn\interASDI{
\tilde\Kc_{ij}=6\ \pmatrix{r&-8\ t\cr -8\ t&48\ t-8\ r}\ .}
With the relation $V_{\tilde \Dc_i}=\fc{3}{4}\tilde\Kc_{ijk} t^jt^k$ we calculate
the divisor volumes $V_{\tilde \Dc_i}$ of the divisors $\tilde \Dc_i$ 
corresponding to the K\"ahler moduli $\tilde t^1=r$ and
$\tilde t^2=t$:
\eqn\DIVfour{
V_{\tilde \Dc_1}=\fc{9}{2}\ (r^2-8t^2)\ \ \ ,\ \ \ V_{\tilde \Dc_2}=72\ t\ (3t-r)\ .}
The divisors $\Dc_i$, which contribute to the superpotential \SUPP, are constructed from the
divisors $\tilde \Dc_i$ through a linear combination \divrel.
The volume of the right  divisors $\Dc_i$, representing also the real parts of the 
holomorphic K\"ahler moduli $T^1,T^2$ become \DenefMM:
\eqn\DIVfourr{\eqalign{
\re(T^1)&:=V_{\Dc_1}=\fc{1}{3}\ V_{\tilde \Dc_1}+\fc{8}{48}\ V_{\tilde \Dc_2}=\fc{3}{2}\ 
(r-4t)^2\ ,\cr
\re(T^2)&:=V_{\Dc_2}=-\fc{1}{48}\ V_{\tilde \Dc_2}=\fc{3}{2}\ t\ (r-3t)\ .}}
In terms of the latter, the K\"ahler potential $K_{KM}=-2\ln Vol(X_6)$ for the K\"ahler
moduli $T^1,T^2$ becomes:
\eqn\Kaehlerres{\eqalign{
K&=2\ln(T^2+\ov T^2)-\ln\lf\{(T^1+\ov T^1)+ 2\ (T^2+\ov T^2)\ri.\cr
&\lf.-(T^1+\ov T^1)^{1/2}\ [T^1+\ov T^1+4\ (T^2+\ov T^2)]^{1/2}\ri\}\cr
&-2\ln\lf\{(T^1+\ov T^1)^2+16\ (T^1+\ov T^1)\ (T^2+\ov T^2)+32\ 
(T^2+\ov T^2)^2\ri.\cr
&+(T^1+\ov T^1)^{3/2}\ [(T^1+\ov T^1)+4\ (T^2+\ov T^2)]^{1/2}\cr
&\lf.+8\ (T^2+\ov T^2)\ (T^1+\ov T^1)^{1/2}\ [(T^1+\ov T^1)+4\ 
(T^2+\ov T^2)]^{1/2}\ri\}\ .}}
The matrix $N$, defined in \divrel, is:
\eqn\matrixN{
N=\pmatrix{\fc{1}{3}& \fc{1}{6}\cr 0&-\fc{1}{48}}\ .}
According to \givennewcoordinates\ the K\"ahler coordinates $t^i$ related to the contributing 
divisors $\Dc_i$ are\foot{According to \Useful\ these coordinates may be read off from the 
first derivatives of the K\"ahler potential $K$:
\eqn\sincei{
t^1=-3V\ K_{T^1}=3\ r\ \ \ ,\ \ \ t^2=-3V\ K_{T^2}=24\ (r-2t)\ .}}:
\eqn\coordASD{
t^1=3\ r>0\ \ \ ,\ \ \ t^2=24\ (r-2t)>0\ .}
The positivity properties follow from the condition, that $r,t$ lie in the K\"ahler cone
$r>4t$ \DenefMM.
Expressed in terms of the coordinates $t^i$, the volume $Vol(X_6)$ reads:
\eqn\Volreads{
\Kc=\fc{2}{9}\ (t^1)^3-\fc{1}{6}\ (t^1)^2\ t^2+\fc{1}{24}\ t^1\ (t^2)^2-\fc{1}{384}\ (t^2)^3\ .}
Finally, the intersection form $\Kc_{ij}=N\tilde \Kc_{ij} N^t$ becomes:
\eqn\newinterASD{
\Kc_{ij}=\pmatrix{\fc{2}{9} t^1-\fc{1}{18}\ t^2& -\fc{1}{18} t^1+\fc{1}{72}\ t^2\cr
\ &\ \cr
-\fc{1}{18} t^1+\fc{1}{72}\ t^2& \fc{1}{72}\ t^1-\fc{1}{384} t^2}\ .}
Now we may apply our condition \Boxi\ to verify whether the ansatz \SUPP\
with the contributing divisors $\Dc_i$ may lead to a stable uplift.
The criterion \Boxi\ is met, since $t^1>0$ and $t^2>0$. Hence, 
no uplift is possible with the ansatz \SUPP.
This means that a stable minimum may be only found in the $\IZ_2\times\IZ_2$ orientifold
if one also includes the minimization w.r.t. the complex structure moduli \DenefMM.

Similarly, the condition \Boxi\ may be verified for the orbifolds $\IZ_3,\IZ_7, \IZ_3\times
\IZ_3,\ \IZ_4\times \IZ_4$,\  
$\IZ_6\times \IZ_6,\ \IZ_2\times \IZ_{6'}$, and $\IZ_8-I$ with $SU(4)^2$
lattice to conclude, that these manifolds are not appropriate for a KKLT scenario.

\newsec{Concluding remarks}

In this article we have discussed  moduli stabilization  in \tb CY orientifold 
compactifications within the approach of the KKLT scenario. In Section 2 we have
derived general conditions for a stable uplift 
following from the $F$--flatness conditions \adsm.
We have investigated the stability criteria of \tb CY orientifold 
compactifications without complex structure moduli, \ie $h_{(2,1)}^{(-)}(X_6)=0$.
For this class of compactifications, we have derived general criteria on the K\"ahler potential,
\cf \eqqs \BoxI, \BoxII, \Boxi\ or \Boxii.
If one of these criteria holds, the axionic mass matrix is not positive definite and 
no stable uplift is possible. In particular, the latter conditions are met 
in the \tb orientifolds of toroidal orbifold compactifications without complex structure moduli: 
$\IZ_3,\IZ_7, \IZ_3\times \IZ_3$,\ $\IZ_4\times \IZ_4,\ \IZ_6\times
\IZ_6,\ \IZ_2\times \IZ_{6'}$, and $\IZ_8-I$ with $SU(4)^2$ lattice.
This is true both at the orbifold point and after the resolution.
Hence, no stable uplift is possible for those orbifolds, \cf Subsection 5.5.
In fact, all the existing \tb orientifolds without complex structure moduli fulfill
one of the conditions \eqqs \BoxI, \BoxII, \Boxi\ or \Boxii\ and therefore compactifications
without complex structure moduli generically seem not to allow a stable uplift. 
Furthermore, in Section 3 we have presented a mechanism to stabilize K\"ahler moduli from the
cohomology $H^{(2)}_-(X_6)$ by analyzing the calibration conditions of $4$--cycles.
Equipped with these results we accomplished  to fix all moduli in some examples of resolved 
orbifolds: 
$\IZ_{6-II}$ on the root lattice of $SU(2)\times SU(6)$
and $\IZ_2\times \IZ_4$ on $SU(2)^2\times SO(5)^2$: We have stabilized 
for the $\IZ_{6-II}$--orbifold all its $27$ moduli fields, \cf Table 4
and \findDIVV\ and for the $\IZ_2\times \IZ_4$ orbifold all 
its $63$ moduli fields, \cf Table 6 and \findDIVI.

In most of the existing literature on flux compactifications, one works in the lowest 
$\ap$--expansion at string tree--level, \ie in the supergravity approximation. Since at this order
in $\ap$ and $g_{\rm string}$ the theory has a no--scale structure, 
which does not fix all K\"ahler moduli, one adds 
some effects which eventually allow the stabilization of all moduli fields. 
As proposed by KKLT \KKLT, one promising possibility is to consider the racetrack 
superpotential \SUP, which we also have used throughout this article.
The  critical points found in Subsection 5.4  
are inert against corrections in $\ap$ and $g_{\rm string}$, since the coupling constant 
$g_{\rm string}$ is small and the volume $V$ of the compactification manifold is large.
However, to discuss also other choices of  minima one must go beyond this approximation 
and include corrections in both $\ap$ and $g_{\rm string}$. 
In particular, there are both perturbative corrections to the K\"ahler potential $K$ to all
orders in $g_{\rm string}$ and 
world--sheet as well as  space--time instanton corrections to the K\"ahler potential $K$
in N=1 CY orientifolds. It is certainly very important to gain control over these 
corrections.

The coefficient $\gamma_j(S,U)$ in the non--perturbative superpotential \SUP\
accounts for gaugino condensation or $D3$--instanton effects.
Generically, this weight factor depends both on the dilaton $S$ 
and the complex structure moduli $U^\lambda$.
Due to the non--renormalization of the gauge kinetic function beyond one--loop, 
for gaugino condensation this dependence is fairly well under control perturbatively, \cf \eqq \ext.
On the other hand, for $D3$--instantons the factor $\gamma_j(S,U)$ represents the one--loop 
determinant of the instanton solution. The latter is hard to compute directly,  
except (in)directly in $F$-- or $M$--theory \WittenBN\ or through some duality arguments 
\doubref\MB\Aspinwall.
In Section 4 we have presented general results (\cf Table 2), 
under which conditions this coefficient $\gamma$ is
non--vanishing in \tb CY orientifolds. However, it is certainly important
to directly calculate the dilaton and complex structure modulus dependence of $\gamma$
by means of an instanton calculation.

We have obtained a fairly complete picture of the critical points of \tb orientifolds of
resolved orbifolds. Indeed, as Tables 6 and 8 show, for one orbifold 
there is a huge number of vacua with the same physical quantities
thus giving rise to a landscape of supersymmetric vacuum solutions in the flux space.
Throughout this article, the fixing of open string moduli is not addressed.
This is legitimate as we only discuss $D3$--branes wrapping internal $4$--cycles and
no space--time filling $D3$--branes. The complete tadpole \totaltadpole\ originating from the $RR$ $4$--form 
is cancelled by curvature and flux.  More general setups would also allow space--time filling 
$D3$--branes and $D7$--branes away from the orientifold planes.
It has been shown in \doubref\LRSi\IBANEZ, 
that even an $ISD$ $3$--form flux implies stabilization of the
$D7$--brane positions and soft--masses for corresponding 
the open string moduli. Certainly, a thorough discussion
of the stabilization of open string moduli would enrich the present picture of the string 
landscape \FM.

\vskip20pt

\centerline{\noindent{\bf Acknowledgments} }

We are indebted to Hans Jockers and Peter Mayr for many valuable
discussions. Moreover we wish to thank
Thomas Grimm, Renata Kallosh, and Alexander Schmidt for useful discussions.

This work is supported in part by the
Deutsche Forschungsgemeinschaft as well as by the
EU-RTN network {\sl Constituents, Fundamental Forces and Symmetries
of the Universe} (MRTN-CT-2004-005104) and is a collaboration between the contractors 
Ludwig--Maximilians 
University at Munich and Universita di Torino to which the Universita 
del Piemonte Orientale, Alessandria, is associated. 

S.R. and W.S. thank the university of Munich for hospitality. The work of E. S. is supported 
by the Marie Curie Grant MERG--CT--2004--006374. E.S. thanks the Erwin Schroedinger Institute 
in Vienna, the Simons Workshop in Mathematics and Physics at Stony Brook, the Arnold Sommerfeld 
Center for Theoretical Physics and the Ludwig--Maximilians University at Munich for hospitality.

\vskip20pt
\break

\appendix\appA{Complex structures of $\IZ_N$-- and $\IZ_N\times\IZ_M$--orbifolds}

In this appendix we introduce a complex basis and complex structures
for the orbifolds $\IZ_{6-II}$ on $SU(2)\times SU(6)$, $\IZ_2\times \IZ_4$ on 
$SU(2)^2\times SO(5)^2$ and $\IZ_4$ on $SU(4)^2$.

\ \br
$\underline{(i)\ \ \IZ_{6-II}-{\rm orbifold\ with\ } SU(2)\times SU(6)\ {\rm lattice}}$
\ \br
\ \br
We introduce the complex coordinates and complex structures \LRSS:
\eqn\worki{\eqalign{
dz^1&=dx^1+e^{2\pi i/6} dx^2+e^{2\pi i/3} dx^3-dx^4+e^{-2\pi i/3}dx^5\ ,\cr
dz^2&=dx^1+e^{2\pi i/3} dx^2+e^{-2\pi i/3}dx^3+dx^4+e^{2\pi i/3}dx^5\ ,\cr
dz^3&=
\fc{1}{\sqrt{12}}\left[\ \fc{1}{3}\ (dx^1-dx^2+dx^3-dx^4+dx^5)-iU\ dx^6\ \right]\ .}}

\eqn\formdef{\eqalign{
\omega_{A0}&=dz^1\wedge d z^2\wedge dz^3 \ ,\cr
\omega_{B0}&=d\bar z^1\wedge d\bar z^2\wedge d\bar z^3 \ ,\cr
\omega_{A3}&=dz^1\wedge dz^2\wedge d\bar z^3 \ ,\cr
\omega_{B3}&=d\bar z^1\wedge d\bar z^2\wedge dz^3 \ .
}}

\eqn\orthonorm{\eqalign{
\int \omega_{A0}\wedge\omega_{B0}&=-i (U+\bar U)\ , \cr
\int \omega_{A3}\wedge\omega_{B3}&=-i (U+\bar U)\ .
}}

\ \br
$\underline{(ii)\ \ 
\IZ_2\times \IZ_4-{\rm orbifold\ with\ } SU(2)^2\times SO(5)^2\ {\rm lattice}:}$
\ \br
\ \br
We introduce the complex coordinates and complex structures:
\eqn\complformen{\eqalign{
dz^1&=dx^1+U\ dx^2 \ , \cr
dz^2&=dx^3-\fc{1}{2}\ (1-i)\ dx^4 \ , \cr
dz^3&=dx^5-\fc{1}{2}\ (1+i)\ dx^6 \ .}}

\ \br
$\underline{(iii)\ \ \IZ_4-{\rm orbifold\ with\ } SU(4)^2\ {\rm lattice}}$
\ \br
\ \br
We introduce the complex coordinates and complex structures:
\eqn\compl{\eqalign{
dz^1&=dx^1-dx^3+i\ dx^2\ ,\cr
dz^2&=dx^4-dx^6+i\ dx^5\ ,\cr
dz^3&=dx^1-dx^2+dx^3+U\ (dx^4-dx^5+dx^6)\ .}}
In the complex basis we have\foot{Note that for a $2$--torus
we have 
$n^tgn=2 \ \re T\ |N|^2$, with 
$N=\fc{1}{\sqrt{2 U_2}}\ (n^1+U\ n^2)$. Sometimes one also introduces $\tilde 
N=n^1+U\ n^2$ such that 
$n^tgn=2 \ \re T\ \lf(\fc{1}{\sqrt{2 U_2}}\ri)^2\ |\tilde N|^2$. Essentially, 
the two coordinates $N$ and $\tilde N$ only affect the normalization of the 
volume ($U_2$ or $1/2$)
of the complex torus with the corners $(0,1,U,1+U)$ or $\fc{1}{\sqrt{2U_2}}\ 
(0,1,U,1+U)$, respectively. At any rate the coordinate $N$ represents an unimodular
transformation from the integer space $n^i$ to the complex space, \ie it preserves
the volume of the integer lattice and there is an one to one map from $n^i$ to $N$.}
\MS:
\eqn\quad{\eqalign{
n^tgn&=2\ \re T^1\ |N^1|^2+2\ \re T^2\ |N^2|^2+2\ \re T^3\ |N^3|^2\cr
&+2\ (\re T^{12}+i\ \re T^{21}) \ N^1\ov N^2+2\ (\re T^{12}-i\ \re T^{21}) \ \ov N^1\ N^2\ ,}}
with the complex (normalized) lattice vectors
\eqn\Compl{\eqalign{
N^1&=\fc{1}{\sqrt 2}\ (n^1-n^3+i\ n^2)\ ,\cr
N^2&=\fc{1}{\sqrt 2}\ (n^4-n^6+i\ n^5)\ ,\cr
N^3&=\fc{1}{\sqrt{2 U_2}}\ \h\ [\ n^1-n^2+n^3+U\ (n^4-n^5+n^6)\ ]}}
following from \compl.
The definition of the complex 
numbers $N^i$ represents  an unimodular transformation, \ie $det Y=1$, on the 
integers $n^i$. Hence it preserves the volume of the integer space $n^i$ when 
moving to the complex space $N^i$. Note the additional factor of $1/2$ for $N^3$. 
The K\"ahler moduli follow
\eqn\Moduli{\eqalign{
T^1&=R^2+x+i\ \alpha\ ,\cr
T^2&=S^2+y+i\ \beta\ ,\cr
T^3&=2\ \sqrt{4 x y -(v+z)^2}+2\ i\ (\delta+\epsilon)\ ,\cr
T^{12}&=u+\h\ (v+z)+\fc{i}{2}\ (\delta-\epsilon)\ ,\cr
T^{21}&=\h\ (v-z)-i\ \gamma-\fc{i}{2}\ (\delta+\epsilon)\ ,}}
with 
$$(\det g)^{1/2}=\fc{1}{8}\ [(T^1+\ov T^1)\ (T^2+\ov T^2)-(T^{12}+\ov T^{12})^2-
(T^{21}+\ov T^{21})^2\ ]\ (T^3+\ov T^3)\ .$$ 
Furthermore, 
the complex structure \MS:
\eqn\CS{U=\fc{1}{2\ x}\ [\ v+z-i\ \sqrt{4 x y -(v+z)^2}\ ]\ .}
Following \LRSS\ we may introduce new K\"ahler moduli:
\eqn\newintro{\eqalign{
\tilde T^{12}&=T^{12}+i\ T^{21}=u+\h(1+i)\ v+\h(1-i)\ z+\gamma+\h(1+i)\ \delta+\h(1-i)\epsilon\ ,\cr
\tilde T^{21}&=T^{12}-i\ T^{21}=u+\h(1-i)\ v+\h(1+i)\ z-\gamma-\h(1-i)\ \delta-\h(1+i)\epsilon}}
such that
$$(\det g)^{1/2}=\fc{1}{8}\ [(T^1+\ov T^1)\ (T^2+\ov T^2)-
(\tilde T^{12}+\ov {\tilde T^{21}})\ (\tilde T^{21}+\ov {\tilde T^{12}})\ ]\ (T^3+\ov T^3)\ .$$ 
Clearly any transformation $n^i\lra n^i +1$ is a symmetry of the 
$T^6$ reflecting the periodicity $x^i\lra x^i +1$ of the lattice 
(in the lattice basis). It is not a priori guaranteed, that our 
three complex tori, introduced through the coordinates \Compl\ share this 
periodicity. The third torus lies within the full six--dimensional lattice
and has been defined such, that the integers $n^i$ are mapped one to one onto
the complex numbers $N^i$. 
By introducing new lattice vectors $\tilde n^i$ in \Compl\ we may disentangle 
the six windings $n^i$ 
such that our three complex tori become independent. We introduce:
\eqn\disentangle{\eqalign{
\tilde n^1&=n^1-n^3\ \ \ \ ,\ \ \ \tilde n^2=n^2\ ,\cr
\tilde n^4&=n^4-n^6\ \ \ \ ,\ \ \ \tilde n^5=n^5\ ,\cr
\tilde n^3&=\h\ (n^1+n^3-n^2)\ \ \ \ ,\ \ \ \tilde n^6=\h\ (n^4+n^6-n^5)\ .}}
Since $\det Y=1$, with $\tilde n^i=Y^i_j\ n^j$, the map \disentangle\ 
represents an one to one transformation from the integers $n^i$ to
the numbers $\tilde n^i$. The latter however may become half--integer. 
This way \Compl\ becomes:
\eqn\Compll{\eqalign{
\tilde N^1&=\fc{1}{\sqrt2}\ (\tilde n^1+i\ \tilde n^2)\ \ ,
\ \ \tilde n^1,\tilde n^2\in \IZ\ ,\cr
\tilde N^2&=\fc{1}{\sqrt2}\ (\tilde n^4+i\ \tilde n^5)\ \ ,
\ \ \tilde n^4,\tilde n^5\in \IZ\ ,\cr
\tilde N^3&=\fc{1}{\sqrt{2 U_2}}\ (\tilde n^3+U\ \tilde n^6)\ \ ,
\ \ \tilde n^3,\tilde n^6\in \h\IZ\ .}}
Now, the complex coordinates $\tilde N^i$
assume the expected form of three independent 
two--tori (\cf the above footnote). Note, that the property 
$\tilde n^3,\tilde n^6\in \h\IZ$ means, that some lattice points $n^i$ 
are mapped into the third torus rather than onto its corners.
E.g. the point $n^6=1$ becomes $\tilde n^6=\h$.
Nevertheless, the third torus with the complex structure $U$ 
is spanned by the points
$(\tilde n^3,\tilde n^6)=(0,0),\ (0,1),\ (0,1)$ and $(1,1)$, with
$\tilde n^1,\tilde n^2,\tilde n^4, \tilde n^5=0$.
This corresponds to the original lattice points $(n^1,n^2,n^3,n^4,n^5,n^6)=
(1,0,1,1,0,1)$. This way, the third torus is decoupled from the other two.

\eqn\workii{\eqalign{
dz^1&=dx^1-dx^3+i\ dx^2 \ ,\cr
dz^2&=dx^4-dx^6+i\ dx^5 \ ,\cr
dz^3&=\fc{1}{4}\lf[\ dx^1-dx^2+dx^3-i\ U\ \lf(dx^4-dx^5+dx^6\ri)\ \ri] \ .}}

\listrefs

\end